\documentclass[aps,pra,twocolumn,showpacs,superscriptaddress]{revtex4}
\usepackage{graphicx}
\usepackage{amsmath}
\usepackage{amssymb,wasysym}
\usepackage{amsfonts}
\usepackage{bm,verbatim}
\usepackage{color}

\newcommand{\bea}{\begin{eqnarray}}
\newcommand{\eea}{\end{eqnarray}}
\newcommand{\be}{\begin{equation}}
\newcommand{\ee}{\end{equation}}

\newcommand{\eee}{\mathbf{e}}
\newcommand{\nn}{\mathbf{n}}
\newcommand{\rr}{\mathbf{r}}
\newcommand{\kk}{\mathbf{k}}
\newcommand{\KK}{\mathbf{K}}

\newcommand{\uu}{\mathbf{u}}
\newcommand{\RR}{\mathbf{R}}

\newcounter{notekappa}
\newcounter{notepaszero}
\newcounter{notespm}
\newcounter{noteylm}
\newcounter{notesimon}

\DeclareMathOperator\ch{ch}
\DeclareMathOperator\sh{sh}
\DeclareMathOperator\thf{th}
\DeclareMathOperator\atan{atan}
\DeclareMathOperator\asin{asin}
\DeclareMathOperator{\Tr}{Tr}

\begin{document}

\title{Absence of a four-body Efimov effect in the $2+2$ fermionic problem}

\author{Shimpei Endo}
\email{shimpei.endo@lkb.ens.fr}
\affiliation{Laboratoire Kastler Brossel, ENS-PSL, CNRS, UPMC-Sorbonne Universit\'es, Coll\`ege de France, Paris, France}

\author{Yvan Castin}
\email{yvan.castin@lkb.ens.fr}
\affiliation{Laboratoire Kastler Brossel, ENS-PSL, CNRS, UPMC-Sorbonne Universit\'es, Coll\`ege de France, Paris, France}

\begin{abstract}
In the free three-dimensional space, we consider a pair of identical $\uparrow$ fermions of some species or in some internal state, and a pair of identical $\downarrow$ fermions of another species or in another state. There is a resonant $s$-wave interaction (that is of zero range and infinite scattering length) between fermions in different pairs, and no interaction within the same pair. We study whether this $2+2$ fermionic system can exhibit (as the $3+1$ fermionic system) a four-body Efimov effect in the absence of three-body Efimov effect, that is the mass ratio $\alpha$ between $\uparrow$ and $\downarrow$ fermions and its inverse are both smaller than 13.6069{\ldots}. For this purpose, we investigate scale invariant zero-energy solutions of the four-body Schr\"odinger equation, that is positively homogeneous functions of the coordinates of degree {$s-7/2$}, where $s$ is a generalized Efimov exponent {that becomes purely imaginary in the presence of a four-body Efimov effect.} Using rotational invariance in momentum space, it is found that the allowed values of $s$ are such that $M(s)$ has a zero eigenvalue; here the operator $M(s)$, that depends on the total angular momentum $\ell$, acts on functions of two real variables (the cosine of the angle between two wave vectors and the logarithm of the ratio of their moduli), and we write it explicitly in terms of an integral matrix kernel. We have performed a spectral analysis of $M(s)$, analytical and for an arbitrary imaginary $s$ for the continuous spectrum, numerical and limited to $s = 0$ and $\ell \le 12$ for the discrete spectrum. We conclude that no eigenvalue of $M(0)$ crosses zero over the mass ratio interval $\alpha \in [1, 13.6069\ldots]$, even if, in the parity sector $(-1)^{\ell}$, the continuous spectrum of $M(s)$ has everywhere a zero lower border. As a consequence, there is no possibility of a four-body Efimov effect for the 2+2 fermions. 

We also enunciated a conjecture for the fourth virial coefficient of the unitary spin-$1/2$ Fermi gas,
inspired from the known analytical form of the third cluster coefficient and involving the integral over the imaginary $s$-axis 
of $s$ times the logarithmic derivative of the determinant of $M(s)$ summed over all angular momenta.
The conjectured value is in contradiction with the experimental results.

\end{abstract}


\date{\today}

\maketitle


\section{Introduction and motivation}
\label{sec:intro_et_motiv}

In three-dimensional cold atomic gases, thanks to magnetic Feshbach resonances, it is now possible to induce resonant $s$-wave interactions between
the particles \cite{manips_resonance}. This means that the $s$-wave scattering length $a$ is in absolute value much larger than
the range (or the effective range) of the interaction. Essentially, one can assume that $1/a=0$, and since the de Broglie atomic wavelength
is also much larger than the range of the interaction, one can replace the interactions by scaling invariant Wigner-Bethe-Peierls
two-body contact conditions on the wavefunction \cite{WBP}: one realizes the long sought unitary limit.

Perhaps the most striking phenomenon that can take place in that regime is the Efimov effect, predicted for three particles with appropriate statistics
and mass ratios \cite{Efimov}. It corresponds to the occurrence of an infinite number of bound states, with an asymptotically geometric spectrum
close to the zero-energy accumulation point. The geometric part of the spectrum is characterized by a ratio, predicted by Efimov's zero-range theory,
and a global energy scale that depends on the microscopic details of the interaction. The mere existence of such an energy scale forces
us to supplement the two-body contact conditions by three-body ones, that involve a length scale, the so-called three-body parameter,
and that {break} the scale invariance at the three-body level. It is at this cost that the zero-range model becomes well defined and leads
to a self-adjoint Hamiltonian. The Efimov effect is now observed experimentally with cold atoms \cite{manips_Efimov}, which gives access to the
value of the three-body parameter \cite{Grimm3bp}.

A natural question is to know whether or not a four-body Efimov effect is possible {\cite{Amado,Gridnev}}, leading to an infinite, asymptotically geometric,
spectrum of tetramers, with an energy ratio predicted by a zero-range theory and a global energy scale fixed by a four-body parameter 
appearing in four-body contact conditions. It is now understood that a prerequisite to the four-body Efimov effect 
is the absence of three-body Efimov effect:  
it is indeed expected that the introduction of three-body contact conditions (in terms of the three-body parameter) imposed by the three-body
Efimov effect is sufficient to also
render the four-body problem well defined, that is without the need for a four-body parameter; as predicted in reference \cite{Amado}, 
no geometric sequence of tetramer states can then be found but,
as shown numerically for four bosons \cite{Deltuva}, sequences of four-body complex energy resonances are expected in general, with the same 
geometric ratio as the trimer Efimov spectrum (see \cite{Greene,Hammer} for early studies not accessing the imaginary part of the energy).
This prerequisite rules out systems with more than one boson \cite{Efimov} as possible candidates for a four-body Efimov effect, and suggests to use fermions
to counterbalance the Efimov effect by the Pauli exclusion principle, at least in three dimensions 
(what happens in lower dimensions {or with resonant interactions in other channels than the $s$-wave} is discussed in {\cite{NishidaTan,NishidaMorozSon}}).

Consider then the so-called $p+q$ fermionic problem: $p$ identical fermions of the same species or spin state resonantly interact
in free space with $q$ identical fermions of another species or spin state. It is assumed that there is no interaction between the identical fermions,
since they cannot scatter in the $s$ wave. It is convenient to adopt a pseudo-spin notation, with $\uparrow$ for the first species and $\downarrow$ for the second.
The two species have in general different masses $m_\uparrow$ and $m_\downarrow$, and the crucial idea is to use their mass ratio
as an adjustable parameter to search for the four-body Efimov effect without triggering the three-body one.

The $3+1$ or $\uparrow\uparrow\uparrow\downarrow$ resonant fermionic problem was investigated in reference \cite{CMP}.
A four-body Efimov effect was predicted for a mass ratio $13.384 < {m_\uparrow/m_\downarrow}<13.6069\ldots$.
Beyond $13.6069\ldots$ the three-body Efimov effect sets in as shown in \cite{Efimov,Petrov,Braaten}, which blocks the four-body Efimov effect as discussed above: 
apart from a finite number of tetramer states, one expects an infinite number of four-body resonances with the same geometric ratio as for the $2+1$
problem.

The main motivation of the present work is to determine the presence or the absence of a four-body Efimov effect 
in the $2+2$ or $\uparrow\uparrow\downarrow\downarrow$ fermionic problem.
To our knowledge, no general and rigorous answer was given to this problem.
{One may think attacking it with the Born-Oppenheimer approximation. We indeed expect (as for the three-body case) that the only possibility
for a four-body Efimov effect is to have a large mass imbalance between the two species, for example the $\uparrow$ fermions are much heavier than the
$\downarrow$ ones. It is found that, in presence of two $\uparrow$ fermions at fixed positions, 
there is a {\sl single} bound state for the $\downarrow$ particle, which creates an effective $\propto -\hbar^2/(m_\downarrow r^2)$ attraction between the $\uparrow$ fermions.
For a large enough $m_\uparrow/m_\downarrow$ mass ratio, this indeed beats the centrifugal barrier $\propto \hbar^2/(m_\uparrow r^2)$ between the $\uparrow$ particles (they are fermions and approach each other with a non-zero angular momentum),
which qualitatively explains the occurrence of a three-body Efimov effect in the $2+1$ problem, as pointed out in 1973 by Efimov \cite{Efimov}. 
However, as there is a {\sl single} bonding orbital, one 
cannot put a {\sl second} $\downarrow$ fermion in that orbital, but one can at best put one in the ground, zero-energy scattering state, which has two consequences:
(i) the Born-Oppenheimer attractive potential between the $\uparrow$ particles is not lowered by the second $\downarrow$ fermion, so no four-body Efimov effect is
predicted at a mass ratio strictly below the three-body Efimov effect threshold, and (ii) as emphasized in \cite{dimdimpeda}, the second $\downarrow$ fermion, being 
in a zero-energy eigenstate, does not have a fast motion as compared to the one of the heavy particles, which sheds doubts on the validity of the Born-Oppenheimer
approximation.
Alternatively, one may expect that this $2+2$ problem was already solved numerically in the literature;}
however no convincingly dense coverage of the mass ratio interval between $1$ and $13.6069\ldots$ {seems to be} available in the numerics \cite{Blume}
considering the narrowness of the above mentioned mass interval. To {obtain a firm answer to the question}, we generalize the method of reference 
{\cite{CMP,CRAS4corps}}, deriving from the 
zero-range model momentum space integral equations for the $2+2$ fermionic problem  at zero energy  (see also the most general formulation of reference
\cite{Ludogen}), and using rotational symmetry and scale invariance to reduce them to a numerically tractable form.

Another motivation is to pave the way for the calculation of the fourth virial coefficient of a two-component unitary Fermi gas:
this would make an interesting bridge between few-body and many-body physics.
For a unit mass ratio $m_\uparrow/m_\downarrow=1$, the value of this virial coefficient was already obtained experimentally from a measurement
of the equation of state of a gas of ultracold atoms \cite{Navon,Zwierlein}.
On the theory side, {there exist two main techniques. First, the diagrammatic technique, used exactly (all diagrams are kept) for the third virial coefficient 
\cite{Leyronas,Kaplan}, and approximately (only some diagrams are kept, those relevant in the perturbative regime of a large effective range or a small scattering length)
for the fourth virial coefficient \cite{JesperLevinsen} leading to a value different from but reasonably close to the experimental value. Second,}  
the harmonic regulator technique \cite{ComtetOuvry}, used with success
for the third virial coefficient \cite{Drummond,Blumeb3,CastinWerner,ChaoEndoCastin},
that requires to determine the spectrum of up to four particles in an isotropic harmonic trap. A first, brute force numerical solution
of this {trapped four-body} problem \cite{Blumeb4} was not able to recover {even the sign of} the experimental value.
In a more analytical way, this spectrum can be deduced from the solutions
of the zero-energy free space problem \cite{Tan,WernerCastinPRA}, due to the $\mathrm{SO(2,1)}$ dynamical symmetry of the unitary Fermi gas 
\cite{CastinCRAS,WernerCastinPRA,ZwergerLivre},
so that the four-body integral equations written here may also be useful for the solution of the virial problem.

Our article is organized as follows. In section \ref{sec:obtention_des_equations_integrales}
we derive the zero-energy momentum-space integral equations in general form. In section \ref{sec:vivent_les_symetries} 
we successively use the rotational
invariance, the scale invariance and the parity invariance to put the integral equations in a maximally reduced form. 
This reduced form, written in section \ref{subsec:forme_explicite}, 
exactly expresses the fact that some operator $M$, depending on the angular momentum $\ell$ and the scaling exponent $s$, has a zero eigenvalue, 
which motivates
its spectral analysis; it allows to show that two components of the continuous spectrum of $M$ can be expressed exactly in terms of the Efimov transcendental
functions appearing in the $\uparrow \uparrow \downarrow$ and $\uparrow \downarrow \downarrow$ three-body problems
(see section \ref{subsec:le3corpsretrouve}) 
and that there is a third, {unexpected} continuum due to a term with no equivalent in the $3+1$ problem
(see section \ref{subsec:der_dritte_continuum}).
The question of the existence of the four-body Efimov effect in the $2+2$ fermionic problem is the subject of 
section \ref{sec:quete}, whereas the secondary motivation of this work, i.e.\ the fourth virial coefficient of the spin $1/2$ unitary Fermi gas,
is relegated to the Appendix \ref{appen:conjecture}, where its expression in terms of the operator $M$ is conjectured from a transposition of the known
analytic expression of the third virial coefficient \cite{CastinWerner,ChaoEndoCastin}, and the conjectured value is compared to the experimental
\cite{Navon,Zwierlein} and theoretical {\cite{Blumeb4,JesperLevinsen}} values. We conclude in section \ref{sec:conclusion}.

\section{Derivation of the general four-body integral equations}
\label{sec:obtention_des_equations_integrales}

Particles 1 and 2, of positions $\rr_1$ and $\rr_2$, belong to species $\uparrow$.  Particles 3 and 4, of positions $\rr_3$ and $\rr_4$, belong to species $\downarrow$.
The four-body wavefunction $\psi$ is subjected to the usual Wigner-Bethe-Peierls contact conditions, for a zero-range interaction of $s$-wave scattering length $a$
between opposite-spin particles. For all $i\in \{1,2\}$ and all $j\in \{3,4\}$, when the distance $r_{ij}$ between particles $i$ and $j$
tends to zero, at fixed position $\RR_{ij}=(m_\uparrow \rr_i +m_\downarrow \rr_j)/(m_\uparrow+m_\downarrow)$ of their center of mass (different from the positions
of the remaining two particles), one imposes
\begin{multline}
\psi_{\uparrow\uparrow\downarrow\downarrow}(\rr_1,\rr_2,\rr_3,\rr_4)\underset{r_{ij}\to 0}{=} \left(\frac{1}{r_{ij}}-\frac{1}{a}\right) \frac{\mu_{\uparrow\downarrow}}{2\pi\hbar^2}\\
\times \mathcal{A}_{ij}((\rr_k-\RR_{ij})_{k\neq i,j})
+O(r_{ij})
\label{eq:BethePeierls}
\end{multline}
where the form of the regular part $\mathcal{A}_{ij}$ supposes that the center of mass of the four particles is at rest, and where $\mu_{\uparrow\downarrow}=m_\uparrow m_\downarrow/(m_\uparrow+
m_\downarrow)$ is the reduced mass of two opposite-spin particles.  Due to the fermionic antisymmetry, the regular parts are not independent functions:
\be
\mathcal{A}_{13}=\mathcal{A}_{24}=-\mathcal{A}_{14}=-\mathcal{A}_{23}\equiv \mathcal{A}
\ee
Schr\"odinger's equation at zero eigenenergy $E=0$, written in the language of distributions, is then
\begin{multline}
H \psi_{\uparrow\uparrow\downarrow\downarrow} = 
 \mathcal{A}(\rr_2-\RR_{13},\rr_4-\RR_{13}) \delta(\rr_1-\rr_3) \\
-\mathcal{A}(\rr_2-\RR_{14},\rr_3-\RR_{14}) \delta(\rr_1-\rr_4) \\
-\mathcal{A}(\rr_1-\RR_{23},\rr_4-\RR_{23}) \delta(\rr_2-\rr_3) \\
+\mathcal{A}(\rr_1-\RR_{24},\rr_3-\RR_{24}) \delta(\rr_2-\rr_4)
\label{eq:Hpsi}
\end{multline}
with the kinetic energy Hamiltonian
\be
H=\sum_{n=1}^{4} -\frac{\hbar^2}{2 m_n} \Delta_{\rr_n} 
\ee
and $\delta(\rr)$ is the Dirac distribution in three dimensions, {stemming from the identity $\Delta_{\rr} (1/r)=-4\pi \delta(\rr)$.}

We now go to momentum space and we take the Fourier transform of Schr\"odinger's equation. In the left-hand side, each Laplace operator gives rise
to a factor $-k_n^2$, where $\kk_n$ is the wave-vector of particle number $n$. In the right-hand side, one obtains for example for the first term:
\begin{multline}
\int \prod_{n=1}^{4} d^3r_n e^{-i\sum_{n=1}^{4} \kk_n\cdot\rr_n} \mathcal{A}(\rr_2-\RR_{13},\rr_4-\RR_{13}) \delta(\rr_1-\rr_3) \\
= (2\pi)^3 \tilde{\mathcal{A}}(\kk_2,\kk_4) \delta(\sum_{n=1}^{4} \kk_n)
\label{eq:un_example}
\end{multline}
where the tilde indicates the Fourier transform. Introducing the function $D\equiv  (2\pi)^3 \tilde{\mathcal{A}}$, we obtain the four-body momentum space ansatz
generalizing to the 2+2 fermionic problem the one of the 3+1 fermionic problem \cite{CMP,CRAS4corps}: 
\begin{multline}
\tilde{\psi}_{\uparrow\uparrow\downarrow\downarrow}(\kk_1,\kk_2,\kk_3,\kk_4) = \frac{\displaystyle\delta\left(\sum_{n=1}^{4} \kk_n\right)}{\displaystyle\sum_{n=1}^{4} \frac{\hbar^2 k_n^2}{2 m_n}}
[D(\kk_2,\kk_4)\\ -D(\kk_2,\kk_3)
-D(\kk_1,\kk_4)+D(\kk_1,\kk_3)]
\label{eq:intdep}
\end{multline}

The ansatz obeys fermionic antisymmetry and Schr\"odinger's equation, not yet the contact condition (\ref{eq:BethePeierls}),
{that it suffices to implement for $(i,j)=(1,3)$.} One thus takes the inverse Fourier transform of 
$\tilde{\psi}$ at $(\rr_1,\rr_2,\rr_3,\rr_4)$, with the parametrization:
\bea
\rr_1 &=&\RR_{13}+\frac{m_3}{m_1+m_3} \rr_{13} \\
\rr_3 &=&\RR_{13}-\frac{m_1}{m_1+m_3} \rr_{13} 
\eea
Only the contribution $\psi_{24}$ of $D(\kk_2,\kk_4)$ to $\psi$ diverges for $r_{13}\to 0$; in that inverse Fourier transform, 
we then take $\KK_{13}=\kk_1+\kk_3$, $\kk_{13}=\mu_{13}(\kk_1/m_1-\kk_3/m_3)$ and $\kk_2, \kk_4$ as integration variables
(clearly $\mu_{13}=\mu_{\uparrow\downarrow}$), 
{so that $\kk_1\cdot \rr_1+\kk_3\cdot \rr_3=\KK_{13}\cdot\RR_{13}+\kk_{13}\cdot\rr_{13}$}
{and $\displaystyle \frac{\hbar^2 k_1^2}{2 m_1} +\frac{\hbar^2 k_3^2}{2 m_3}=\frac{\hbar^2 k_{13}^2}{2\mu_{13}}+\frac{\hbar^2 K_{13}^2}{2(m_1+m_3)}$;}
integration over $\KK_{13}$ is straightforward, due to the momentum conservation, and integration over $\kk_{13}$ also can be done using
\be
u(\rr) = \int \frac{d^3k_{13}}{(2\pi)^3} \frac{e^{i\kk_{13}\cdot\rr}}{k_{13}^2+q_{13}^2} = \frac{e^{- q_{13}r}}{4\pi r}
\ee
One obtains
\begin{multline}
\psi_{24}(\rr_1,\rr_2,\rr_3,\rr_4)=
\int \frac{d^3k_2 d^3k_4}{(2\pi)^9}
{\frac{2\mu_{13}}{\hbar^2} u(r_{13})} \\
\times e^{i[\kk_2\cdot(\rr_2-\RR_{13})+\kk_4\cdot(\rr_4-\RR_{13})]} D(\kk_2,\kk_4)
\label{eq:psi24}
\end{multline}
with $q_{13}\geq 0$ such that
\be
\frac{\hbar^2 q_{13}^2}{2\mu_{13}} = \frac{\hbar^2 (\kk_2+\kk_4)^2}{2(m_1+m_3)} + \frac{\hbar^2 k_2^2}{2m_2} + \frac{\hbar^2 k_4^2}{2 m_4}
\label{eq:defq13}
\ee
Taking $r_{13}\to 0$ in $\psi_{24}$ is then elementary. In the contribution to $\psi$ of $D(\kk_2,\kk_3)$, $D(\kk_1,\kk_4)$ and $D(\kk_1,\kk_3)$, {noted as $\psi_{\neq 24}$}, one can directly take
$\rr_{13}=\mathbf{0}$. {Thanks to momentum conservation one can replace $\kk_1+\kk_3$ by $-(\kk_2+\kk_4)$ within the position-dependent
phase factor, so that the positions $\rr_2-\RR_{13}$ and $\rr_4-\RR_{13}$ appear as in Eq.~(\ref{eq:psi24}):}
{
\begin{multline}
\psi_{\neq 24}(\rr_1=\RR_{13},\rr_2,\rr_3=\RR_{13},\rr_4)= \\
\int \frac{d^3k_2 d^3k_4}{(2\pi)^9} e^{i[\kk_2\cdot(\rr_2-\RR_{13})+\kk_4\cdot(\rr_4-\RR_{13})]} 
\int\frac{d^3k_1 d^3k_3}{(2\pi)^3}\\
\frac{\displaystyle\delta\left(\sum_{n=1}^{4} \kk_n\right)}{\displaystyle\sum_{n=1}^{4} \frac{\hbar^2 k_n^2}{2 m_n}} [-D(\kk_2,\kk_3)-D(\kk_1,\kk_4)+D(\kk_1,\kk_3)]
\end{multline}
}
Finally the contact condition at the unitary limit, that is for $1/a=0$, leads to the following integral equation for $D$:
\begin{multline}
0=\frac{\mu_{\uparrow\downarrow}^{3/2}}{2\pi\hbar^2}\left[\frac{(\kk_2+\kk_4)^2}{m_\uparrow+m_\downarrow}
+\frac{k_2^2}{m_\uparrow}+\frac{k_4^2}{m_\downarrow}\right]^{1/2}
\!\!\!\!\! D(\kk_2,\kk_4) \\
+\!\int\!\!\frac{d^3k_1 d^3k_3}{(2\pi)^3} \frac{\displaystyle\delta\!\!\left(\sum_{n=1}^{4} \kk_n\!\!\right)}{\displaystyle\sum_{n=1}^{4} \frac{\hbar^2 k_n^2}{2 m_n}}
[D(\kk_2,\kk_3)+D(\kk_1,\kk_4)-D(\kk_1,\kk_3)]
\label{eq:equint}
\end{multline}
where the first term is simply $\displaystyle\frac{q_{13}\mu_{13}}{2\pi\hbar^2} D(\kk_2,\kk_4)$.
Contrarily to the $3+1$ fermionic case \cite{CMP,CRAS4corps}, $D$ is not subjected to any condition of exchange symmetry.

\section{Taking advantage of symmetries}
\label{sec:vivent_les_symetries}

\subsection{Overview}
\label{subsec:overview}

The unknown function $D(\kk_2,\kk_4)$ in the integral equation (\ref{eq:equint}) depends on six real variables. 
This is already a strong reduction, as compared to the twelve real variables of the original four-body wavefunction, 
but still this makes a numerical solution challenging. 

Fortunately one can use rotational invariance as in section \ref{subsec:rotinvar}: the unknown function $D$ 
can be considered for example as being the $m_z=0$ component of a spinor of angular momentum $\ell$. 
Then it is known how the various $2\ell+1$ components of the spinor transform under an arbitrary common rotation
of $\kk_2$ and $\kk_4$, in terms of rotation matrices having spherical harmonics as matrix elements,
so that it suffices to know the value of the $2\ell+1$ component of the spinor in the particular
configuration where vector $\kk_2$ points along $x$ axis in the positive direction and $\kk_4$ lies in the $xy$ upper half-plane
$y\geq 0$, at
an angle $\theta_{24} \in [0,\pi]$ with respect to {$\kk_2$}. As this particular configuration is characterized
by the cosine of the angle $\theta_{24}$ and the two moduli $k_2$ and $k_4$, the unknown function $D(\kk_2,\kk_4)$ can be represented
in terms of $2\ell +1$ unknown functions $f^{(\ell)}_{m_z}$ of these three real variables \cite{CRAS4corps}:
\begin{multline}
D(\kk_2,\kk_4)=
\sum_{m_z=-\ell}^{\ell}[Y_\ell^{m_z}(\eee_2\cdot\eee_z,\eee_{4\bot 2}\cdot\eee_z,
\eee_{24}\cdot\eee_z)]^* \\ \times f^{(\ell)}_{m_z}(k_2,k_4,u_{24})
\label{eq:ansatzrot}
\end{multline}
In this expression we have introduced the unit vectors
\bea
\eee_2 &=& \frac{\kk_2}{k_2} \label{eq:base1}\\
\eee_{4\perp2} &=& \frac{1}{v_{24}}\left(\frac{\kk_4}{k_4}-u_{24}\eee_2\right) \label{eq:base2} \\
\eee_{24} &=& \frac{\kk_2\wedge\kk_4}{|\kk_2\wedge\kk_4|} \label{eq:base3}
\eea
Here $\theta_{24}\in [0,\pi]$ is the angle between $\kk_2$ and $\kk_4$, and the notations
\be
u_{24}\equiv\cos\theta_{24} \ \ \mbox{and}\ \ v_{24}\equiv\sin\theta_{24} \label{eq:defuv}
\ee
will be used throughout the paper. It is apparent that $\eee_{4\perp2}$ is obtained by 
projecting $\eee_4=\kk_4/k_4$ orthogonally to $\eee_2$ and by renormalizing the result to unity. Then $(\eee_2, \eee_{4\perp2},
\eee_{24})$ forms a direct orthonormal basis. In that basis, an arbitrary (unit) vector $\nn$ has uniquely defined spherical coordinates,
that is polar angle $\theta_\nn\in [0,\pi]$ with respect to axis $\eee_{24}$ and azimuthal angle $\phi_\nn\in [0,2\pi[$ in the
$\eee_2-\eee_{4\perp2}$ plane with respect to axis $\eee_2$. Then
\be
Y_\ell^{m_z}(\eee_2\cdot\nn,\eee_{4\bot 2}\cdot\nn,\eee_{24}\cdot\nn) \equiv Y_\ell^{m_z}(\theta_\nn,\phi_\nn) 
\ee
where the right-hand side is the standard notation for the spherical harmonics \cite{WuKiTung}.
Integral equations can then be obtained for the $f_{m_z}^{(\ell)}$, see section \ref{subsec:rotinvar}.

For an infinite $s$-wave scattering length the Wigner-Bethe-Peierls contact conditions (\ref{eq:BethePeierls}) are scale invariant.
As the integral equation (\ref{eq:equint}) was further specialized to the zero energy case, its solution can be taken as scale 
invariant, which allows one to eliminate one more variable \cite{CMP}:
\begin{multline}
f_{m_z}^{(\ell)} (k_2,k_4,u_{24}) = (k_2^2+k_4^2)^{-(s+7/2)/2} (\ch x)^{s+3/2}  \\
\times 
e^{im_z\theta_{24}/2} \Phi_{m_z}^{(\ell)}(x,u_{24})
\label{eq:scaleinv}
\end{multline}
with 
\be
x\equiv \ln\frac{k_4}{k_2}
\label{eq:defx}
\ee
The first factor contains the scaling exponent of the solution, which involves the unknown quantity $s$. 
By inserting the ansatz (\ref{eq:scaleinv}) into the linear integral equations of section \ref{subsec:rotinvar},
one obtains linear integral equations for the unknown functions $\Phi_{m_z}^{(\ell)}(x,u)$, represented by a matrix
$M^{(\ell)}(s)$ that depends parametrically on $s$, see section \ref{subsec:scalinvar};  
requiring that the functions $\Phi_{m_z}^{(\ell)}(x,u)$ are not identically zero, one gets an implicit equation
for $s$, in the form \footnote{In principle, $M^{(\ell)}(s)$ is an operator, with an infinite number of eigenvalues,
and its determinant is infinite.  The mathematically correct form is to require 
that one of the eigenvalues of $M^{(\ell)}(s)$ vanishes. In practice, in the numerical calculation, one discretizes
and truncates the variables, so that the determinantal form can be used.}
\be
\det M^{(\ell)}(s) = 0
\label{eq:detM}
\ee
The way the first factor in Eq.~(\ref{eq:scaleinv}) is parametrized by the quantity $s$ ensures compatibility
with the notation used by Efimov for the three-body problem \cite{Efimov}. In the three-body problem, the Efimov effect takes
place if and only if one of the scaling exponents $s$ is purely imaginary, and the geometric trimer energy spectrum has 
then a ratio $\exp(-2\pi/|s|)$.
In the four-body problem, with our definition of $s$, the four-body Efimov effect occurs if and only if there is
a purely imaginary $s$ solving Eq.~(\ref{eq:detM}), in which case there exists a geometric sequence of tetramer eigenenergies
with a ratio $\exp(-2\pi/|s|)$. A justification is given in \cite{CMP}.
The second factor in the ansatz (\ref{eq:scaleinv}) ensures that the matrix $M(s)$ is hermitian for purely imaginary $s$,
with bounded diagonal matrix elements, which is both mathematically and numerically advantageous; as compared to 
\cite{CMP} it contains a additional term $s$ in the exponent, that for purely imaginary $s$ 
suppresses phase oscillations in the matrix elements of $M(s)$ at large $|x|$
\footnote{This additional term leads to the factor $(\ch x)^s$ in the ansatz (\ref{eq:scaleinv}) and to
an additional factor $(\ch x'/\ch x)^s$ in the kernel $K$ of Eq.~(\ref{eq:noyau_matriciel}). This does not change the determinant
of the matrix $M(s)$. This also preserves the hermitian nature of $M(s)$ for $s\in i\mathbb{R}$.}.
The third factor in Eq.~(\ref{eq:scaleinv}) is a phase factor taking into account the fact that exchanging $\kk_2$ and $\kk_4$
in Eq.~(\ref{eq:ansatzrot}) transforms the spherical harmonics $Y_\ell^{m_z}$ into  $(-1)^{\ell}e^{im_z\theta_{24}}
{Y_\ell^{-m_z}}$  {with the same values of the variables} \cite{CRAS4corps}; it ensures that the matrix $M(s)$
transforms in the simplest way under the exchange of $m_\uparrow$ and $m_\downarrow$, which must leave our $2+2$ problem invariant.

A last reduction of the problem can be obtained from parity invariance. It turns out that, under the transformation
$(\kk_2,\kk_4)\rightarrow (-\kk_2,-\kk_4)$,  the term of index $m_z$ in the sum (\ref{eq:ansatzrot}) acquires a factor 
$(-1)^{m_z}$ \cite{CRAS4corps}. This shows that the odd-parity functions $\Phi_{m_z}^{(\ell)}$ (that is with $m_z$ odd) are decoupled from the even-parity functions
$\Phi_{m_z}^{(\ell)}$ (that is with $m_z$ even) in the integral equations, and that $M^{(\ell)}(s)$ has zero matrix elements
between the odd and the even channels.

\subsection{Rotational invariance}
\label{subsec:rotinvar}

To obtain the integral equations for the unknown functions $f_{m_z}^{(\ell)}$ in Eq.~(\ref{eq:ansatzrot}) we use a variational
formulation: The integral equation ({\ref{eq:equint}}) is equivalent to
\be
\partial_{D^*(\kk_2,\kk_4)} \mathcal{E}[D,D*]=0
\ee
where $D$ and its complex conjugate $D^*$ are taken as independent variables, $\partial_{D^*}$ is the functional derivative
with respect to $D^*$ and the functional $\mathcal{E}$ is given by
\be
\mathcal{E}=\mathcal{E}_{\rm diag}+\mathcal{E}_{\rm 24,23}+\mathcal{E}_{\rm 24,14}-\mathcal{E}_{\rm 24,13}
\label{eq:fonc}
\ee
with the diagonal part
\begin{multline}
\mathcal{E}_{\rm diag}= \int d^3k_2 d^3 k_4  D^*(\kk_2,\kk_4) D(\kk_2,\kk_4) \\
\times \frac{\mu_{\uparrow\downarrow}^{3/2}}{2\pi\hbar^2} 
\left[\frac{(\kk_2+\kk_4)^2}{m_\uparrow+m_\downarrow}
+\frac{k_2^2}{m_\uparrow}+\frac{k_4^2}{m_\downarrow}\right]^{1/2} 
\end{multline}
and the generic off-diagonal part
\begin{multline}
\mathcal{E}_{24,ij}=\int \frac{d^3k_2 d^3k_4 d^3k_1 d^3k_3}{(2\pi)^3} D^*(\kk_2,\kk_4) D(\kk_i,\kk_j)   \\
\times \frac{\delta(\kk_1+\kk_2+\kk_3+\kk_4)}{\frac{\hbar^2}{2m_\uparrow}
(k_1^2+k_2^2) + \frac{\hbar^2}{2m_\downarrow} (k_3^2+k_4^2)} 
\end{multline}
Then one inserts the ansatz (\ref{eq:ansatzrot}) into these functionals. Assuming that one is able to integrate over all
variables other than $k_2,k_4,\theta_{24}$ and $k_i,k_j,\theta_{ij}$, one obtains a functional of the $f_{m_z}^{(\ell)}$ and
$f_{m_z}^{(\ell)*}$, which it remains to differentiate with respect to $f_{m_z}^{(\ell)*}$ to obtain the integral equations
for the $f_{m_z}^{(\ell)}$.

Integration is simplified by the following remark: The final integral equations and their solutions $f_{m_z}^{(\ell)}$
cannot depend on the specific vector $\eee_z$ introduced in Eq.~(\ref{eq:ansatzrot}). One can then replace $\eee_z$ by an arbitrary
unit vector $\nn$ in the ansatz (\ref{eq:ansatzrot}), and one can
average the resulting functional $\mathcal{E}$ over $\nn$ uniformly on the unit sphere, for fixed $f_{m_z}^{(\ell)}$. 
The result of this average is particularly simple when the orthonormal basis of Eqs.~(\ref{eq:base1},\ref{eq:base2},\ref{eq:base3})
reduces to the usual Cartesian basis:
\be
(\eee_2,\eee_{4\bot 2},\eee_{24})=(\eee_x,\eee_y,\eee_z)
\ee
Then
\footnote{With the notation $Y_{\ell}^{m_z}(\nn)=Y_\ell^{m_z}(\theta_\nn,\phi_\nn)$, {where $\theta_\nn$ and $\phi_\nn$ are the polar and azimuthal
angles of $\nn$ in the usual spherical coordinates attached to the $xyz$ Cartesian coordinates}, one faces the integral
over the unit sphere $\int \frac{d^2n}{4\pi} Y_{\ell}^{m_z}(\nn) [Y_{\ell}^{m_z'}(\mathcal{R}^{-1}\nn)]^*$. 
Since $Y_{\ell}^{m_z'}(\mathcal{R}^{-1}\nn)=\langle \nn|R|\ell,m_z'\rangle=
\sum_{m_z''} Y_{\ell}^{m_z''}(\nn) \langle \ell,m_z''|R|\ell,m_z'\rangle$ and the spherical harmonics form an orthonormal
basis, {$\int d^2n\, Y_{\ell}^{m_z}(\nn) [Y_{\ell}^{m_z''}(\nn)]^*=\delta_{m_z,m_z''}$,}
we get (\ref{eq:moyang}). {Here $\mathcal{R}$ is a rotation in the three-dimensional space and the operator $R$ is its representation in the Hilbert space.}}
\setcounter{noteylm}{\thefootnote}
\begin{multline}
\langle Y_\ell^{m_z}(\eee_2\cdot\nn,\eee_{4\bot 2}\cdot\nn,\eee_{24}\cdot\nn)
[Y_\ell^{m_z'}(\eee_i\cdot\nn,\eee_{j\bot i}\cdot\nn,\eee_{ij}\cdot\nn)]^*\rangle_{\nn} \\
=\frac{1}{4\pi}\left(\langle \ell,m_z |R^{(ij)}|\ell,m_z'\rangle\right)^*
\label{eq:moyang}
\end{multline}
where $\langle\ldots\rangle_{\nn}$ indicates the average over the direction of $\nn$ and
the quantum operator $R^{(ij)}$ represents (in the usual spin-$\ell$ irreducible representation, with vectors
$|\ell,m_z\rangle$ of angular momentum $m_z \hbar$ along $z$)
the unique real space rotation $\mathcal{R}^{(ij)}$ that maps the Cartesian
basis onto the basis $(\eee_i,\eee_{j\bot i},\eee_{ij})$:
\be
(\eee_i,\eee_{j\bot i},\eee_{ij})= \mathcal{R}^{(ij)} (\eee_x,\eee_y,\eee_z)
\label{eq:defR}
\ee
After average over $\nn$, and integration over $\kk_1$ and $\kk_3$ in $\mathcal{E}_{24,ij}$, it remains an integral over
$\kk_2$ and $\kk_4$, with an integrand invariant by common rotation of $\kk_2$ and $\kk_4$. To evaluate that integrand,
one can then indeed assume that $\kk_2$ is along $x$ (in the positive direction) and that $\kk_4$ lies in the plane $xy$ in the 
upper half $y\geq 0$:
\bea
\label{eq:k2fixe}
\kk_2 &=& k_2 \eee_x \\
\label{eq:k4fixe}
\kk_4 &=& k_4 (\cos\theta_{24} \eee_x+\sin\theta_{24} \eee_y)\ \mbox{with} \ \theta_{24}\in [0,\pi]
\eea
so that
\be
(\eee_2,\eee_{4\bot 2},\eee_{24})=(\eee_x,\eee_y,\eee_z)
\ee
and one can use Eqs.~(\ref{eq:moyang},\ref{eq:defR}).
Then one pulls out a factor $4\pi$ (resulting from integration over the solid angle of $\kk_2$) compensated 
by the $4\pi$ denominator in Eq.~(\ref{eq:moyang}), and an uncompensated factor $2\pi$ (resulting
from the integration over the azimuthal angle of $\kk_4$ for the spherical coordinates of {polar} axis $\kk_2/k_2=\eee_x$ for
$\kk_4$), and one is left with an integration over the moduli $k_2$ and $k_4$ and over the angle $\theta_{24}$.

For the functional $\mathcal{E}_{\rm diag}$, this gives a simple result: Since $i=2$ and $j=4$, 
the matrix $\mathcal{R}^{(ij)}$ is the identity matrix, $R^{(ij)}$ reduces to the identity operator; also there is no
$\kk_1$ or $\kk_3$ integration. One obtains
\begin{multline}
\mathcal{E}_{\rm diag}=\!\!\sum_{m_z=-\ell}^{\ell}\!2\pi\!\int_0^{\infty}\!\!\!\! dk_2 k_2^2 dk_4 k_4^2 \int_{-1}^{1}\!\!\!\!du_{24} 
|f_{m_z}(k_2,k_4,u_{24})|^2 \\
\times \frac{\mu_{\uparrow\downarrow}^{3/2}}{2\pi\hbar^2}\left(\frac{k_2^2+k_4^2+2 k_2 k_4 u_{24}}{m_\uparrow+m_\downarrow}
+\frac{k_2^2}{m_\uparrow}+\frac{k_4^2}{m_\downarrow}\right)^{1/2} 
\label{eq:foncrotdiag}
\end{multline}
with the same notation as in Eq.~(\ref{eq:defuv}).
For the generic off-diagonal part this leads to
\begin{multline}
\mathcal{E}_{24,ij}=\!\!\sum_{m_z,m_z'=-\ell}^{\ell}\!2\pi\!\int_0^{\infty}\!\!\!\! dk_2 k_2^2 dk_4 k_4^2\int_{-1}^{1}\!\!\!\!du_{24}
\int \frac{d^3k_1 d^3k_3}{(2\pi)^3} \\
\left(\langle\ell,m_z |R^{(ij)}|\ell,m_z'\rangle\right)^* f^{(\ell)*}_{m_z}(k_2,k_4,u_{24}) f^{(\ell)}_{m_z'}(k_i,k_j,u_{ij}) \\
\times \frac{\delta(\kk_1+\kk_2+\kk_3+\kk_4)}{\frac{\hbar^2(k_1^2+k_2^2)}{2m_\uparrow}+\frac{\hbar^2(k_3^2+k_4^2)}{2m_\downarrow}}
\end{multline}
The way to proceed with the integration over the directions of $\kk_1$ and $\kk_3$ depends on the indices $i$ and $j$.

\subsubsection{Case $(i,j)=(2,3)$}

For $(i,j)=(2,3)$, one trivially integrates over $\kk_1$ using the Dirac distribution, that imposes $\kk_1=-(\kk_2+\kk_3+\kk_4)$,
and one integrates over $\kk_3$ using spherical coordinates of polar axis $\eee_x$ and of azimuthal axis $\eee_y$; the azimuthal angle
is called $\phi$, and the polar angle is called $\theta_{23}$ since it is the angle between $\kk_2$ and $\kk_3$
[see Fig.~\ref{fig:geom}(a)]. Then $\mathcal{R}^{(ij)}$ in Eq.~(\ref{eq:defR}) is the rotation of axis $x$ and of angle $\phi$:
\be
\mathcal{R}^{(23)}=\mathcal{R}_x(\phi) \ \mbox{and}\ R^{(23)}=e^{-i\phi L_x/\hbar}
\ee
where $L_x$ is the angular momentum operator along $x$. Also
\begin{multline}
k_1^2=k_2^2+k_3^2+k_4^2+2 k_2 k_3 u_{23} + 2 k_2 k_4 u_{24} \\
+2 k_3 k_4 (u_{23} u_{24} + v_{23} v_{24} \cos \phi)
\label{eq:valk1}
\end{multline}
with $u_{23}=\cos\theta_{23}$ and $v_{23}=\sin \theta_{23}$ as in Eq.~(\ref{eq:defuv}).  This gives
\begin{multline}
\mathcal{E}_{24,23}=\!\!\!\!\!\!\sum_{m_z,m_z'=-\ell}^{\ell}\!\!\!\!\!2\pi\!\int_0^{\infty}\!\!\!\! dk_2 dk_3 dk_4 k_2^2k_3^2k_4^2
\int_{-1}^{1}\!\!\!\! du_{23} du_{24}\!\int_0^{2\pi}\!\!\!\! d\phi \\
\frac{\langle \ell,m_z|e^{i\phi L_x/\hbar}|\ell,m_z'\rangle f^{(\ell)*}_{m_z}(k_2,k_4,u_{24})f^{(\ell)}_{m_z'}(k_2,k_3,u_{23})}
{(2\pi)^3\left[\frac{\hbar^2(k_1^2+k_2^2)}{2m_\uparrow} +\frac{\hbar^2(k_3^2+k_4^2)}{2m_\downarrow}\right]}
\label{eq:foncrot23}
\end{multline}
where $k_1$ is given by Eq.~(\ref{eq:valk1}) 
and we used the fact that $L_x$ has real matrix elements in the standard $|\ell,m_z\rangle$ basis.

\subsubsection{Case $(i,j)=(1,4)$}

\begin{figure}[tb]
\centerline{\includegraphics[width=4cm,clip=]{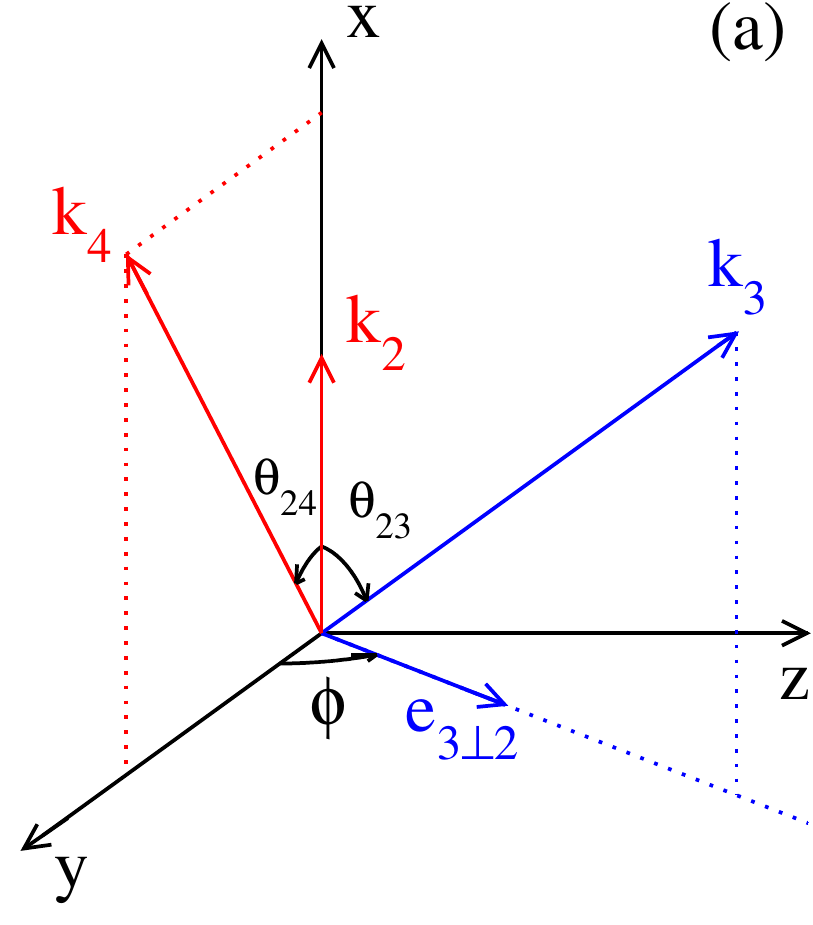} \includegraphics[width=4.5cm,clip=]{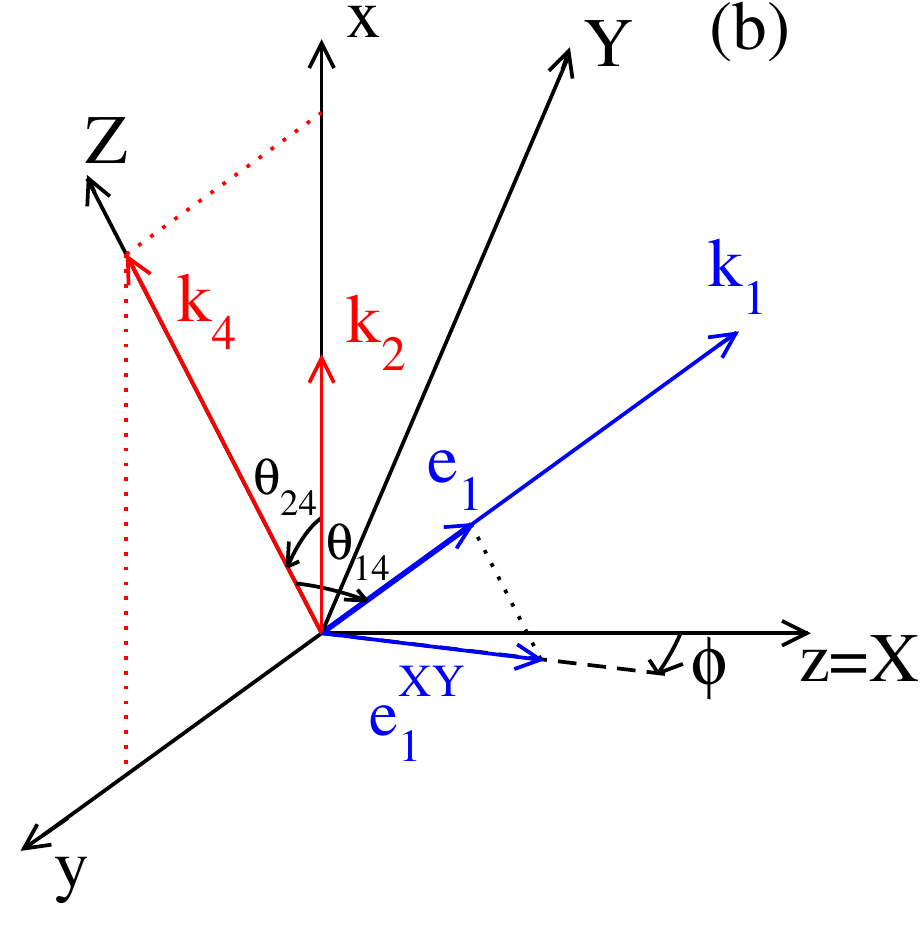}}
\centerline{\includegraphics[width=6cm,clip=]{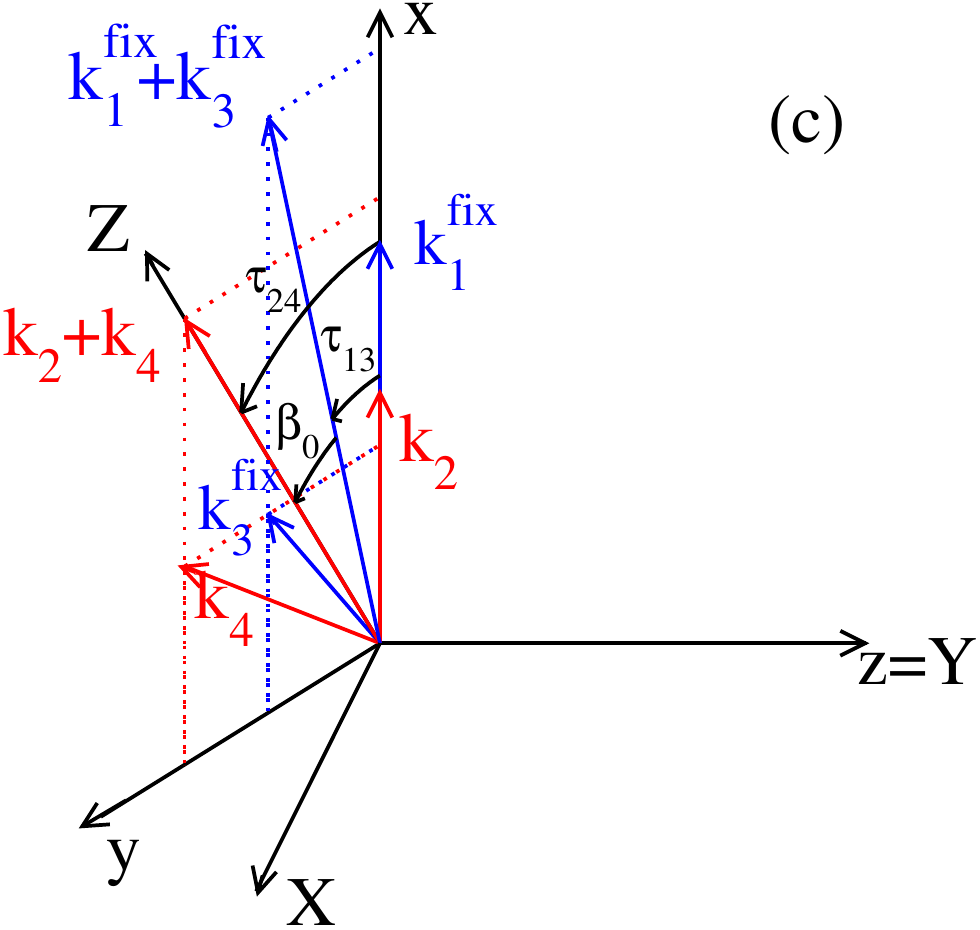}}
\caption{(Color online)
Positions and parametrisations of the wavevectors appearing in the angular integration in the functionals
$\mathcal{E}_{24,ij}$. The vectors $\kk_2$ and $\kk_4$ are given by Eqs.~(\ref{eq:k2fixe},\ref{eq:k4fixe}).  
(a) For $(i,j)=(2,3)$, $\kk_1=-(\kk_2+\kk_3+\kk_4)$ and one integrates over $\kk_3$ using spherical coordinates
of polar axis $x$ and azimuthal axis $y$. (b) For $(i,j)=(1,4)$, $\kk_3=-(\kk_1+\kk_2+\kk_4)$ and one integrates
over $\kk_1$ using the polar axis $Z$ (direction of $\kk_4$) and the azimuthal axis $X$ 
(direction of $\eee_z$) as defined by Eq.~(\ref{eq:basetournee}), leading to
the polar angle $\theta_{14}$ and the azimuthal angle $\phi$ {($<0$ in the figure)}. The dashed line gives the direction
of the component $\eee_1^{\mathrm{XY}}$ of $\eee_1=\kk_1/k_1$ in the $XY$ plane. (c) For $(i,j)=(1,3)$, 
one integrates over the rotation $\mathcal{R}$, such that $\kk_1$ and $\kk_3$ are given by the action of $\mathcal{R}$
on vectors $\kk_1^{\rm fix}$ and $\kk_3^{\rm fix}$ in the $xy$ plane as in Eqs.~(\ref{eq:k1fix},\ref{eq:k3fix}), using the 
parametrisation in terms of Euler angles associated to the convenient axes $X$, $Y$ and $Z$ of Eq.~(\ref{eq:baseeuler}).}
\label{fig:geom}
\end{figure}

For $(i,j)=(1,4)$, one integrates over $\kk_3$ using the Dirac distribution, that imposes
$\kk_3=-(\kk_1+\kk_2+\kk_4)$ and one integrates over $\kk_1$ using spherical coordinates in a rotated basis 
\be
\label{eq:basetournee}
(\eee_X,\eee_Y,\eee_Z)=(\eee_z,\eee_4\wedge \eee_z,\eee_4)\ \mbox{with}\ \eee_4=\frac{\kk_4}{k_4}.
\ee
The direction $\eee_Z$ of $\kk_4$ is taken as the polar axis, so that the polar angle is $\theta_{14}$; $\eee_X$ is taken
as the azimuthal axis, with the azimuthal angle called $\phi$, see Fig.~\ref{fig:geom}(b). Then the real space rotation
$\mathcal{R}^{(ij)}$ in Eq.~(\ref{eq:defR}) is
\begin{multline}
\mathcal{R}^{(14)}=\mathcal{R}_Z(\phi-\frac{\pi}{2})\mathcal{R}_z(\theta_{24}-\theta_{14}) \\
=\mathcal{R}_z(\theta_{24}) \mathcal{R}_x(\phi-\frac{\pi}{2})\mathcal{R}_z(-\theta_{14})
\end{multline}
and the corresponding operator has matrix elements
\begin{multline}
\langle \ell,m_z| R^{(14)}|\ell,m_z'\rangle = e^{-i m_z \theta_{24}} \\
\times \langle \ell,m_z| e^{-i(\phi-\frac{\pi}{2})L_x/\hbar}|\ell,m_z'\rangle 
e^{i m_z' \theta_{14}}
\end{multline}
Using $\eee_x=[\mathcal{R}^{(14)}]^{-1}\eee_1= u_{24} \eee_Z+v_{24}\eee_Y$
and $\eee_1=u_{14}\eee_Z+v_{14}(\cos\phi\, \eee_X+\sin\phi\, \eee_Y)$
we get 
\begin{multline}
k_3^2=k_1^2+k_2^2+k_4^2+2 k_1 k_2 [u_{14}u_{24}+v_{14}v_{24}\cos(\phi-\pi/2)] \\
+2 k_1 k_4 u_{14} + 2 k_2 k_4 u_{24}.
\label{eq:valk3}
\end{multline}
This gives:
\begin{multline}
\mathcal{E}_{24,14}=\!\!\!\!\!\!\sum_{m_z,m_z'=-\ell}^{\ell}\!\!\!\!\!2\pi\!\int_0^{\infty}\!\!\!\! dk_1 dk_2 dk_4 k_1^2k_2^2k_4^2
\int_{-1}^{1}\!\!\!\! du_{14} du_{24}\!\int_0^{2\pi}\!\!\!\! d\phi \\
\frac{e^{im_z\theta_{24}} \langle \ell,m_z|e^{i(\phi-\pi/2)L_x/\hbar}|\ell,m_z'\rangle e^{-im_z'\theta_{14}}}
{(2\pi)^3\left[\frac{\hbar^2(k_1^2+k_2^2)}{2m_\uparrow} +\frac{\hbar^2(k_3^2+k_4^2)}{2m_\downarrow}\right]} \\
\times f^{(\ell)*}_{m_z}(k_2,k_4,u_{24})f^{(\ell)}_{m_z'}(k_1,k_4,u_{14})
\label{eq:foncrot14}
\end{multline}
where $k_3$ is given by Eq.~(\ref{eq:valk3}).

\subsubsection{Case $(i,j)=(1,3)$}

For $(i,j)=(1,3)$, we find it convenient to replace the integration over $\kk_1$ and $\kk_3$
by an integration over the moduli $k_1$ and $k_3$, over the angle $\theta_{13}\in[0,\pi]$ and over a rotation matrix $\mathcal{R}$
uniformly distributed over the rotation group $\mathrm{SO(3)}$, the vectors $\kk_1$ and $\kk_3$ being given by the action of
$\mathcal{R}$ on vectors fixed in the $xy$ plane:
\bea
\label{eq:k1fix}
\kk_1=\mathcal{R} \kk_1^{\rm fix} &\mbox{with}& \kk_1^{\rm fix} = k_1 \eee_x \\
\label{eq:k3fix}
\kk_3=\mathcal{R} \kk_3^{\rm fix} &\mbox{with}& \kk_3^{\rm fix} = k_3 (u_{13}\eee_x+v_{13}\eee_y)
\eea
Then $\mathcal{R}$ is precisely the rotation matrix $\mathcal{R}^{(ij)}$ of Eq.~(\ref{eq:defR}) and
\begin{multline}
\mathcal{E}_{24,13}=\!\!\sum_{m_z,m_z'=-\ell}^{\ell}\!2\!\int_0^{\infty}\!\!\!\left(\prod_{n=1}^4 dk_n k_n^2\right)\!
\int_{-1}^{1}\!\!\!\!du_{13} du_{24}\!\int_{\mathrm{SO(3)}}\!\!\!\!\! d\mathcal{R} \\
\left(\langle\ell,m_z |R|\ell,m_z'\rangle\right)^* f^{(\ell)*}_{m_z}(k_2,k_4,u_{24}) f^{(\ell)}_{m_z'}(k_1,k_3,u_{13}) \\
\times \frac{\delta(\kk_2+\kk_4+\mathcal{R}(\kk_1^{\rm fix}+\kk_3^{\rm fix}))}{\frac{\hbar^2(k_1^2+k_2^2)}{2m_\uparrow}+\frac{\hbar^2(k_3^2+k_4^2)}{2m_\downarrow}}
\label{eq:E2413}
\end{multline}
where the factor $2$ originates from $(4\pi\times 2\pi)^2/[4\pi (2\pi)^3]$,
$R$ is the operator representing $\mathcal{R}$ and $d\mathcal{R}$ is the invariant measure over the group 
$\mathrm{SO(3)}$ normalized to unity (see \S 8.2 of reference \cite{WuKiTung})
\footnote{For an arbitrary function $\phi$, one defines $I\equiv\int d^3k_1 d^3k_3 \phi(\kk_1,\kk_3)$. Then  one has
$I=(4\pi\times 2\pi)\int_{\mathrm{SO(3)}}
d\mathcal{R}\int_0^{+\infty} dk_1 dk_3 k_1^2 k_3^2 \int_{-1}^{1} du_{13} 
\phi(\mathcal{R} \kk_1^{\rm fix}, \mathcal{R} \kk_3^{\rm fix}).$
To show this one makes in the integral defining $I$ the unit Jacobian change of variable
$\kk_1\rightarrow \mathcal{R} \kk_1$ and $\kk_3\rightarrow \mathcal{R} \kk_3$, where $\mathcal{R}$ is any rotation:
$I=\int d^3k_1 d^3k_3 \phi(\mathcal{R}\kk_1,\mathcal{R}\kk_3)$. As the result does not depend on $\mathcal{R}$, 
we can average it over $\mathrm{SO(3)}$, with the normalized invariant measure. Exchanging the order of integration
over $\mathcal{R}$ and over $\kk_1$ and $\kk_3$, we get $I=\int d^3k_1 d^3k_3 J_{\kk_1,\kk_3}$
with $J_{\kk_1,\kk_3}\equiv\int_{\mathrm{SO(3)}} d\mathcal{R} \phi(\mathcal{R}\kk_1,\mathcal{R}\kk_3)$.
In $J_{\kk_1,\kk_3}$ one then performs the change of variable $\mathcal{R}\rightarrow \mathcal{R} \rho$, where
$\rho$ is any rotation. As the measure is invariant, 
$J_{\kk_1,\kk_3}=\int_{\mathrm{SO(3)}} d\mathcal{R} \phi(\mathcal{R} \rho \kk_1,\mathcal{R}\rho \kk_3)$.
Then for any given $\kk_1$ and $\kk_3$, one chooses $\rho$ such that $\rho \kk_1=\kk_1^{\rm fix}$ and 
$\rho \kk_3=\kk_3^{\rm fix}$, so that $J_{\kk_1,\kk_3}=\int_{\mathrm{SO(3)}} d\mathcal{R}
\phi(\mathcal{R}\kk_1^{\rm fix},\mathcal{R}\kk_3^{\rm fix})$. Inserting this expression of $J_{\kk_1,\kk_3}$
into $I$ and exchanging again the order of integration gives
$I=\int_{\mathrm{SO(3)}} d\mathcal{R} \int d^3k_1 d^3k_3 \phi(\mathcal{R}\kk_1^{\rm fix},\mathcal{R}\kk_3^{\rm fix})$.
At fixed $\kk_1$ one integrates over $\kk_3$ in spherical coordinates of polar axis $\kk_1$; as the integrand does not depend
on the azimuthal angle, we pull out a factor $2\pi$. The resulting integral over {$k_3$} and $\theta_{13}$
does not depend on the direction of $\kk_1$ so, after integration over $\kk_1$ in spherical coordinates of arbitrary
polar axis, one pulls out an additional factor $4\pi$ and gets the desired relation. 
}.  
To integrate over $\mathcal{R}$, we use the Euler parametrisation as in Eq.~(7.1-12) of reference \cite{WuKiTung}:
\be
\mathcal{R}=\mathcal{R}_Z(\alpha) \mathcal{R}_Y(\beta) \mathcal{R}_Z(\gamma)
\ee
where the Euler angles $\alpha$ and $\gamma$ run over an interval of length $2\pi$ and the Euler angle $\beta$ runs over
$[0,\pi]$, so that the invariant measure is (see \S8.2 of reference \cite{WuKiTung})
\be
d\mathcal{R}=\frac{d\alpha \sin\beta d\beta d\gamma}{8\pi^2}
\ee
Due to the occurrence of $\kk_2+\kk_4$ in the argument of the Dirac distribution in Eq.~(\ref{eq:E2413}), 
the convenient direct orthonormal basis defining the rotation axes $X$, $Y$ and $Z$ is now [see Fig.~\ref{fig:geom}(c)]
\be
\label{eq:baseeuler}
(\eee_X,\eee_Y,\eee_Z)=(\eee_z\wedge \frac{\kk_2+\kk_4}{|\kk_2+\kk_4|}, \eee_z, \frac{\kk_2+\kk_4}{|\kk_2+\kk_4|}).
\ee
Then the Dirac distribution can be written as
\footnote{A three-dimensional $\delta(\kk)$ is the product of three one-dimensional $\delta(\uu_n\cdot\kk)$ where $(\uu_n)$
is an orthonormal basis. As explained in the text one can take $\kk=\kk_2+\kk_4+\mathcal{R}_Y(\beta) \mathcal{R}_Z(\gamma)(\kk_1^{\rm fix}
+\kk_3^{\rm fix})$. We first take $\uu_1=\eee_Y$ so that $\uu_1\cdot\kk=(\mathcal{R}_Z(-\gamma)\eee_Y)\cdot(\kk_1^{\rm fix}+\kk_3^{\rm fix})
=\sin\gamma\, \eee_X \cdot(\kk_1^{\rm fix} +\kk_3^{\rm fix})=-\sin\gamma \sin \beta_0 |\kk_1^{\rm fix} +\kk_3^{\rm fix}|$ 
where we used Eq.~(\ref{eq:defbeta0}). This gives the factor $\delta(\sin\gamma)$ in Eq.~(\ref{eq:Dirac}).
As explained in the text we can restrict to the case $\gamma=0$ (up to a change $\beta \leftrightarrow -\beta$)
and we are left with a two-dimensional Dirac $\delta(\kk_\perp)$ in the plane orthogonal to $\eee_Y$.
In principle $\kk_\perp=\kk_2+\kk_4+\mathcal{R}_Y(\beta)(\kk_1^{\rm fix}+\kk_3^{\rm fix})$ but we can equivalently
take $\kk_\perp=\mathcal{R}_Y(-\beta)(\kk_2+\kk_4)+\kk_1^{\rm fix}+\kk_3^{\rm fix}$ due to the rotational invariance of the Dirac distribution.
Using $\mathcal{R}_Y(-\beta)\eee_Z=\cos\beta\,\eee_Z-\sin\beta\,\eee_X$ and taking $\uu_2=\frac{\kk_1^{\rm fix}+\kk_3^{\rm fix}}
{|\kk_1^{\rm fix}+\kk_3^{\rm fix}|}=\cos\beta_0\, \eee_Z-\sin\beta_0\,\eee_X$ and its dual $\uu_3=\sin\beta_0\, \eee_Z+\cos\beta_0\,\eee_X$
in the $ZX$ plane, we justify Eq.~(\ref{eq:Dirac}).}
\begin{multline}
\delta(\kk_2+\kk_4+\mathcal{R}(\kk_1^{\rm fix}+\kk_3^{\rm fix}))=\delta(\sin\gamma) \delta(\sin(\beta_0-\beta\cos\gamma)) \\
\times \frac{\delta(|\kk_2+\kk_4| \cos(\beta_0-\beta\cos\gamma)+|\kk_1^{\rm fix}+\kk_3^{\rm fix}|)}
{|\sin\beta_0|\, |\kk_1^{\rm fix}+\kk_3^{\rm fix}|\, |\kk_2+\kk_4|}
\label{eq:Dirac}
\end{multline}
where we have introduced the oriented angle $\beta_0$ between $\kk_1^{\rm fix}+\kk_3^{\rm fix}$ and $\kk_2+\kk_4$
such that [see Fig.~\ref{fig:geom}(c)]
\be
\kk_1^{\rm fix}+\kk_3^{\rm fix}=|\kk_1^{\rm fix}+\kk_3^{\rm fix}|(-\sin\beta_0\, \eee_X + \cos\beta_0\, \eee_Z).
\label{eq:defbeta0}
\ee
There is no dependence on $\alpha$ in the right-hand side of Eq.~(\ref{eq:Dirac}): in the argument of $\delta$, one can write
$\kk_2+\kk_4$ as $\mathcal{R}_Z(\alpha)(\kk_2+\kk_4)$ and, due to the rotational invariance of the
three-dimensional Dirac distribution, one can factor out and remove the rotation $\mathcal{R}_Z(\alpha)$.
The integration over $\alpha$ in Eq.~(\ref{eq:E2413}) then pulls out in the matrix element of $R$
the orthogonal projector on the state of total angular momentum $\ell$ and of vanishing angular momentum along $Z$:
\be
\int_0^{2\pi} \!\!\! d\alpha \,e^{-i\alpha L_Z/\hbar}=2\pi |\ell,m_Z=0\rangle \langle \ell,m_Z=0|
\ee
In the integral over $\gamma$, for example over the interval $[-\pi/2,3\pi/2]$, only the points $\gamma=0$ and $\gamma=\pi$ contribute.
The contribution of $\gamma=\pi$ can be deduced from the one of $\gamma=0$ by changing $\beta$ into $-\beta$,
due to $\mathcal{R}_Z(\pi)\mathcal{R}_Y(\beta)\mathcal{R}_Z(\pi)=\mathcal{R}_Y(-\beta)$
and to the invariance of $\kk_1^{\rm fix}+\kk_3^{\rm fix}$ and $|m_Z=0\rangle$ by rotation of axis $Z$.
In the integral over $\beta\in [0,\pi]$, the $\gamma=\pi$ contribution can then be taken into account by extending the 
integration of the $\gamma=0$ contribution to $\beta\in [-\pi,0]$: one can take $\gamma=0$ in Eq.~(\ref{eq:Dirac}) and one faces
\begin{multline}
\int_{-\pi}^{\pi} d\beta |\sin\beta| e^{i\beta m_z'} \delta(\sin(\beta_0-\beta)) 
\delta(|\kk_2+\kk_4| \cos(\beta_0-\beta) \\
+|\kk_1^{\rm fix}+\kk_3^{\rm fix}|)=
\delta(|\kk_2+\kk_4|-|\kk_1^{\rm fix}+\kk_3^{\rm fix}|) \\
\times \int_{-\pi}^{\pi} d\beta |\sin\beta| e^{i\beta m_z'} 
\sum_{n\in\mathbb{Z}} \delta(\beta-\beta_0-\pi-2\pi n) \\
{= \delta(|\kk_2+\kk_4|-|\kk_1^{\rm fix}+\kk_3^{\rm fix}|) |\sin \beta_0| (-1)^{m_z'} e^{i\beta_0 m_z'} }
\end{multline}
Due to the $2\pi$ periodicity of the integrand {we have shifted} the domain of integration so as to only keep for example the term $n=0$ of
the Dirac comb.
Finally, using $\beta_0=\tau_{24}-\tau_{13}$, where $\tau_{24}$ {is the angle $\in [0,\pi]$ between $\kk_2$ and $\kk_2+\kk_4$ 
and $\tau_{13}$ is the angle $\in [0,\pi]$ between $\kk_1^{\rm fix}$ and $\kk_1^{\rm fix}+\kk_3^{\rm fix}$} 
so that (up to a phase factor)
\be
|\ell,m_Z=0\rangle = e^{-i \tau_{24}L_z/\hbar} |\ell,m_x=0\rangle,
\label{eq:aunfact}
\ee
and {using} the property that
\footnote{\label{note:spm} Similarly to Eq.~(\ref{eq:aunfact}), $|\ell,m_x=0\rangle=s_+ e^{-i(\pi/2)L_y/\hbar}$ $|\ell,m_z=0\rangle=
s_-e^{-i(-\pi/2)L_y/\hbar}|\ell,m_z=0\rangle$ where $s_\pm$ are just signs since $|\ell,m_x=0\rangle$
can be taken with real components in the $|\ell,m_z\rangle$ basis ($L_y$ has purely imaginary matrix elements). 
Then $s_- s_+ \langle\ell,m_z=0|e^{-i\pi L_y/\hbar}|\ell,m_z=0\rangle=1$. The action of $e^{-i\pi L_y/\hbar}$ in Cartesian coordinates is 
$(x,y,z)\rightarrow (-x,y,-z)$; in {spherical} coordinates of {polar} axis $z$ it is $(\theta,\phi)\rightarrow (\pi-\theta,\pi-\phi)$.
For $Y_\ell^{m_z=0}(\theta,\phi)$, which does not depend on $\phi$, this is equivalent to the action of parity
$(\theta,\phi)\rightarrow (\pi-\theta,\pi+\phi)$ and it pulls out a factor $(-1)^l$ so that $s_-=(-1)^{\ell} s_+$
and $|\ell,m_x=0\rangle=(s_+/2)(e^{-i(\pi/2)L_y/\hbar}+(-1)^\ell e^{-i(-\pi/2)L_y/\hbar})|\ell,m_z=0\rangle$.
Series expanding the exponentials in this last expression, and using the fact that $L_y$ only couples states of
different $m_z$ parity, one gets Eq.~(\ref{eq:propriete}).}
\setcounter{notespm}{\thefootnote}
\be
\langle \ell,m_x=0|\ell,m_z'\rangle=0 \ \ \mbox{if}\ \ \ell+m_z'\ \mbox{is odd},
\label{eq:propriete}
\ee
allowing one to replace $(-1)^{m_z'}$ with $(-1)^\ell$, we obtain
\begin{multline}
\mathcal{E}_{24,13}=\!\!\sum_{m_z,m_z'=-\ell}^{\ell}\!\frac{(-1)^\ell}{2\pi}\!\int_0^{\infty}\!\!\!\left(\prod_{n=1}^4 dk_n k_n^2\right)\!
\int_{-1}^{1}\!\!\!\!du_{13} du_{24}\\
\frac{e^{im_z\tau_{24}}\langle \ell,m_z|\ell,m_x=0\rangle\langle \ell,m_x=0|\ell,m_z'\rangle e^{-im_z'\tau_{13}}}
{|\kk_2+\kk_4|\left[\frac{\hbar^2(k_1^2+k_2^2)}{2m_\uparrow}+\frac{\hbar^2(k_3^2+k_4^2)}{2m_\downarrow}\right]|\kk_1^{\rm fix}+\kk_3^{\rm fix}|} \\
\times \delta(|\kk_2+\kk_4|-|\kk_1^{\rm fix}+\kk_3^{\rm fix}|) f^{(\ell)*}_{m_z}(k_2,k_4,u_{24}) f^{(\ell)}_{m_z'}(k_1,k_3,u_{13}) 
\label{eq:E2413rot}
\end{multline}
knowing that $|\ell,m_x=0\rangle$ has real components in the basis $|\ell,m_z\rangle$ up to a global phase, and that $\kk_2$ and $\kk_4$
are given by Eqs.~(\ref{eq:k2fixe},\ref{eq:k4fixe}) and $\kk_1^{\rm fix}$ and $\kk_3^{\rm fix}$ by Eqs.~(\ref{eq:k1fix},\ref{eq:k3fix}).

\subsection{Scale invariance}
\label{subsec:scalinvar}

To take advantage of the scale invariance of the zero-energy solution, one uses the ansatz (\ref{eq:scaleinv}) with $s\in i\mathbb{R}$, 
as physically explained
in section \ref{subsec:overview}, and one inserts it in the various terms 
(\ref{eq:foncrotdiag},\ref{eq:foncrot23},\ref{eq:foncrot14},\ref{eq:E2413rot}) of the functional (\ref{eq:fonc}).
In Eq.~(\ref{eq:foncrotdiag}) one performs in the integral over $k_4$ the change of variable $k_4=e^x k_2$, where $x$ ranges from
$-\infty$ to $+\infty$, and one sets $u_{24}=u$ for conciseness, also introducing the mass ratio
\be
\alpha\equiv \frac{m_\uparrow}{m_\downarrow}.
\label{eq:defalpha}
\ee
One pulls out a constant factor $\mathcal{F}$, that will be given and discussed later, to obtain
\begin{multline}
\label{eq:findiag}
\mathcal{E}_{\rm diag}=\mathcal{F} \!\!\sum_{m_z=-\ell}^{\ell}\int_\mathbb{R}\! dx \int_{-1}^{1}\!\!\!du
\left[\frac{\alpha}{(1+\alpha)^2}\left(1+\frac{u}{\ch x}\right)\right. \\
+ \left.\frac{e^{-x}+\alpha e^x}{2(\alpha+1)\ch x}\right]^{1/2}
|\Phi_{m_z}^{(\ell)}(x,u)|^2
\end{multline}
In Eq.~(\ref{eq:foncrot23}) one performs the change of variable $k_4=e^x k_2$ and $k_3=e^{x'}k_2$ in the integrals over $k_4$ and $k_3$,
also setting $\theta_{24}=\theta$, $u_{24}=u$, $v_{24}=v$
and $\theta_{23}=\theta'$, $u_{23}=u'$, $v_{23}=v'$ for concision. One then pulls out the same factor $\mathcal{F}$ to obtain
\begin{multline}
\label{eq:fin2423}
\mathcal{E}_{24,23}=\mathcal{F}\!\!\sum_{m_z,m_z'=-\ell}^{\ell} \int_\mathbb{R}\! dx dx' \int_{-1}^{1}\!\!\!du du'
\left(\frac{e^x\ch x'}{e^{x'}\ch x}\right)^{s/2} \\
\times \left(\frac{e^{x+x'}}{4\ch x\ch x'}\right)^{1/4} \Phi_{m_z}^{(\ell)*}(x,u) \Phi_{m_z'}^{(\ell)}(x',u') \\
\times \int_0^{2\pi} \frac{d\phi}{(2\pi)^2} 
\frac{e^{-im_{z}\theta/2}\langle l,m_{z}|e^{i\phi L_x/\hbar}|l,m'_{z}\rangle e^{im'_{z}\theta'/2}}{\mathcal{D}_{24,23}(\phi;x,u;x',u';\alpha)}
\end{multline}
In the denominator, we have introduced the notation
\be
\mathcal{D}_{24,23}= \frac{\frac{\hbar^2(k_1^2+k_2^2)}{2m_\uparrow}+\frac{\hbar^2(k_3^2+k_4^2)}{2m_\downarrow}}
{\frac{\hbar^2 k_3 k_4}{\mu_{\uparrow\downarrow}}}
\label{eq:D2423}
\ee
where $k_1$ is given by Eq.~(\ref{eq:valk1}) so that
\begin{multline}
\mathcal{D}_{24,23}(\phi;x,u;x',u';\alpha)= \ch(x-x') \\
+\frac{1}{1+\alpha}(e^{-x-x'}+ e^{-x'} u + e^{-x} u' +u u' + v v'\cos\phi).
\label{eq:D2423bis}
\end{multline}
In Eq.~(\ref{eq:foncrot14}) one performs the change of variables $k_4=e^x k_2$ and $k_1=e^{x-x'} k_2$ (so that $k_4/k_1=e^{x'}$)
in the integrals over $k_4$ and $k_1$, and the change of variable $\phi=\frac{\pi}{2}+\phi'$ in the integral
over $\phi$ \footnote{One can keep $[0,2\pi]$ as the range of integration {over} $\phi'$ since the integrand is a periodic function
of $\phi'$ of period $2\pi$.}, also setting $\theta_{24}=\theta$, $u_{24}=u$, $v_{24}=v$
and $\theta_{14}=\theta'$, $u_{14}=u'$, $v_{14}=v'$. Again pulling out the factor $\mathcal{F}$ one gets
\begin{multline}
\label{eq:fin2414}
\mathcal{E}_{24,14}=\mathcal{F}\!\!\sum_{m_z,m_z'=-\ell}^{\ell} \int_\mathbb{R}\! dx dx' \int_{-1}^{1}\!\!\!du du'
\left(\frac{e^{-x}\ch x'}{e^{-x'}\ch x}\right)^{s/2} \\
\times \left(\frac{e^{-(x+x')}}{4\ch x\ch x'}\right)^{1/4} \Phi_{m_z}^{(\ell)*}(x,u) \Phi_{m_z'}^{(\ell)}(x',u') \\
\times \int_0^{2\pi} \frac{d\phi'}{(2\pi)^2} 
\frac{e^{im_{z}\theta/2}\langle l,m_{z}|e^{i\phi' L_x/\hbar}|l,m'_{z}\rangle e^{-im'_{z}\theta'/2}}{\mathcal{D}_{24,14}(\phi';x,u;x',u';\alpha)}
\end{multline}
In the denominator we have introduced the notation
\be
\label{eq:D2414}
\mathcal{D}_{24,14}= \frac{\frac{\hbar^2(k_1^2+k_2^2)}{2m_\uparrow}+\frac{\hbar^2(k_3^2+k_4^2)}{2m_\downarrow}}
{\frac{\hbar^2 k_1 k_2}{\mu_{\uparrow\downarrow}}}  
\ee
with $k_3$ given by Eq.~(\ref{eq:valk3}) so that
\begin{multline}
\label{eq:D2414bis}
\mathcal{D}_{24,14}(\phi';x,u;x',u';\alpha) = \ch(x-x') \\
+\frac{\alpha}{1+\alpha} (e^{x+x'} +e^{x'} u + e^{x} u'+u u' + v v'\cos\phi').
\end{multline}
Finally, in Eq.~(\ref{eq:E2413rot}), one performs the change of variables $k_4=e^x k_2$ and $k_3=e^{x'}k_1$ in the integrals over
$k_4$ and $k_3$, also setting $\theta_{24}=\theta$, $u_{24}=u$, $\tau_{24}=\tau$
and $\theta_{13}=\theta'$, $u_{13}=u'$, $\tau_{13}=\tau'$. 
The integration over $k_1$ is straightforward due to the occurrence of a Dirac distribution in Eq.~(\ref{eq:E2413rot}).
Due to the phase factor in the ansatz (\ref{eq:scaleinv}), there naturally appear the angles $\gamma\equiv \tau-\theta/2$
and $\gamma'=\tau'-\theta'/2$. Since $\tau$ is the angle between {$\kk_2$} and $\kk_2+\kk_4$
{[see Fig.~\ref{fig:geom}(c)]}, one has according
to Eqs.~(\ref{eq:k2fixe},\ref{eq:k4fixe}) and using the usual representation of vectors in the $xy$ plane by complex numbers:
\be
e^{i\gamma} = \frac{1+e^x e^{i\theta}}{|1+e^x e^{i\theta}|} e^{-i\theta/2} = \frac{e^{(x+i\theta)/2}+e^{-(x+i\theta)/2}}
{|e^{(x+i\theta)/2}+e^{-(x+i\theta)/2}|}
\ee
As $\theta\in [0,\pi]$, the real part $\cos\gamma$ of this expression is non negative so that one can choose
$\gamma\in [-\frac{\pi}{2},\frac{\pi}{2}]$. Then forming the ratio of the imaginary part to the real part of the same expression
gives the value of $\tan \gamma$ and
\be
\gamma=\atan \left[\thf\left(\frac{x}{2}\right) \tan\left(\frac{\theta}{2}\right)\right] \ \mbox{with}\ \tan\left(\frac{\theta}{2}\right)
=\left(\frac{1-u}{1+u}\right)^{1/2}
\label{eq:defgamma}
\ee
One has the same expressions for $\gamma'$, replacing the variables $x$, $\theta$ and $u$ by $x'$, $\theta'$, $u'$. This leads to
\begin{multline}
\label{eq:fin2413}
\mathcal{E}_{24,13}=\mathcal{F}\!\!\!\!\!\!\sum_{m_z,m_z'=-\ell}^{\ell}\int_\mathbb{R}\! dx dx' \!\!\int_{-1}^{1}\!\!\!du du'
\!\left[\frac{(u'+\ch x')\ch x'}{(u+\ch x)\ch x}\right]^{s/2} \\
\times \frac{(-1)^{\ell} \Phi_{m_z}^{(\ell)*}(x,u) \Phi_{m_z'}^{(\ell)}(x',u')}{4\pi[(u+\ch x)(u'+\ch x')\ch x \ch x']^{1/4}} \\
\times 
\frac{e^{im_z\gamma} \langle \ell,m_z|\ell,m_x=0\rangle \langle \ell,m_x=0|\ell,m_z'\rangle e^{-im_z'\gamma'}}
{\left(\frac{e^{-x'}+\alpha e^{x'}}{1+\alpha}\right) (u+\ch x)+\left(\frac{e^{-x}+\alpha e^x}{1+\alpha}\right)(u'+\ch x')}
\end{multline}

In all the results (\ref{eq:findiag},\ref{eq:fin2423},\ref{eq:fin2414},\ref{eq:fin2413}) there appears a factor 
\be
\mathcal{F} = \frac{\mu_{\uparrow\downarrow}}{8\hbar^2} \int_0^{+\infty} \frac{dk_2}{k_2}.
\ee
This factor contains a diverging integral, making these last calculations not entirely rigorous. We have
checked however that always the same diverging integral is pulled out, even if one singles out a wavenumber other than $k_2$
(performing for example the change of variables $k_2=e^{-x}k_4$ and $k_1=e^{-x'} k_4$ in the integrals over $k_2$ and
$k_1$ in Eq.~(\ref{eq:foncrot14})). This is certainly due to the scale invariance of $dk_2/k_2=d(\ln k_2)$. 
Alternatively, one can write the integral equation for the $f_{m_z}^{(\ell)}$ deduced from the functional
derivatives of Eqs.~(\ref{eq:foncrotdiag},\ref{eq:foncrot23},\ref{eq:foncrot14},\ref{eq:E2413rot}) of the functional Eq.~(\ref{eq:fonc})
with respect to $f_{m_z}^{(\ell)*}$; at this stage, one has only used rotational invariance. Then,
one inserts the scale invariant ansatz (\ref{eq:scaleinv}), and
one obtains exactly the same integral equations for the $\Phi_{m_z}^{(\ell)}$ as those derived from the functional
derivatives of Eqs.~(\ref{eq:findiag},\ref{eq:fin2423},\ref{eq:fin2414},\ref{eq:fin2413}) with respect to $\Phi_{m_z}^{(\ell)*}$.

\subsection{Parity invariance}
\label{subsec:parite}

The term of index $m_z$ in the ansatz (\ref{eq:ansatzrot}) is simply multiplied by $(-1)^{m_z}$ under the action of parity $(\kk_2,\kk_4)
\rightarrow (-\kk_2,-\kk_4)$ \cite{CRAS4corps}. This means that the odd-$m_z$ components of $\Phi_{m_z}^{(\ell)}$ are
decoupled from the even-$m_z$ components of $\Phi_{m_z}^{(\ell)}$ in the integral equation. This property can also be obtained 
by an explicit calculation: first, for $m_z$ and $m_z'$ of different parities, the coupling amplitude between $|\ell,m_z\rangle$ and $|\ell,m_z'\rangle$ 
must vanish in Eq.~(\ref{eq:E2413rot}); this can be seen from Eq.~(\ref{eq:propriete}). 
Second, it also vanishes in Eqs.~(\ref{eq:fin2423},\ref{eq:fin2414}) after integration
over $\phi$ or $\phi'$: $L_x$ obeys the selection rule $\Delta m_z=\pm 1$, and
in an expansion of $e^{i\phi L_x}$ in powers of $\phi$, only even powers of $\phi$ and $L_x$ 
survive due to the parity of the denominator $\mathcal{D}$. In what follows, at a given angular momentum $\ell$, 
we shall distinguish the manifold of parity $(-1)^{\ell+1}$, where $\mathcal{E}_{24,13}$ and the contribution of $D(\kk_1,\kk_3)$
in Eq.~(\ref{eq:equint}) are zero, and the manifold of parity $(-1)^{\ell}$ where they are {\sl a priori} non zero. Note that, in the particular
case $\ell=0$, there exists only the manifold of parity $(-1)^{\ell}$.

\section{Final form and asymptotic analysis}

\subsection{Explicit form of the integral equation}
\label{subsec:forme_explicite}

By taking the functional derivative of $\mathcal{E}$ of Eq.~(\ref{eq:fonc}) with respect to $\Phi_{m_z}^{(\ell)}$, using
the forms (\ref{eq:findiag},\ref{eq:fin2423},\ref{eq:fin2414},\ref{eq:fin2413}) of the various terms
{and not forgetting the minus sign in front of the last contribution in Eq.~(\ref{eq:fonc})}, we obtain the form
of the integral equation (\ref{eq:equint}) maximally reduced by use of the rotational symmetry and of the scale invariance:
\begin{widetext}
\be
0=\left[\frac{\alpha}{(1+\alpha)^2}\left(1+\frac{u}{\ch x}\right)+\frac{e^{-x}+\alpha e^x}{2(\alpha+1)\ch x}\right]^{1/2} 
\Phi^{(\ell)}_{m_z}(x,u)
+ \int_\mathbb{R} dx' \int_{-1}^{1} du' \sum_{m_z'=-\ell}^{\ell} {K}_{m_z,m'_z}^{(\ell)} 
(x,u ; x',u' ; \alpha)\Phi^{(\ell)}_{m_z'}(x',u')
\label{eq:check}
\ee
with the following expression for the matrix kernel $K^{(\ell)}$:
\begin{multline}
{K}_{m_z,m'_z}^{(\ell)}(x,u ; x',u' ; \alpha)=\\
\left(\frac{e^x\ch x'}{e^{x'}\ch x}\right)^{s/2} \left(\frac{e^{x+x'}}{4\ch x\ch x'}\right)^{1/4}
\int_0^{2\pi} \frac{d\phi}{(2\pi)^2}
\frac{e^{-im_z\theta/2}\langle \ell,m_z|e^{i\phi L_x/\hbar}|\ell,m_z'\rangle e^{im_z'\theta'/2}}{\ch(x-x')+\frac{1}{1+\alpha}
[
{(u+e^{-x})(u'+e^{-x'})} + v v'\cos\phi]} \\
+\left(\frac{e^{-x}\ch x'}{e^{-x'}\ch x}\right)^{s/2} \left(\frac{e^{-x-x'}}{4\ch x\ch x'}\right)^{1/4}
\int_0^{2\pi} \frac{d\phi}{(2\pi)^2}
\frac{e^{im_z\theta/2}\langle \ell,m_z|e^{i\phi L_x/\hbar}|\ell,m_z'\rangle e^{-im_z'\theta'/2}}{\ch(x-x') +\frac{\alpha}{1+\alpha}
[
{(u+e^{x})(u'+e^{x'})}
+ v v'\cos\phi]} \\
- \frac{(-1)^{\ell}}{4\pi[(u+\ch x) (u'+\ch x')\ch x\ch x']^{1/4}}
\left(\frac{(u'+\ch x')\ch x'}{(u+\ch x)\ch x}\right)^{s/2}
\frac{e^{im_z\gamma}\langle \ell,m_z|\ell,m_x=0\rangle\langle \ell,m_x=0|\ell,m_z'\rangle e^{-im_z'\gamma'}}
{\left(\frac{e^{-x'}+\alpha e^{x'}}{1+\alpha}\right) (u+\ch x)
+\left(\frac{e^{-x}+\alpha e^x}{1+\alpha}\right)(u'+\ch x')}
\label{eq:noyau_matriciel}
\end{multline}
\end{widetext}
Here, the scaling exponent $s$ is purely imaginary, so that a four-body Efimov takes place in our $2+2$ fermionic problem
if Eq.~(\ref{eq:check}) has a non-identically zero solution $\Phi^{(\ell)}$ for some non-zero $s$. We recall that
the angle $\theta\in [0,\pi]$ is such that $u=\cos \theta$ and $v=(1-u^2)^{1/2}=\sin \theta$, and that the angle $\gamma$
is given by Eq.~(\ref{eq:defgamma}); the same relations hold among the primed variables.

The first, second and third contributions in Eq.~(\ref{eq:noyau_matriciel}) originate respectively from the terms
$D(\kk_2,\kk_3)$, $D(\kk_1,\kk_4)$ and $D(\kk_1,\kk_3)$ in the unreduced integral equation (\ref{eq:equint}); 
the diagonal term in Eq.~(\ref{eq:check}) emanates from the diagonal term of that equation.
The integrals over $\phi$ can be evaluated after insertion of a closure 
relation in the eigenbasis of $L_x$
\footnote{For $b_0>b_1>0$,
$\int_0^{2\pi} \frac{d\phi}{2\pi} \frac{e^{i m_x \phi}}
{b_0+b_1\cos\phi}=z_0^{|m_x|}/[(b_0-b_1)(b_0+b_1)]^{1/2}$
with $z_0=-b_1/\{b_0+[(b_0-b_1)(b_0+b_1)]^{1/2}\}$.}.
Importantly, the third contribution in Eq.~(\ref{eq:noyau_matriciel}) vanishes when $\ell+m_z$ or $\ell+m_z'$ are odd,
i.\,e.\ in the parity channel $(-1)^{\ell+1}$,
as shown by the property (\ref{eq:propriete}) and as already pointed out in section \ref{subsec:parite}.

It is interesting to note the decoupled form of the prefactors in each contribution of Eq.~(\ref{eq:noyau_matriciel}),
of the form $[f(x,u)]^{s/2+1/4}[f(x',u')]^{-s/2+1/4}$ with a function $f$ given by
$e^x/(2\ch x)$, $e^{-x}/(2\ch x)$ and $1/[(u+\ch x) \ch x]$, respectively. The fact that this function
$f(x,u)$ is not common to all contributions prevents one from suppressing the $s$ dependence of the matrix kernel $K^{(\ell)}$
by a simple gauge transform on $\Phi^{(\ell)}$: As expected, the $s$-dependence of the problem (\ref{eq:check}) is non trivial.

Our results (\ref{eq:check},\ref{eq:noyau_matriciel}) must obey the symmetry of the $2+2$ problem
under the exchange of $\uparrow$ and $\downarrow$. First, this exchange 
has the effect of changing the mass ratio $\alpha$ into its inverse $1/\alpha$,
see Eq.~(\ref{eq:defalpha}). Second, the momenta $\kk_2$ and $\kk_4$ in $D(\kk_2,\kk_4)$ are exchanged,
so that $x$ of Eq.~(\ref{eq:defx}) is changed into its opposite; this also reverts the direction of the quantization axis $\kk_2 \wedge \kk_4$
along which the angular momentum $m_z$ is measured in Eq.~(\ref{eq:ansatzrot}): it changes $m_z$ into $-m_z$ according to the identity \cite{CRAS4corps}
\be
e^{-i\pi L_x/\hbar} |\ell,m_z\rangle=(-1)^{\ell} |\ell, -m_z\rangle;
\ee
on the contrary, the non-oriented angle
$\theta_{24} \in [0,\pi]$ between $\kk_2$ and $\kk_4$ is unchanged, so that the variable $u$ is unaffected. Hence,
one must have
\be
K_{m_z,m_z'}^{(\ell)} (x,u ; x',u' ; \alpha) = K_{-m_z,-m_z'}^{(\ell)} (-x,u ; -x',u' ; \alpha^{-1})
\label{eq:belle_symetrie}
\ee
for all values of the argument and of the indices of the kernel. It is clear that Eq.~(\ref{eq:noyau_matriciel})
indeed obeys the symmetry requirement (\ref{eq:belle_symetrie}): the first and second contributions are interchanged, 
whereas the third one is invariant since
$\gamma$ is changed into $-\gamma$, see Eq.~(\ref{eq:defgamma}).
Note that our results also respect the parity invariance, see section \ref{subsec:parite}, and that the matrix kernel is hermitian
as our variational derivation guarantees:
\be
K_{m_z,m'_z}^{(\ell)} (x,u ; x',u' ; \alpha) = \left[K_{m_z',m_z}^{(\ell)} (x',u' ; x,u ; \alpha) \right]^*
\ee

\subsection{Recovering the three-body problem from four-body asymptotics}
\label{subsec:le3corpsretrouve}

The right-hand side of the integral equation (\ref{eq:check}) defines an operator $M^{(\ell)}(s)$ acting on the spinor functions
$\Phi_{m_z}^{(\ell)}(x,u)$. The spectrum of this operator is physically relevant, since a four-body Efimov effect takes place 
with an Efimov scaling exponent $s \in i\mathbb{R}$ if and only if one of the eigenvalue $\Omega$ of $M^{(\ell)}(s)$ is zero. As $M^{(\ell)}(s)$ is a hermitian operator,
since $s$ is here purely imaginary,
its spectrum is real and in general includes a discrete part and a continuous part.
The discrete spectrum corresponds to localized, square integrable eigenfunctions; we are able to determine it only numerically. 
The {expected contribution to the} continuous spectrum corresponds to extended functions, that explore arbitrarily large values of $|x|$; as we now explain, 
it can be determined analytically from the asymptotic analysis of the kernel (\ref{eq:noyau_matriciel}) when
$x$ and $x'$ tend to $\pm \infty$, by a generalisation of the discussion of reference \cite{CMP}.
{There is also an unexpected contribution to the continuous spectrum, whose analysis is deferred to section \ref{subsec:der_dritte_continuum}.}

\begin{figure}[tb]
\centerline{\includegraphics[width=8cm,clip=]{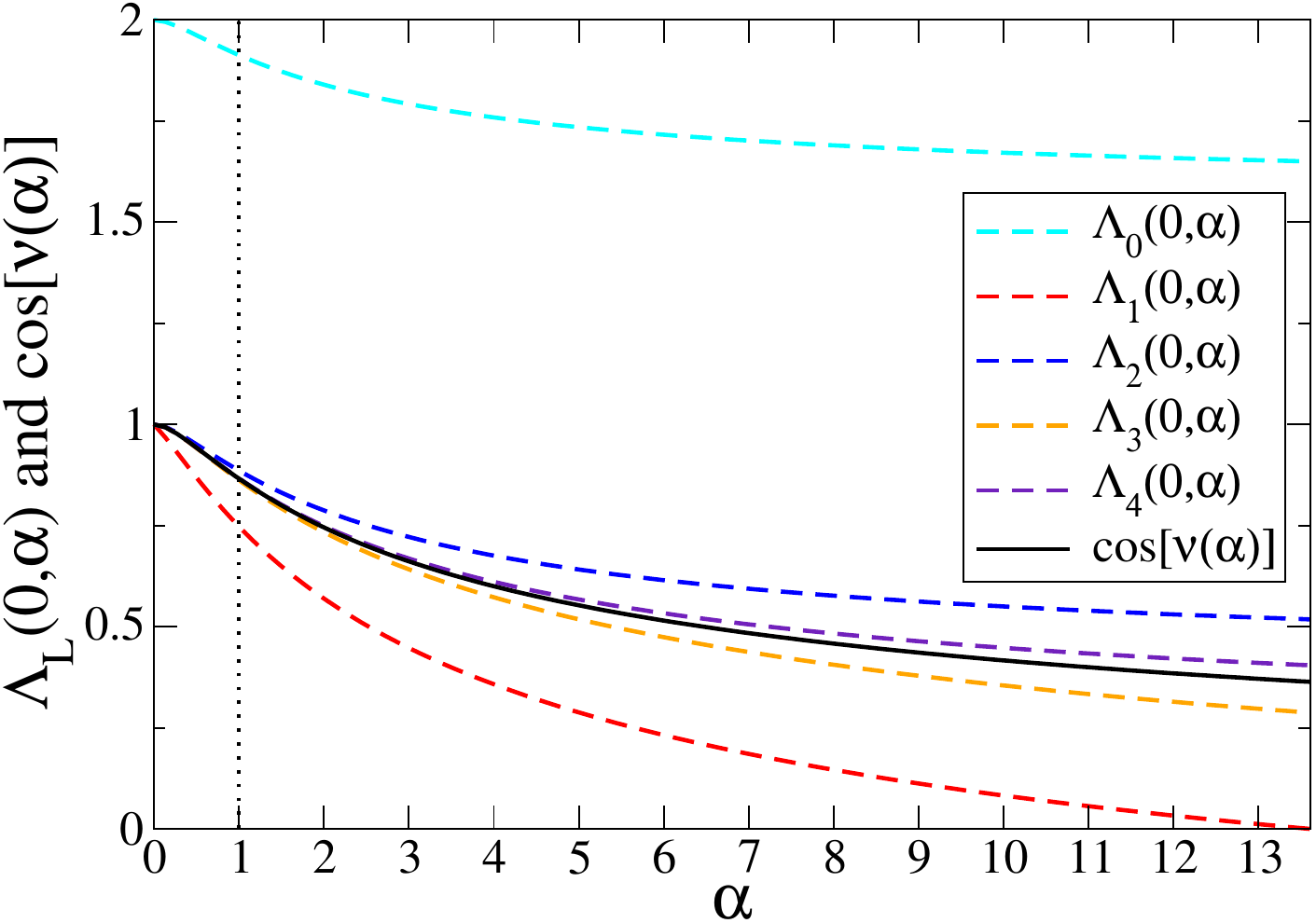}}
\caption{(Color online) Analytically obtained borders of the continuous spectrum of $M(s)$ ($s\in i\mathbb{R}^+$) corresponding
to the limits $x\to\pm\infty$.  It is a collection of components characterized by an angular momentum
quantum number $L$ of a three-body asymptotic problem (not to be confused with the total angular momentum $\ell$
of the four-body states), leading to a continuous set of eigenvalues ranging
from $\Lambda_L(0,\alpha)$ to $\cos[\nu(\alpha)]$ for $x\to -\infty$, and ranging from $\Lambda_L(0,1/\alpha)$ to
$\cos[\nu(1/\alpha)]$ for $x\to +\infty$, where $\alpha$ is the mass ratio
given by Eq.~(\ref{eq:defalpha}), the $\Lambda_L$ function is given by Eq.~(\ref{eq:defLam}) and
the mass angle $\nu$ is given by Eq.~(\ref{eq:def_angle_masse}). We plot $\cos[\nu(\alpha)]$
(black solid line) and  $\Lambda_L(0,\alpha)$ (color dashed lines, 
with $L=0,2,4$ from top to bottom above the solid line, and $L=1,3$ from bottom to top below the solid line)
as functions of $\alpha\in [0,\alpha_c(2;1)]$ where $\alpha_c(2;1)$ is the critical mass
ratio (\ref{eq:alphac21}) for the Efimov effect in the $\uparrow\uparrow\downarrow$ three-body problem. Due to the $\alpha
\leftrightarrow 1/\alpha$ symmetry of the $2+2$ fermionic problem, one can restrict to $\alpha\geq 1$ (that is to the
right of the vertical dotted line); the borders of the $x\to-\infty$ continuum can then be directly read on the figure,
and the ones of the $x\to+\infty$ continuum can be obtained by mentally folding back the $\alpha\leq 1 $ part
of the figure into the $\alpha\geq 1$ part.}
\label{fig:bordevin}
\end{figure}

\noindent{\bf\boldmath Sector $x\to +\infty, x'\to +\infty$:} Clearly, the diagonal part of $M^{(\ell)}(s)$ in Eq.~(\ref{eq:check}) 
tends exponentially rapidly to a finite and non-zero value, and the second and third contributions to the kernel in Eq.~(\ref{eq:noyau_matriciel})
tend exponentially fast to zero. In the first contribution in Eq.~(\ref{eq:noyau_matriciel}), the prefactor tends exponentially to unity
since $e^x/(2\ch x)\to 1$, and in the denominator of the integrand, all the $x$ or $x'$-dependent
terms are exponentially suppressed, except the first one $\ch(x-x')$ 
since no hypothesis must be made on the magnitude of the difference $x-x'$. The eigenvalue problem then asymptotically
reduces to
\begin{multline}
\Omega^{\rightarrow (\ell)} \tilde{\Phi}^{(\ell)}_{m_z}(x,u)=\left[\frac{\alpha(2+\alpha)}{(1+\alpha)^2}\right]^{1/2} \!\!\!\!\!\!\tilde{\Phi}^{(\ell)}_{m_z}(x,u)
+\int_\mathbb{R} dx'\!\!\!\sum_{m_z'=-\ell}^{\ell} \\
\int_{-1}^{1} \!\! du' \int_0^{2\pi}\!\!\!\!\frac{d\phi'}{(2\pi)^2} \frac{\langle \ell,m_z|e^{i\phi' L_x/\hbar}|\ell,m_z'\rangle}
{\ch(x-x')+\frac{uu'+vv'\cos\phi'}{1+\alpha}} {\tilde{\Phi}}^{(\ell)}_{m_z'}(x',u')
\end{multline}
where the arrow in the exponent of $\Omega$ indicates that $x$ and $x'$ tend to positive infinity, and
the phase factors $e^{-im_z\theta/2}$ and $ e^{i m_z'\theta'/2}$ have been eliminated by a gauge transform on the spinor, 
$\Phi_{m_z}^{(\ell)}(x,u)=e^{-i m_z\theta/2}\tilde{\Phi}_{m_z}^{(\ell)}(x,u)$. Then, one performs a spin rotation, by moving to the internal basis
of the eigenstates $|\ell, m_x\rangle$ of $L_x$, in which $e^{i\phi L_x/\hbar}$ is diagonal: the components 
$\tilde{\Phi}_{m_x}^{(\ell)}(x,u)$ are all decoupled. For a given $m_x$, with $|m_x|\leq \ell$,
the trick is to extend $\tilde{\Phi}_{m_x}^{(\ell)}(x,u)$ 
into a function of the real variable $x$ and of the vector $\nn$ on the two-dimensional unit sphere:
\be
F_{m_x}^{(\ell)}(x,\nn) \equiv \tilde{\Phi}_{m_x}^{(\ell)}(x,\cos\theta) e^{i m_x \phi}
\label{eq:extension}
\ee
where $\theta\in [0,\pi]$ and $\phi\in [0,2\pi]$ are the polar and azimuthal angles of $\nn$ in spherical coordinates, e.g.\
with respect to $x$ and $y$ axes, $\nn=(\cos\theta,\sin\theta\cos\phi, \sin\theta\sin\phi)$. 
In the phase factor $e^{i m_x\phi'}$ in the numerator and in $\cos\phi'$ in the denominator, 
one can then replace $\phi'$ by $\phi'-\phi$: the integrand is a periodic function of $\phi'$ of period $2\pi$ and
its integral has the same value whatever the interval of length $2\pi$ over which $\phi'$ runs. Then, one recognizes the scalar product
$\nn\cdot \nn' = u u'+v v'\cos(\phi-\phi')$ where $\nn'=(\cos\theta',\sin\theta'\cos\phi', \sin\theta'\sin\phi')$.
The eigenvalue problem is now
\begin{multline}
\Omega^{\rightarrow(\ell)}_{m_x} F_{m_x}^{(\ell)}(x,\nn) = \frac{[\alpha(2+\alpha)]^{1/2}}{1+\alpha} F_{m_x}^{(\ell)}(x,\nn) \\
+\int_\mathbb{R} dx' \int_{|\nn|=1}\frac{d^2n}{(2\pi)^2} \frac{F_{m_x}^{(\ell)}(x',\nn')}{\ch(x-x')+\frac{\nn\cdot \nn'}{1+\alpha}}
\end{multline}
The corresponding operator is invariant by translation along $x$ and by rotation of $\nn$ over the unit sphere. Its eigenfunctions
$F_{m_x}^{(\ell)}(x,\nn)$ can therefore be taken as plane waves of the variable $x$ and spherical harmonics of the variables $(\theta,\phi)$, 
with the {\sl same} quantum number $m_x$ (this is imposed by the form (\ref{eq:extension})) but with {\sl any}
integer quantum number $L\geq |m_x|$ for the total angular momentum:
\be
F_{m_x}^{(\ell)}(x,\nn)=e^{i k x} Y_{L}^{m_x}(\theta,\phi)
\ee
As usual for a rotationally invariant operator, the eigenvalue does not depend on $m_x$. It only depends on $L$, so it suffices 
to specialize to $m_x=0$, where $Y_{L}^{0}(\theta,\phi)\propto P_L(\cos\theta)$, where $P_L(X)$ is the Legendre polynomial of
degree $L$. Then, one gets the continuous spectrum ``to the right" ($x,x'\to +\infty$):
\be
\Omega_{m_x}^{\rightarrow(\ell)}(\alpha) \in \{\Lambda_L(ik,\alpha^{-1}),\forall k\in \mathbb{R},\forall L\geq |m_x|\}
\label{eq:continuum_droite}
\ee
The function $\Lambda_L$ of $s\in i\mathbb{R}$ and of the mass ratio 
was introduced and analytically calculated in \cite{Rittenhouse,Tignone}, generalizing previous results \cite{Werner3corps,Gasaneo}:
\begin{multline}
\label{eq:defLam}
\Lambda_L(s,\beta)\equiv  \frac{(1+2\beta)^{1/2}}{1+\beta} +
\int_{-1}^{1} \!\! du\, \! \int_{\mathbb{R}}\frac{dx}{2\pi} \frac{e^{-s x}P_L(u)}{\ch x
+\frac{\beta}{1+\beta} u}  \\
= \cos \nu(\beta) + \frac{1}{\sin \nu(\beta)} \int_{\frac{\pi}{2}-\nu(\beta)}^{\frac{\pi}{2}+\nu(\beta)}
\!\!d\theta\, P_L\left(\frac{\cos\theta}{\sin \nu(\beta)}\right) \frac{\sin (s\theta)}{\sin(s\pi)}
\end{multline}
where, in the second expression obtained after integration over $x$ \cite{Tignone}, 
\be
\nu(\beta)=\asin \frac{\beta}{1+\beta}
\label{eq:def_angle_masse}
\ee
is a mass angle.
For all $\beta>0$, 
it is found numerically for even $L$ that the maximal value of $\Lambda_L(s,\beta)$ over $s\in i\mathbb{R}^+$
is reached at $s=0$, and the minimal value is reached for $|s|\to+\infty$ (where $\Lambda_L(s,\beta)$ tends to
$\cos\nu(\beta)$). For odd $L$, the situation is found to be reversed: $\Lambda_L(s,\beta)$ is minimal at $s=0$
and maximal at infinity. To summarize, we expect that
\bea
\cos\nu(\beta) \leq \Lambda_L(s,\beta) \leq \Lambda_L(0,\beta) && \forall s\in i\mathbb{R}, L\ \mbox{even}, \nonumber \\
\Lambda_L(0,\beta)\leq \Lambda_L(s,\beta) \leq \cos\nu(\beta) && \forall s\in i\mathbb{R},  L\ \mbox{odd}
\label{eq:ordreconti}
\eea
This allows to determine the borders of the continuous component of
quantum number $L$ in Eq.~(\ref{eq:continuum_droite}), see Fig.~\ref{fig:bordevin}.
A physical explanation for the emergence of the function $\Lambda_L$  is postponed to the end of the section.

\noindent{\bf\boldmath Sector $x\to -\infty, x'\to -\infty$:} The calculation closely resembles the previous one, except that it is now
the second contribution in the right-hand side of Eq.~(\ref{eq:noyau_matriciel}) that survives. This was expected from the symmetry
relation (\ref{eq:belle_symetrie}). We arrive at the continuous spectrum ``to the left" ($x,x'\to-\infty$):
\be
\Omega_{m_x}^{\leftarrow(\ell)}(\alpha) \in \{\Lambda_L(ik,\alpha),\forall k\in \mathbb{R},\forall L\geq |m_x|\}
\label{eq:continuum_gauche}
\ee
that differs from (\ref{eq:continuum_droite}) by the occurrence of $\alpha$ (rather than $1/\alpha$) in the argument of
the $\Lambda_L$ function \footnote{If $x\to -\infty, x'\to +\infty$ or $x\to +\infty, x'\to-\infty$, the matrix 
kernel (\ref{eq:noyau_matriciel}) entirely tends exponentially to zero, which does not bring any significant new information.}.
The borders of the $L$-components of that continuum are plotted in Fig.~\ref{fig:bordevin} for the first few values
of $L$, using the numerically checked property (\ref{eq:ordreconti}).

\noindent{\bf Parity considerations:}
At fixed $\ell$, the results (\ref{eq:continuum_droite},\ref{eq:continuum_gauche}) are expressed in terms of the quantum number $m_x$, whereas the original problem only 
distinguishes between an even parity manifold ($m_z$ is even) and an odd parity manifold ($m_z$ is odd).
In practice, due to the property (\ref{eq:propriete}), the continua (\ref{eq:continuum_droite},\ref{eq:continuum_gauche}) with $L=0$ can be realized only in the manifold
of parity $(-1)^\ell$, at any considered total angular momentum $\ell$ (obviously one must then take $m_x=0$). 
The other continua (with $L\geq 1$) can all be realized, in both odd and even manifolds, for all values
of $\ell\geq 0$ 
\footnote{This is trivial for $\ell=0$. For $\ell\geq 1$, this
results from the fact that, for $L\geq 1$, one can take $m_x=1$. Then $\langle \ell, m_z|\ell,m_x=1\rangle\neq 0$,
except if $\ell$ is even and $m_z=0$ (in agreement with Eq.~(\ref{eq:propriete}), considering the $x\leftrightarrow y$ symmetry),
in which case one may return to the choice $m_x=0$ and use the fact that $\langle \ell,m_z=0|\ell, m_x=0\rangle \neq 0$ for even $\ell$.}.

\noindent{\bf Physical discussion:} 
The function $\Lambda_L(s,\beta)$ appears in the unitary three-body problem of two fermionic particles interacting
with a single distinguishable particle, $\beta$ being the mass ratio of the majority-to-minority species.
For $s\in i\mathbb{R}$ this function is given by the first form in Eq.~(\ref{eq:defLam}); 
it can be analytically extended to real values of $s$ using
e.g.\ the second form in Eq.~(\ref{eq:defLam}) \cite{Tignone}.
The zero-energy solutions of this three-body problem have an Efimov scaling exponent $s$: the three-body wavefunction 
scales as $R^{s-2}$, $R$ being the three-particle hyperradius, and the allowed values of $s$ at total angular momentum $L$ must solve 
\be
\Lambda_L(s,\beta)=0.
\label{eq:cond_s_3corps}
\ee
This three-body system exhibits an Efimov effect if and only if this equation has a purely imaginary solution $s \in i\mathbb{R}^*$.
This occurs only at odd $L$, starting from a mass ratio \cite{Petrov}
\be
\beta > \alpha_c(2;1)=13.60696\ldots
\label{eq:alphac21}
\ee
for $L=1$, and at increasingly larger critical mass ratios for $L=3, 5, \ldots$ \cite{Kartavtsev,Tignone}.

It is thus apparent that the asymptotic analysis of the $2+2$ fermionic problem brings up the three-body problem.
This is intuitive in position space: imagine that at fixed position $\rr_4\neq \mathbf{0}$ of the fourth particle (of spin $\downarrow$), 
the positions $(\rr_i)_{1\leq i\leq 3}$ of the other particles (of spin $\uparrow\uparrow\downarrow$) 
simultaneously tend to zero; then the four-body wavefunction
$\psi(\rr_1,\rr_2,\rr_3,\rr_4)$ must reproduce the behavior of the zero-energy scattering state of two $\uparrow$ and one $\downarrow$
particles, characterized by a mass ratio $\beta=m_\uparrow/m_\downarrow=\alpha$, in particular it must exhibit the same scaling exponents $s$
as the $2+1$ problem (see \S 5.3.6 in reference \cite{ZwergerLivre}); as these scaling exponents solve Eq.~(\ref{eq:cond_s_3corps}) with $\beta=\alpha$,
this explains the occurrence of $\Lambda_L(s,\alpha)$ in the spectrum (\ref{eq:continuum_gauche}) 
{\footnote{{Let us explain more physically why the $\Lambda_L$ function appears in the expression of the continuous spectrum.
The idea is to consider a physical state of the $2+2$ fermionic system, corresponding to a non-zero square integrable solution $\Phi_{m_z}^{(\ell)}(x,u)$
of Eq.~(\ref{eq:check}),
and to see how the four-body wavefunction scales when three particles, say $1,2$ and $3$, converge to the same location, the fourth particle being at some other 
fixed location.
As we have seen, an extended eigenstate of the continuum that varies for $x\to -\infty$ as $e^{ikx}e^{i m_z \theta/2} \langle \ell,m_z|\ell,m_x=0\rangle P_L(u)$, 
$k\in \mathbb{R}$, has an eigenvalue $\Omega=\Lambda_L(ik,\alpha)$.
According to the analytic continuation argument of reference \cite{CMP}, this implies that the $\Omega=0$ localized eigenstate
$\Phi_{m_z}^{(\ell)}(x,u)$ vanishes for $x\to -\infty$ as $e^{\kappa x} e^{i m_z \theta/2} \langle \ell,m_z|\ell,m_x=0\rangle P_L(u)$,
where the real quantity $\kappa$ is the minimal positive root of 
\[
\Lambda_L(\kappa,\alpha)=0
\]
with $L$ chosen so as to minimize $\kappa$ (minimizing $\kappa$ amounts to selecting the most slowly decreasing exponential function $e^{\kappa x}$,
that is the leading contribution for $x\to-\infty$). 
This implies that $\kappa$ is one of the possible scaling exponents $s_3$ of the $2+1$ fermionic problem, see Eq.~(\ref{eq:cond_s_3corps}).
In order to determine the limit of Eq.~(\ref{eq:psi24}) when $r_{13}$ and $|\rr_2-\RR_{13}|\to 0$ tend to zero with the same scaling
law, that is both are proportional to the vanishing hyperradius $R_{123}$ of particles $1$, $2$ and $3$, we determine the large
$\kk_2$ limit of the integrand of $\psi_{24}$ in Eq.~(\ref{eq:psi24}) at fixed $\kk_4$: Omitting to write the angular part for simplicity, we find that
$D(\kk_2,\kk_4)$ scales as $k_2^{-2-\kappa} k_4^{\kappa-s-3/2}$, so that its Fourier transform, according to the usual
power-law counting argument, scales as $|\rr_2-\RR_{13}|^{\kappa-1} |\rr_4-\RR_{13}|^{s-\kappa-3/2}$. {
The same reasoning applies to $\psi_{14}$.}
At fixed $|\rr_4-\RR_{13}|>0$,
the four-body wavefunction therefore scales as $R_{123}^{\kappa-2}=R_{123}^{s_3-2}$, exactly as predicted by Eq.~(5.179) of 
reference \cite{ZwergerLivre}. This whole discussion is formal for the $2+2$ fermionic problem since, as we shall see, there is
no four-body Efimov effect, but it explicitly applies to the $3+1$ fermionic problem and nicely completes reference \cite{CMP}.}}}. 
\setcounter{notekappa}{\thefootnote}
Even if $\ell=0$ for the four-body system, $L$ can take
any value, as the angular momentum can be distributed among particle $4$ and the first three particles.
The equivalent in momentum space of the considered limit is to have divergent $(\kk_i)_{1\leq i\leq 3}$ at fixed $\kk_4$, which, due
to scale invariance, is
equivalent to having $\kk_4\to \mathbf{0}$ at fixed $(\kk_i)_{1\leq i\leq 3}$, that is $x\to -\infty$ according to Eq.~(\ref{eq:defx}).
This is why $\beta=\alpha$ corresponds to the spectrum (\ref{eq:continuum_gauche}). A similar reasoning with $\rr_2$ fixed
with $(\rr_i)_{i\neq 2}$ tending to zero leads to $\beta=1/\alpha$ and $x\to +\infty$, as for the spectrum (\ref{eq:continuum_droite}).

\subsection{A third, unexpected continuum}
\label{subsec:der_dritte_continuum}

The first two contributions in Eq.~(\ref{eq:noyau_matriciel}) are innocuous: the denominator in their integrands cannot vanish,
see Eqs.~(\ref{eq:D2423},\ref{eq:D2423bis}) and Eqs.~(\ref{eq:D2414},\ref{eq:D2414bis}), and, as we have seen, they have a short range 
in the $(x,x')$ space. On the contrary, the third contribution in Eq.~(\ref{eq:noyau_matriciel}),
which is non-zero only in the $(-1)^\ell$ parity sector, diverges when $(x,u)\to (0,-1)$
or $(x',u')\to (0,-1)$. This creates doubt about the bounded nature of the eigenvalues of $M(s)$,
$s\in i\mathbb{R}$, for that parity. We investigate this problem mathematically in the appendix \ref{app:Mborne} and we
conclude that $M(s)$ is bounded.

Physically, this divergence of the kernel leads to a quite interesting effect: the emergence of a third component 
of the continuous spectrum of $M(s)$, different from the previously discussed $x\to \pm \infty$ continua.
The intuitive idea is that one can turn the eigenvalue problem $\Omega \Phi=M(s) \Phi$
into an integral equation with a bounded kernel through an appropriate change of variables, with the consequence 
that one of the new variables, that we shall call $t$, can tend to $-\infty$, in which case the eigenvector $\Phi$
takes a plane wave structure $\propto \exp(i k t)$, $k\in \mathbb{R}$,  with a spectrum:
\begin{equation}
\Omega^{\odot (\ell)}\in \left\{\frac{1}{\sqrt{2}}[1-\frac{(-1)^{\ell}}{\ch(k\pi/2)}], \forall k\in \mathbb{R}\right\}
\ [\mbox{parity} (-1)^{\ell}]
\label{eq:dritte}
\end{equation}
This is an unexpected feature of the $2+2$ fermionic problem, absent in the $3+1$ fermionic case \cite{CMP}.

To obtain this result, we construct a local approximation to the integral equation in the vicinity
of $(x,u)=(0,-1), (x',u')=(0,-1)$, keeping only the leading diverging contributions.
We use 
\be
y\equiv\pi-\theta
\ee
rather than $u=\cos\theta$ as integration variable, so that $y,y'\to 0$ when $u, u'\to -1$.
This pulls out a Jacobian $\sin y'$ that we absorb 
(in a way preserving the hermiticity of the problem) with a change of function. We also take into account
the fact that the third, diverging contribution in Eq.~(\ref{eq:noyau_matriciel}) involves a projector onto the
$|\ell, m_x=0\rangle$ state, and that the phase factors $e^{i m_z \gamma}$ and $e^{-i m_z' \gamma'}$,
$[(u+\ch x)\ch x]^{-s/2}$ and $[(u'+\ch x')\ch x']^{s/2}$
can be eliminated by a change of gauge. Hence the ansatz
\be
\Phi_{m_z}^{(\ell)}(x,u)=\frac{(x^2+y^2)^{-s/2}}{y^{1/2}}  e^{i m_z \gamma} \langle \ell, m_z| \ell, m_x=0\rangle \Phi(x,y)
\label{eq:ansatz_pas1}
\ee
where $\sin y$ was linearized, $u+\ch x$ was quadratized.  The resulting local eigenvalue problem is
\begin{multline}
\!\!\!\!\!\!{\left(\Omega-\frac{1}{\sqrt{2}}\right)}\Phi(x,y)= -\frac{(-1)^\ell}{2^{1/2}\pi} 
\int_{D} \frac{dx'dy' (yy')^{1/2}}{[(x^2+y^2)(x'^2+y'^2)]^{1/4}}  \\
\times \frac{\Phi(x',y')}{x^2+y^2+x'^2+y'^2}
\end{multline}
where we have conveniently restricted the integration to the {upper half ($y>0$)} of the disk $D$ of radius $\rho_0\ll 1$ centered in
$(0,0)$. In polar coordinates
\be
(x,y) = (\rho \cos \psi, \rho \sin \psi)
\label{eq:def_rho_psi}
\ee
only $(yy')^{1/2}$ depends on $\psi$ in the kernel.  It depends on $\psi$ in a factorised way so that
$\Phi$ is also factorised:
\be
\Phi(x,y)=\rho^{-1/2} \Phi(\rho) (\sin\psi)^{1/2}
\label{eq:ansatz_pas2}
\ee
and, since $\int_0^\pi d\psi' \sin \psi'=2$,
\be
{\left(\Omega-\frac{1}{\sqrt{2}}\right)} \Phi(\rho)=-\frac{(-1)^\ell 2^{1/2}}{\pi}
\int_0^{\rho_0} d\rho' \frac{(\rho\rho')^{1/2}}{\rho^2+\rho'^2} \Phi(\rho')
\ee
The scale invariance of this kernel motivates the logarithmic change of variable
\be
t=\ln \frac{\rho}{\rho_0}\ \ \mbox{and}\ \ \phi(t)=\left(\frac{\rho}{\rho_0}\right)^{1/2}\Phi(\rho)
\label{eq:deft}
\ee
The resulting eigenvalue problem
\be
\Omega\phi(t)=\frac{1}{2^{1/2}}\phi(t)-\frac{(-1)^\ell}{2^{1/2}\pi} \int_{-\infty}^{0}
\frac{dt' \phi(t')}{\ch(t-t')}
\label{eq:enfin_borne}
\ee
admits Eq.~(\ref{eq:dritte}) as a continuous spectrum {with eigenfunctions $\phi(t)$ that are for $t\to-\infty$ linear
superpositions of $e^{ikt}$ and $e^{-ikt}$},
since $\int_{\mathbb{R}} dt e^{ikt}/\ch t=\pi/\ch(k\pi/2)$; we have checked numerically that it
has no discrete eigenvalue
\footnote{{The continuous spectrum $\Omega^{\odot(\ell)}$ can be recovered by keeping only but exactly 
the last contribution in Eq.~(\ref{eq:noyau_matriciel}) to the matrix kernel
$K_{m_z,m_z'}^{(\ell)}$ of Eq.~(\ref{eq:check}), that is without resorting to a local approximation of this contribution around $(x,u)=(0,-1)$. The explicit calculation remains simple for a unit mass ratio $\alpha=1$.
The eigenvectors of the resulting operator are then of the form $\Phi_{m_z}^{(\ell)}(x,u)=e^{i m_z \gamma} \langle \ell,m_z|\ell,m_x=0\rangle \Phi(x,u)$ 
with the ansatz $\Phi(x,u)=(\ch x)^{-(s-1/2)/2}(u+\ch x)^{-(s+7/2)/2} \chi(\sqrt{2}k_{24}/K_{24})$, $k_{24}$ and $K_{24}$ being the relative and center-of-mass wave 
numbers of particles $2$ and $4$,
so that $2k_{24}/K_{24}=(\frac{\ch x-u}{\ch x+u})^{1/2}$.
One then obtains the integral equation for $\chi(k)$:
$\Omega (2k^2+1)^{1/2}\chi(k)=(k^2+1)^{1/2} \chi(k) -\frac{2(-1)^{\ell}}{\pi} \int_0^{+\infty} dk' k'^2\chi(k')/(1+k^2+k'^2)$. Further setting 
$\chi(k)=k^{-3/2}(1+2k^2)^{-1/4}\psi(t)$ with $k=\exp(-t)$, one obtains $\Omega \psi(t)=\left(\frac{1+e^{-2t}}{1+2 e^{-2t}}\right)^{1/2} \psi(t)
-\frac{2(-1)^{\ell}}{\pi}\int_{\mathbb{R}} dt' \frac{\psi(t') \exp[-3(t+t')/2]}{(1+e^{-2t}+e^{-2t'}) [(1+2e^{-2t})(1+2e^{-2t'})]^{1/4}}$.
The $t\to-\infty$ continuum of that eigenvalue problem solves $(\sqrt{2}\Omega-1)\psi_\infty(t)=-\frac{(-1)^\ell}{\pi}\int_\mathbb{R} dt' \frac{\psi_\infty(t')}{\ch(t-t')}$,
with plane wave solutions $\psi_\infty(t')=e^{ikt}$ reproducing (\ref{eq:dritte}). The ansatz for $\Phi(x,u)$ results from the fact that $D(\kk_2,\kk_4)$ is of the form
$K_{24}^{-(s+7/2)} P_\ell(\eee_z\cdot\KK_{24}/K_{24}) \chi(\sqrt{2}k_{24}/K_{24})$ ($P_\ell$ is a Legendre polynomial), 
which is apparent if one turns back to Eq.~(\ref{eq:equint}) and realizes that its last contribution conserves the total wave vector $\KK_{24}$ (up to a sign).}}.
\setcounter{notesimon}{\thefootnote}

\noindent{\bf Physical interpretation:}
We collect Eqs.~(\ref{eq:ansatz_pas1},\ref{eq:ansatz_pas2},\ref{eq:deft}), {taking as a particular
solution of Eq.~(\ref{eq:enfin_borne}) at $t$ large and negative} 
the function $\phi(t)=1$, {which corresponds to an asymptotic plane wave in $t$ space with
a vanishing wave vector, that is to $k=0$ in Eq.~(\ref{eq:dritte})}
\footnote{{In reality, the physical solution for the tensor $\Phi_{m_z}^{(\ell)}(x,u)$ must correspond to a zero-eigenvalue of the matrix $M^{(\ell)}(s)$. In this respect, $k=0$ is acceptable only for $\ell$ even.
Furthermore,  as we shall see in Appendix \ref{appen:conjecture} [see the footnote 
called above Eq.~(\ref{eq:cond_quantif})], when $\Omega=0$, $\phi(t)$ actually scales linearly in $t$
for $t\to -\infty$, so that $D(\kk_2,\kk_4)$ actually diverges as $\ln (|\kk_2+\kk_4|/k_{24})/(|\kk_2+\kk_4|/k_{24})^{s+3/2}$ when $\kk_2+\kk_4\to \mathbf{0}$. 
This would result in a $\ln (r_{24}/|\RR_{24}-\RR_{13}|)/r_{24}$ divergence of the function $\mathcal{A}_{13}(\rr_2-\RR_{13},\rr_4-\RR_{13})$ at
$r_{24}=0$, which is physically opaque. Let us keep in mind however that, according to section \ref{sec:quete} there is no four-body Efimov effect in the $2+2$
problem, so that Eq.~(\ref{eq:check}) does not actually support any non-identically zero solution $\Phi^{(\ell)}$ if $s\in i\mathbb{R}$. }}.
Restricting for simplicity to a zero total angular momentum $\ell=0$
\footnote{{For $\ell>0$, one can use the identity:
$\sum_{m_z=-\ell}^{\ell} [Y_\ell^{m_z}(\eee\cdot \eee_z,\eee_{4\bot 2}\cdot\eee_z, \eee_{24}\cdot\eee_z)]^* e^{im_z(\gamma_{24}+\theta_{24}/2)}
\langle \ell,m_z|\ell,m_x=0\rangle=s_+ Y_\ell^{0}(\hat{\KK}_{24})$ where
$\hat{\KK}_{24}=(\kk_2+\kk_4)/|\kk_2+\kk_4|$, we recall that $\gamma_{24}=\tau_{24}-\theta_{24}/2$ and $\tau_{24}$ is the angle between 
$\kk_2$ and $\kk_2+\kk_4$, the spherical harmonics notation is as in footnote [\thenoteylm] and the sign $s_+$ is defined in footnote [\thenotespm]. This identity is also useful for
footnote [\thenotesimon].}}, 
we then find that
\be
\Phi^{(0)}_0(x,u) \underset{(x,u)\to (0,-1)}{\propto} \frac{1}{\rho^{s+3/2}} 
\ee
A more inspiring writing is obtained in terms of the center-of-mass and relative wavevectors $\KK_{24}=\kk_2+\kk_4$ and
$\kk_{24}=(\kk_2-\alpha\kk_4)/(1+\alpha)$ of particles 2 and 4: 
\be
\Phi^{(0)}(x,u) \underset{K_{24}/k_{24}\to 0} \propto \left(\frac{k_{24}}{K_{24}}\right)^{s+3/2}
\ee
One has indeed $\rho^2\simeq 2(u+\ch x)$ and $K_{24}^2=2 k_2^2 e^x(u+\ch x)$, so that $K_{24}$ and $\rho$ vanish
in the same way when $(x,u)\to (0,-1)$; also the ratio $K_{24}/k_{24}$ tends to zero if and only if $u+\ch x\to 0$
\footnote{{If $k_{24}\to +\infty$, $|x|$ necessarily tends to $+\infty$ since $|u|\leq 1$, in which case $K_{24}$ and $k_{24}$ both
diverge as $e^{|x|}$ and have a non-vanishing ratio.}}. 
Restricting to a small neighbourhood of the singularity, $K_{24} < \epsilon k_{24}$, where $\epsilon \ll 1$, 
we can in the ansatz (\ref{eq:scaleinv}) approximate the factor $(\ch x)^{s+3/2}$ by one and, in the denominator,
approximate $k_2^2+k_4^2=2 k_{24}^2+2\frac{\alpha-1}{\alpha+1} \kk_{24}\cdot \KK_{24} + \frac{1+\alpha^2}{(1+\alpha)^2} K_{24}^2$
by its leading order approximation $2 k_{24}^2$ to isolate the singular behavior of $D(\kk_2,\kk_4)$:
\be
D_{\rm sing}(\kk_2,\kk_4) \propto \frac{1}{k_{24}^{s+7/2}} \left(\frac{k_{24}}{K_{24}}\right)^{s+3/2}
\ee

The key idea is then to see how this translates into a singularity of the regular part $\mathcal{A}_{13}$ of the four-body
wavefunction that appears in the Wigner-Bethe-Peierls contact condition (\ref{eq:BethePeierls}). As we have seen below
Eq.~(\ref{eq:un_example}), $\mathcal{A}_{13}=\mathcal{A}$ is related to $D(\kk_2,\kk_4)$ by a Fourier transform; using $(\kk_{24},\KK_{24})$
rather than $(\kk_2,\kk_4)$ as integration variables, and the fact that $\kk_2\cdot \rr_2 +\kk_4\cdot \rr_4=\kk_{24}\cdot \rr_{24}+\KK_{24}\cdot \RR_{24}$,
where $\rr_{24}=\rr_2-\rr_4$ and $\RR_{24}=(m_2\rr_2+m_4\rr_4)/(m_2+m_4)$ are the relative and center-of-mass coordinates of the 
particles $2$ and $4$, we obtain for the contribution to $\mathcal{A}$ of the singularity of $D$:
\begin{multline}
\mathcal{A}_{\rm sing}(\rr_2-\RR_{13},\rr_4-\RR_{13}) \propto \int_{K_{24}< \epsilon k_{24}} \!\!\!\!\!\!\!\! \!\!\!\!\!\!\!\!\!\!\!\!
 d^3k_{24} d^3K_{24} e^{i\KK_{24}\cdot (\RR_{24}-\RR_{13})}  \\
\times \frac{e^{i\kk_{24}\cdot \rr_{24}}}{k_{24}^{s+7/2}} \left(\frac{k_{24}}{K_{24}}\right)^{s+3/2}
\end{multline}
Integrating over the solid angles for $\kk_{24}$ and $\KK_{24}$, performing the change of variable $K_{24}=q k_{24} r_{24}/|\RR_{24}-\RR_{13}|$ at fixed $k_{24}$, 
changing the order of integration over $k_{24}$ and $q$ and finally integrating over $k_{24}$
\footnote{{We used $\int_0^{+\infty} dy \sin(ay) \sin(by)/y^{s+1/2}=\frac{1}{2} \Gamma(\frac{1}{2}-s)$ $ \sin[\frac{\pi}{4}(1-2s)]
[|a-b|^{s-\frac{1}{2}}-|a+b|^{s-\frac{1}{2}}]$, for $s\in i\mathbb{R}$, $a$ and $b$ real numbers that differ in absolute value.}}
we obtain
\begin{multline}
\mathcal{A}_{\rm sing}(\rr_2-\RR_{13},\rr_4-\RR_{13}) \propto \frac{|\RR_{24}-\RR_{13}|^{s-3/2}}{r_{24}} \\
\times
\int_0^{\epsilon |\RR_{24}-\RR_{13}|/{r_{24}}} \frac{dq}{q^{s+1/2}} [ |q-1|^{s-1/2}-(q+1)^{s-1/2}]
\label{eq:Asingjol}
\end{multline}
It then becomes obvious that the singularity in $D(\kk_2,\kk_4)$ at $\kk_2+\kk_4=\mathbf{0}$
is linked to a $1/r_{24}$ divergence of the regular part $\mathcal{A}_{13}$ of the four-body wavefunction at $r_{24}=0$
\footnote{{The integrand in Eq.~(\ref{eq:Asingjol}) is $O(1/q^2)$ when $q\to +\infty$, so that the integral converges when $r_{24}\to 0$
at fixed non-zero $\RR_{24}-\RR_{13}$. Furthermore it converges to a non-zero value (e.g.\ to $\pi$ for $s=0$).}}
\footnote{{If one takes a function $D(\kk_2,\kk_4)$ with no singularity in $\kk_2+\kk_4=\mathbf{0}$, for example $\Phi^{(0)}_0(x,u)=\exp(-\kappa |x|)$
in the ansatz (\ref{eq:scaleinv}) as in footnote {[\thenotekappa]} 
for $\ell=L=0$, one finds by an explicit calculation that $\mathcal{A}(\rr_2-\RR_{13},\rr_4-\RR_{13})$ is finite for
$\rr_2=\rr_4=\RR_{24}$.}}.
This was physically expected: $\mathcal{A}_{13}(\rr_2-\RR_{13},\rr_4-\RR_{13})$ is essentially the wavefunction of particles $2$ and $4$
knowing that particles $1$ and $3$ have converged to the same location in the $s$-wave; since $2$ and $4$ {are} in different spin states, they interact in the $s$-wave
and must be sensitive to the $2-4$ Wigner-Bethe-Peierls contact conditions, which implies a $1/r_{24}$ divergence when $r_{24}\to 0$.
Such a divergence of $\mathcal{A}_{13}$ was already pointed out in the scattering problem of two $\uparrow\downarrow$ dimers in reference
\cite{dimdim}, and in the general $N_\uparrow+N_\downarrow$ fermionic problem when $N_\uparrow\geq 2$ and $N_\downarrow \geq 2$
in reference \cite{relgen} (see footnote 20 of that reference) 
\footnote{{There is a paradox here. As shown by Eq.~(\ref{eq:Hpsi}), the action of the Hamiltonian $H$ on $\psi_{24}(\rr_1,\rr_2,\rr_3,\rr_4)$
leads to a $\delta(\rr_1-\rr_3)$ distribution, not to a $\delta(\rr_2-\rr_4)$ distribution. This shows that $\psi_{24}$ is the so-called $1-3$ Faddeev
component, and it cannot have any $1/r_{24}$ singularity. How can $\mathcal{A}_{13}$ then have such a singularity?
The answer as usual lies in the order of the limits. At fixed non-zero $r_{13}$, it is apparent that the function $u(r_{13})$ in Eq.~(\ref{eq:psi24}),
through its dependence on $q_{13}$ given by Eq.~(\ref{eq:defq13}),
provides an ultraviolet cut-off of order $1/r_{13}$ in the $(\kk_2,\kk_4)$ wavevector space, so that $\psi_{24}$ cannot diverge when $r_{24}\to 0$.
But if one first takes the $r_{13}\to 0$ limit, the function $u(r_{13})$ is replaced by its equivalent $1/(4\pi r_{13})$ which has no momentum dependence:
The wavevector cut-off is set to infinity and a $1/r_{24}$ divergence in $\lim_{r_{13}\to 0} (r_{13} \psi_{24})$ can now take place
at $r_{24}=0$.}}.

This interpretation of the matrix kernel singularity at $(x,u)=(0,-1)$ has a simple, though illuminating implication: The $1/r_{24}$ divergence in $\mathcal{A}_{13}$
can take place only when the particles $2$ and $4$ converge to the same location in the relative partial wave of zero angular momentum, since
$\uparrow$ and $\downarrow$ particles resonantly interact only in the $s$-wave. In such a configuration the angular momentum $\ell$ of the function
$\mathcal{A}_{13}$ (that is of the whole system) is carried out by the center of mass motion of particles $2$ and $4$ with respect to $\RR_{13}$;
in this case there is a univocal link between the angular momentum $\ell$ and the parity, as for single particle systems,
and the parity of $\mathcal{A}_{13}$ must be $(-1)^\ell$.
This explains why the singularity at $(x,u)=(0,-1)$, and ultimately the third continuum (\ref{eq:dritte}), can appear only in that parity
channel
\footnote{{This reasoning can be transposed to the case of four identical bosons, when three of them converge to the same location 
in the relative three-body channel where the Efimov effect takes place. As this channel has a zero angular momentum and an even parity, 
this implies that, in such a configuration, the total internal angular momentum  $\ell$
of the four-body system is carried by the relative motion of the fourth boson with respect to the center of mass of the first three bosons, leading
to a global parity $(-1)^\ell$. This indicates that, for $\ell\neq 0$, 
the four-boson unitary system in an isotropic harmonic trap should have interacting states in the $(-1)^{\ell+1}$ parity sector that are immune to the three-body Efimov
effect. Such ``universal" states have indeed been observed numerically in reference \cite{HammerCR}, but for a total internal angular momentum $\ell=0$: 
this observation cannot be explained by our reasoning. }}.

\section{Search for the four-body Efimov effect}
\label{sec:quete}

\begin{figure*}[tb]
\centerline{\includegraphics[width=9cm,clip=]{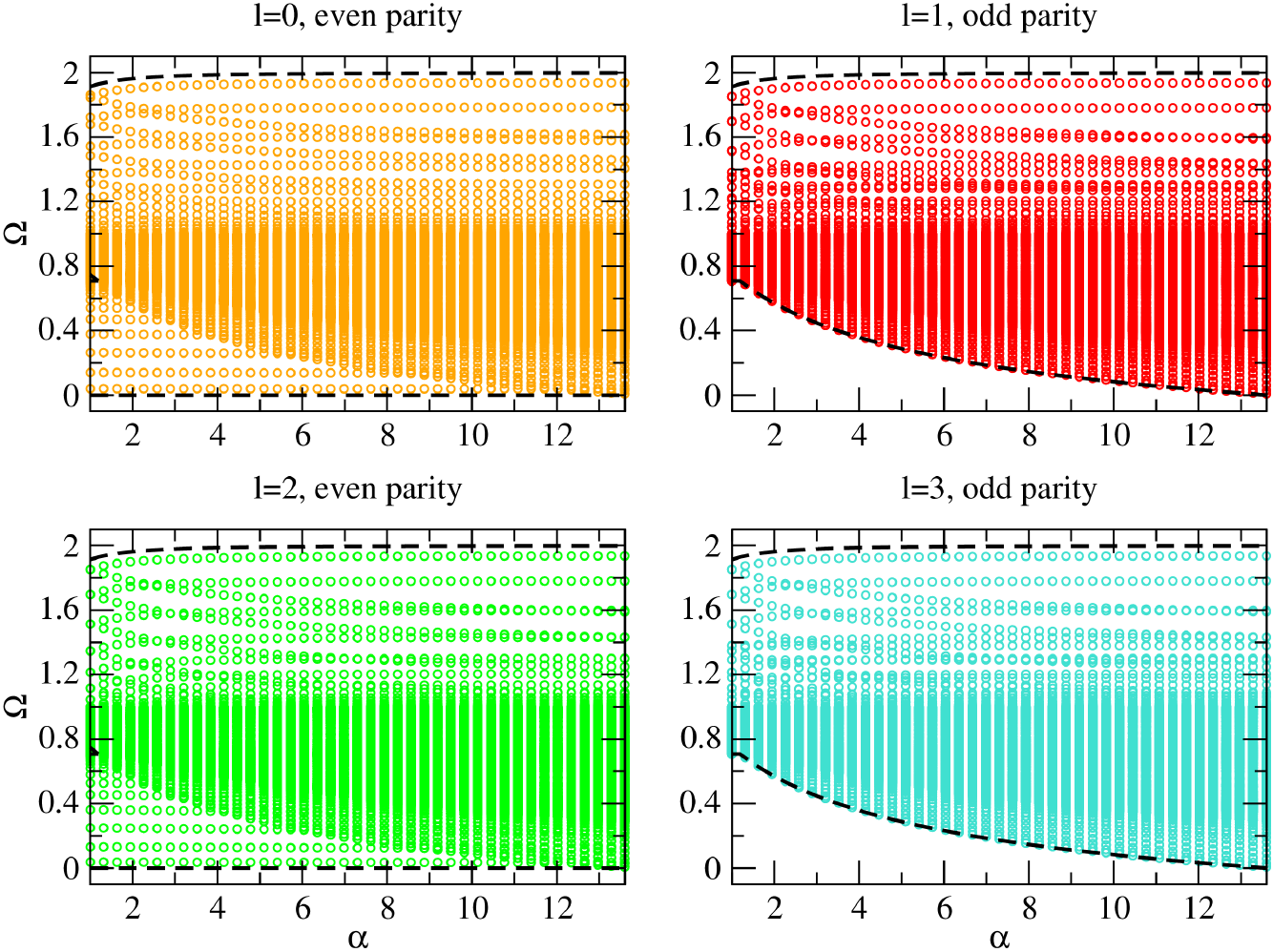} 
\includegraphics[width=9cm,clip=]{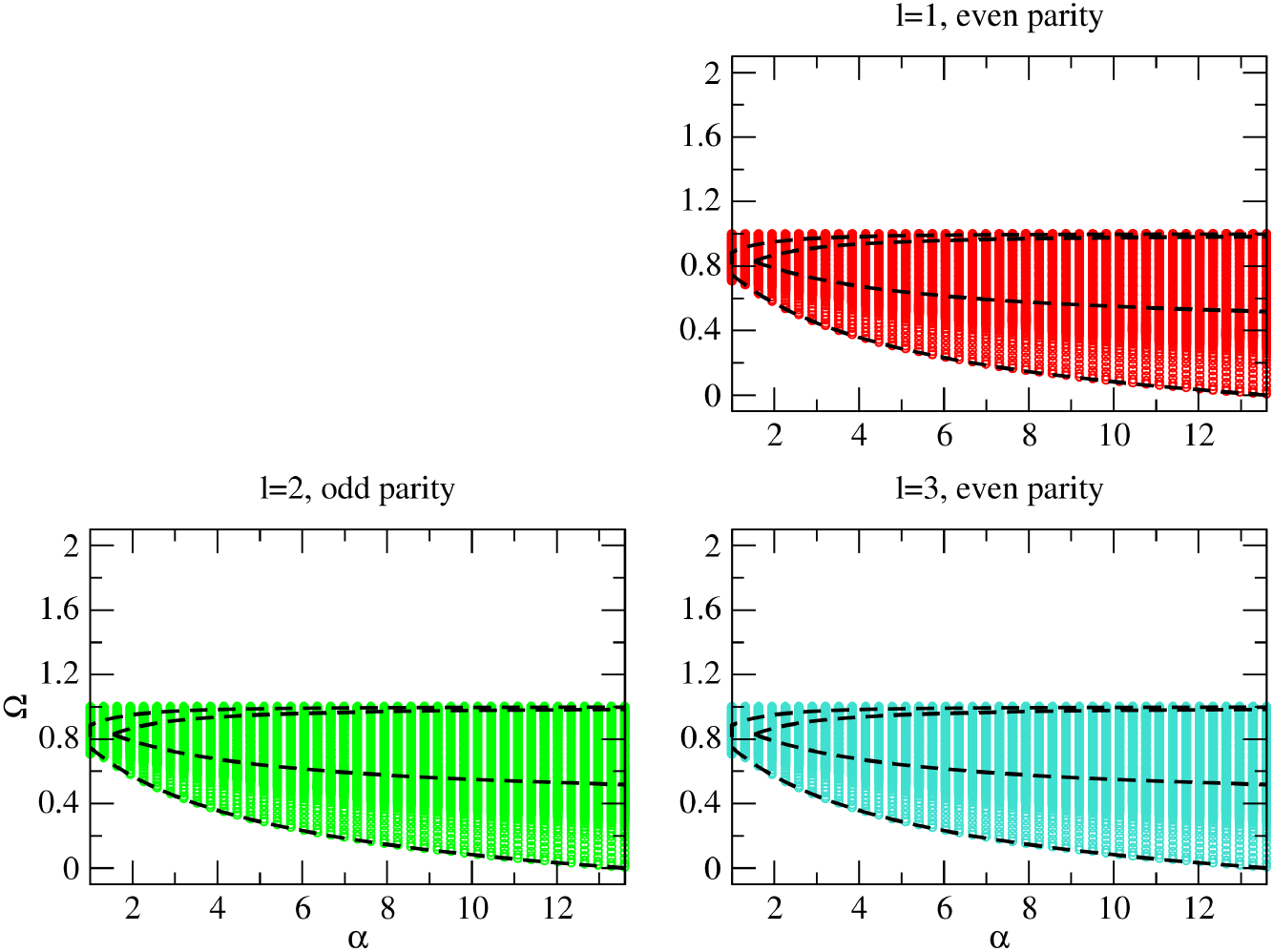}}
\caption{{(Color online)} Eigenvalues of $M(s=0)$ as functions of the mass ratio $\alpha\in [1,13.6]$, 
for various values of the angular momentum $\ell$ and the parity  {[left half for the parity $(-1)^\ell$,
right half for the parity $(-1)^{\ell+1}$]}, obtained numerically after discretisation and truncation 
of the variables $x$ and $\theta$ in the zone $\rho\equiv[x^2+(\pi-\theta)^2]^{1/2}>\rho_0$ 
and of their log-polar versions $t\equiv \ln(\rho/\rho_0)$ and $\psi$
in the zone $\rho<\rho_0$ (see text): $x_{\rm max}=-x_{\rm min}=12$, $dx=1/5$, $d\theta \simeq \pi/15$,
$\rho_0=2/5$, $t_{\rm min}=-12$, $dt=1/5$, $d\psi=\pi/15$, with the Gauss-Legendre integration scheme for the integrals
over $\theta$ and $\psi$, and the mid-point integration formula for the integrals over $x$ and $t$.
For the $(-1)^{\ell+1}$ parity channels the $\rho<\rho_0$ zone is not useful and is not included in the numerics.
The boundaries of the continuous spectrum of $M(s)$ are given by black thick dashed curves: 
for the $(-1)^{\ell+1}$ parity channels, this corresponds to the $x\to +\pm\infty$ continua 
of Eqs.~(\ref{eq:continuum_droite},\ref{eq:continuum_gauche})
with all $L\geq 1$; for the $(-1)^{\ell}$ parity channels, this corresponds to the $x\to +\pm\infty$ continua 
of Eqs.~(\ref{eq:continuum_droite},\ref{eq:continuum_gauche}) with all $L\geq 0$ and to the $(x,\theta)\to (0,\pi)$ 
[that is $t\to -\infty$] continuum of Eq.~(\ref{eq:dritte}).
Contrarily to the $3+1$ case, no eigenvalue of $M(s=0)$ 
is found to cross zero for $\alpha<\alpha_c(2;1)=13.6069\ldots$: no four-body Efimov effect is found for the $2+2$ fermionic problem.}
\label{fig:spectre}
\end{figure*}

\begin{figure}[tb]
\centerline{\includegraphics[width=8cm,clip=]{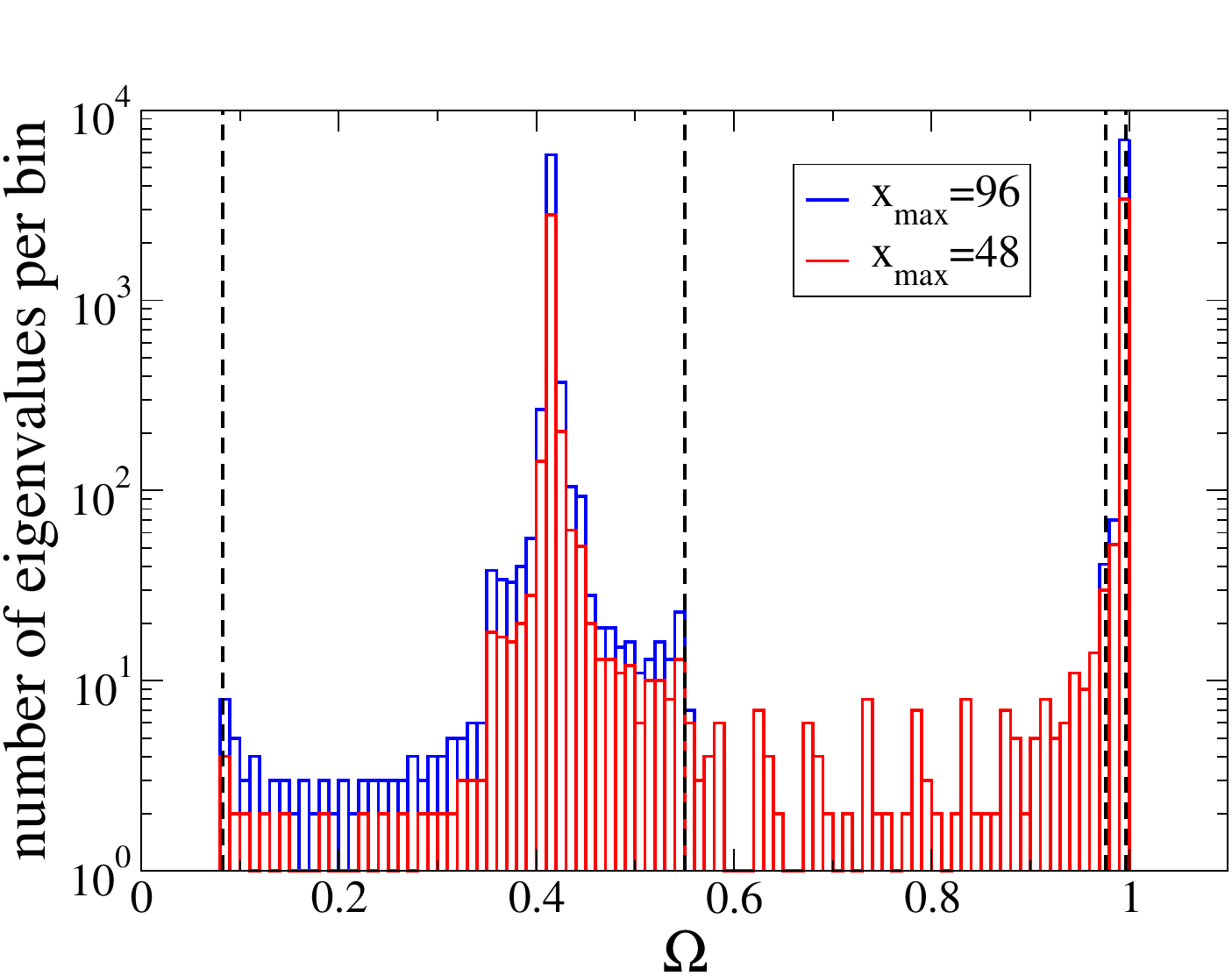}}
\caption{(Color online) Histogram of the eigenvalues $\Omega$ of $M(s=0)$ for a mass ratio $\alpha=10$ and an angular momentum $\ell=1$ in the $(-1)^{\ell+1}$ parity channels, 
obtained numerically after discretisation and truncation of the $x$ and $\theta$ variables. The numerical grid is the same as in 
Fig.~\ref{fig:spectre},
except that much larger values of $x_{\rm max}=-x_{\rm min}$ are used, to reveal the emergence of the continuous part of the spectrum in the $x_{\rm max}\to +\infty$
limit: $x_{\rm max}=48$ (red bars in the foreground) and $x_{\rm max}=96$ (blue bars in the background). The black vertical dashed lines 
indicate the analytically predicted borders of the continuous spectrum (as in Fig.~\ref{fig:spectre}); in between the first two ones and in between the last two 
ones, it is indeed observed that the number of eigenvalues per bin is approximately multiplied by two when $x_{\rm max}$ is doubled.
On the contrary, the histogram is unaffected by the change of $x_{\rm max}$ in the bins strictly in between the second and third dashed lines, indicating
that the corresponding eigenvalues belong to the discrete spectrum of $M(s=0)$, with localized eigenfunctions in $x$ space. }
\label{fig:histo}
\end{figure}

\begin{figure}[tb]
\centerline{\includegraphics[width=8cm,clip=]{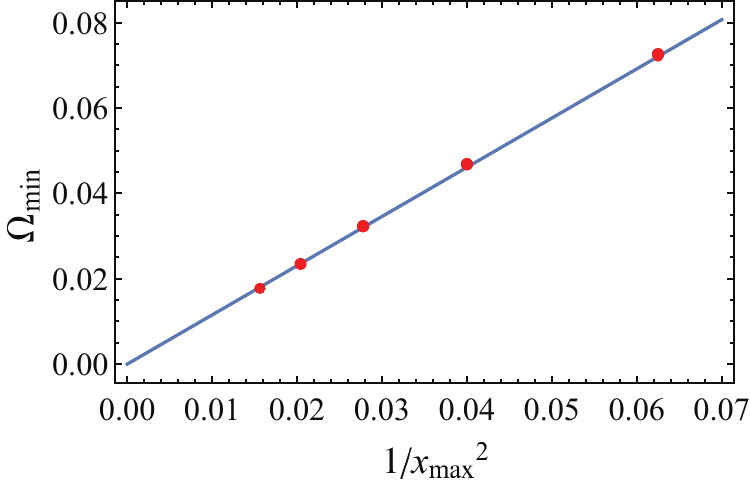}}
\caption{(Color online) Numerically determined minimal eigenvalue $\Omega_{\rm min}$ of $M^{(\ell)}(s=0)$ almost at the three-body critical mass ratio,
$\alpha=13.6069\simeq \alpha_c(2;1)$, as a function of the numerical cut-off $t_{\rm min}=-x_{\rm max}$. For each given cut-off value,
each angular momentum $\ell$ from $0$ to $12$ and parity sector $(-1)^\ell$ and $(-1)^{\ell+1}$ contributes as a point in the figure: The fact that the
points {(in red)} are superimposed and cannot be distinguished shows that $\Omega_{\rm min}$ does not depend on $\ell$ nor on the parity.
Furthermore, $\Omega_{\rm min}$ is always positive, it is linear in $1/x_{\rm max}^2$ and extrapolates to zero for infinite cut-off {(see blue line)}: This is perfectly consistent
with the fact that $\Omega_{\rm min}$ corresponds to the lower border of the $2+1$ continuum $\Lambda_{L=1}(ik,\alpha)$, where
$k$ has a minimal, discrete value scaling as $1/x_{\rm max}$ in presence of the numerical cut-off, and $\Lambda_{L=1}(ik,\alpha_c(2;1))$ vanishes
quadratically at $k=0$. In other words, there is no negative $\Omega_{\rm min}$ and no four-body Efimov effect for $2+2$ fermions.}
\label{fig:dediee}
\end{figure}

In the $3+1$ fermionic problem, the signature of a four-body Efimov effect was that an eigenvalue of the corresponding $M(s=0)$
operator crosses zero for some value of $\alpha$ below $\alpha_c(2;1)\simeq 13.6069$, specifically for $\alpha=\alpha_c(3;1)\simeq 13.384$
\cite{CMP}. The question here is to know whether or not such a crossing can occur for the $2+2$ fermionic problem, that is for $M(s=0)$
corresponding to Eqs.~(\ref{eq:check},\ref{eq:noyau_matriciel}). We answer this question by a numerical calculation of the eigenvalues of $M(s=0)$.

\noindent {\bf Numerical implementation:}
In the $(-1)^{\ell+1}$ parity sector, we truncate the $x$ variable in a symmetric way, that is to $[x_{\rm min}=-x_{\rm max},x_{\rm max}]$,
and we discretize it with a uniform step $dx$ according to the usual mid-point integration method. 
We use $\theta$ rather than $u=\cos \theta$ as integration variable, so that we 
use $\check{\Phi}(x,\theta)=(\sin \theta)^{1/2} \Phi(x,u)$
rather than $\Phi(x,u)$ as the unknown function; multiplying the eigenvalue problem $\Omega \Phi = M(s=0) \Phi$ by $(\sin\theta)^{1/2}$,
we get an hermitian eigenvalue problem with the same eigenvalues, the same diagonal part and a kernel 
$\check{K}^{(\ell)}_{m_z,m_z'}(x,\theta;x',\theta')=(\sin\theta \sin\theta')^{1/2} K^{(\ell)}_{m_z,m_z'}(x,u;x',u')$ with
$u=\cos\theta$, $u'=\cos\theta'$. For better accuracy, we use the Gauss-Legendre integration scheme \cite{Recipes} 
for $\theta'$ as this angular variable is naturally bounded to $[0,\pi]$
\footnote{If for simplicity one omits to write the other variables, 
the integral over $\theta'$, of the form $\int_0^\pi d\theta' \check{K}(\theta,\theta') \check{\Phi}(\theta')$,
is approximated by $\sum_{j=1}^{n_\theta} w(\theta_j) \check{K}(\theta_i,\theta_j) \check{\Phi}(\theta_j)$, where $(\theta_i)_{1\leq i\leq n_\theta}$
is the set of (non-equispaced) discrete values of $\theta$ proposed by the $n_\theta$-points Gauss-Legendre method and $w(\theta_i)$ the corresponding weights.
To render the resulting discretised form of $\check{M}(s=0)$ hermitian, it suffices to take as unknowns
$w(\theta_j)^{1/2} \check{\Phi}(\theta_j)$ and to multiply the eigenvalue equation by $w(\theta_i)^{1/2}$, which leads to the kernel
$[w(\theta_i) w(\theta_j)]^{1/2} \check{K}(\theta_i,\theta_j)$ without modifying the spectrum.}.

In the $(-1)^\ell$ parity sector, there is the additional complication that the kernel diverges at the point $(x,\theta)=(0,\pi)$, making the previous
$(x,\theta)$ discretisation inefficient. At a small distance from this point, say less than $\rho_0$, 
the optimal set of variables is $(t,\psi)$, where $\psi\in [0,\pi]$ is defined
by Eq.~(\ref{eq:def_rho_psi}) and $t\in \mathbb{R}^-$ is given by Eq.~(\ref{eq:deft}), since the resulting kernel is bounded after a 
convenient change of the unknown function $\tilde{\Phi}_{m_z}^{(\ell)}(t,\psi)=\rho\,\check{\Phi}_{m_z}^{(\ell)}(x,\theta)$, see Eq.~(\ref{eq:enfin_borne}).
So we resort to a mixed scheme: for $\rho=[x^2+(\pi-\theta)^2]^{1/2}> \rho_0$,
we use the $(x,\theta)$ set of variables, with $x$ uniformly discretised with a step $dx$ submultiple of $\rho_0$ and $\theta$ discretised according to
a Gauss-Legendre scheme over the interval $[0, \theta_{\rm max}]$ where $\theta_{\rm max}=\pi$ for $|x|>\rho_0$ and $\theta_{\rm max}=
\pi-(\rho_0^2-x^2)^{1/2}$ for $|x|\leq \rho_0$, the number of angular points being scaled linearly with $\theta_{\rm max}$; for $\rho<\rho_0$, we use the
$(t,\psi)$ set of variables, with $t$ truncated to $[t_{\rm min},0]$ and discretized with a uniform step $dt$ according to the mid-point integration
method, and $\psi\in [0,\pi]$ discretised according to the Gauss-Legendre scheme
\footnote{With the notations of the previous footnote, 
one takes $[dx\, w_x(\theta_i)]^{1/2} \check{\Phi}^{(\ell)}_{m_z}(x,\theta_i)$ as unknowns in the zone $\rho>\rho_0$, and 
$[dt\, w(\psi_i)]^{1/2} \rho_0 e^t \check{\Phi}^{(\ell)}_{m_z}(x(t,\psi_i),\theta(t,\psi_i))$ as unknowns in the zone $\rho<\rho_0$. In this
way, after multiplication of the eigenvalue equation by $[dx w_x(\theta_i)]^{1/2}$ or $[dt w(\psi_i)]^{1/2} \rho_0 e^t$, one obtains
an hermitian matrix. The $x$-dependence of the weight $w_x(\theta_i)$ results from the $x$-dependence of $\theta_{\rm max}$.}.

\noindent{\bf Results:} 
The numerically obtained spectrum of $M(s=0)$ for the $2+2$ fermionic problem is plotted in Fig.~\ref{fig:spectre} {[left half for the $(-1)^\ell$ parity 
channels, right half for the $(-1)^{\ell+1}$ parity channels]}, for the first values $0\leq \ell \leq 3$ of the four-body internal angular momentum
quantum number $\ell$, as a function of the mass ratio $\alpha=\frac{m_\uparrow}{m_\downarrow}$. It is a symmetric function under the exchange $\alpha 
\leftrightarrow 1/\alpha$ so the figure is restricted to $\alpha\geq 1$; as the starting integral equation (\ref{eq:equint}) assumes scale invariance,
which is broken by the three-body Efimov effect beyond the threshold $\alpha_c(2;1)=13.6069\ldots$, the figure is also restricted to $\alpha<\alpha_c(2;1)$.

For the ${(-1)^{\ell}}$ parity channels, the spectrum is entirely within the limits of the analytically predicted continuous spectrum, which are
shown as dashed lines, except for even $\ell$ in a barely visible small triangle 
\footnote{The triangle corresponds to the zone $1/\sqrt{2}\leq \Omega \leq \Lambda_{L=1}(0,\alpha)$.} 
close to $\Omega=0.75$ with $1\leq \alpha \lesssim 1.2$: this means that 
the necessarily discrete numerical spectrum must tend to a continuum when the truncations $x_{\rm max}=-x_{\rm min}$ 
and $t_{\rm min}$ tend to $+\infty$ and $-\infty$ respectively.
The lower border of the continuum corresponds for even $\ell$ to the $k\to 0$ limit in Eq.~(\ref{eq:dritte}), that is zero,
and for odd $\ell$ to the smallest of the two quantities,
$1/\sqrt{2}$ [this is the $k\to+\infty$ limit of Eq.~(\ref{eq:dritte})] and $\Lambda_{L=1}(0,\alpha)$ 
[this is the minimal value of Eq.~(\ref{eq:continuum_gauche}), {see Fig.~\ref{fig:bordevin}}].
The upper border of the continuum corresponds, whatever the parity of $\ell$, to $\Lambda_{L=0}(0,1/\alpha)$ [this is the maximal value of 
Eq.~(\ref{eq:continuum_droite}), {see Fig.~\ref{fig:bordevin}}].

For the ${(-1)^{\ell+1}}$ parity channels, there are three differences. First, the continuum (\ref{eq:dritte}) cannot be realized,
so that the lower border of the continuous spectrum of $M(s=0)$, now given by $\Lambda_{L=1}(0,\alpha)$, 
reaches zero only at $\alpha=\alpha_c(2;1)$. Second, the $L=0$ continua in Eqs.~(\ref{eq:continuum_droite},
\ref{eq:continuum_gauche}) cannot be realized so that the upper border of the continuum of $M(s=0)$, given by $\Lambda_{L=2}(0,1/\alpha)$, is everywhere below 
$1=\lim_{\alpha\to +\infty} \Lambda_{L=2}(0,1/\alpha)$. Third, the continuum presents, in the $\alpha-\Omega$ plane for $1.53 \lesssim \alpha$, 
a large void internal area, corresponding to $\Lambda_{L=2}(0,\alpha)<\Omega<\Lambda_{L=1}(0,1/\alpha)$. Still, many numerically found eigenvalues
lay in this internal area: these eigenvalues must correspond to the discrete spectrum of $M(s=0)$, with localized (square-integrable) eigenfunctions
\footnote{The same conclusion must hold for the eigenvalues $\Omega$ in Fig.~\ref{fig:spectre} [parity $(-1)^{\ell+1}$]
that are above the upper external border of the continuum, as well as for those below the lower external border of the continuum 
(there are some for $\alpha$ close to unity).}.
We checked this numerically by calculating the density of states of $M(s=0)$, in practice the histogram of its eigenvalues, for increasing
values of the numerical truncation $x_{\rm max}=-x_{\rm min}$: by doubling the values of $x_{\rm max}$ and $x_{\rm min}$, 
the spacing $\approx \pi/x_{\rm max}$ between successive $k$
in Eqs.~(\ref{eq:continuum_droite},\ref{eq:continuum_gauche}) is approximately divided by two so that the density of states of the numerical quasi-continuum is
approximately multiplied by two, whereas the density of states of the discrete spectrum is essentially not (only exponentially weakly) affected as soon as $x_{\rm max}$
is much larger than the localized eigenfunctions width in $x$ space. This is what is observed in Fig.~\ref{fig:histo}, knowing that the locations
of the internal and external borders of the continuum (corresponding to the dashed lines in Fig.~\ref{fig:spectre}) are indicated by vertical
dashed lines in Fig.~\ref{fig:histo}.

\noindent{\bf Synthesis:} 
Contrarily to the $3+1$ case, no discrete eigenvalue of $M(s=0)$
(necessarily discrete because it would be below the lower border of the continuum)
crosses zero for $\alpha<\alpha_c(2;1)=13.6069\ldots$, that is below the three-body Efimov effect threshold: 
no four-body Efimov effect is found for the $2+2$ fermionic problem
{\footnote{{It is explicitly supposed here that a potential purely imaginary root $s_4$ of Eq.~(\ref{eq:detM}) would exist above some threshold
value $\alpha_c(2;2)$ of the mass ratio $\alpha$, with $s_4=0$ at threshold. One can however imagine another scenario, with $s_4$ still a continuous function
of $\alpha$: $s_4$ would exist for all mass ratio
$\alpha\in [1,\alpha_c(2;1)]$, with $\alpha_c(2,1)=13.6069\ldots$ the three-body Efimov effect threshold, in which case $s_4$ would not need to cross
zero for some $\alpha$. This scenario is however excluded (i) by the experimental results for the 
spin $1/2$ unitary Fermi gas, which has a mass ratio $\alpha=1$ (no significant four-body losses are observed) 
and (ii) by the numerical calculations in the conjecture on the fourth cluster coefficient $b_4$ of that unitary Fermi gas in Appendix \ref{appen:conjecture}
[it is found for $\alpha=1$ that the operator $M(s)$ is positive for all purely imaginary $s$, which excludes the existence of a root $s_4$].}}}.
This conclusion is apparent in Fig.~\ref{fig:spectre}, obtained for all the internal angular momentum quantum numbers $0\leq \ell \leq 3$.
It extends to all the angular momentum values that we were able to numerically explore, $4\leq \ell \leq 12$, as we have shown with a dedicated
careful spectral analysis almost perfectly at the three-body critical mass ratio, $\alpha=13.6069$, see Fig.~\ref{fig:dediee}.

\section{Conclusion}
\label{sec:conclusion}
We have studied in three dimensions 
a four-body $2+2$ fermionic system with resonant interactions and we have derived its momentum space integral equations
at zero energy. By using rotational invariance and scale invariance, we have reduced them to a numerically tractable 
two-dimensional form (the unknown function depends on two variables only). 
With these equations we have numerically shown that no four-body Efimov effect occurs for the $2+2$ fermionic system
in angular momentum channels $0\leq \ell \le 12$. The $3+1$ fermionic system thus remains the only known one exhibiting
a four-body Efimov effect \cite{CMP}.

A detailed treatment of the second motivation for deriving these integral equations, that is 
a calculation of the already measured \cite{Navon,Zwierlein} fourth cluster coefficient $b_4$
of the spin $1/2$ unitary Fermi gas, is beyond the scope of this paper.
Still we have numerically calculated an educated guess for $b_4$ inspired from the analytical form of the third cluster coefficient
$b_3$ \cite{CastinWerner,ChaoEndoCastin}: it is found that this guess does not reproduce the experimental value (see Appendix \ref{appen:conjecture})
so that a dedicated work is needed {and left for the future}.

\begin{acknowledgments}
We acknowledge useful discussions with J.\ Levinsen and D.\ Gridnev.
S.E.\ thanks the Japan Society for the Promotion of Science (JSPS) for support. The group of Y.C.\ is affiliated to IFRAF. 
\end{acknowledgments}

\appendix
\section{Is the spectrum of $M(s)$ bounded ?}
\label{app:Mborne}

In the parity sector $(-1)^\ell$, the third contribution in Eq.~(\ref{eq:noyau_matriciel}) diverges
when $(x,u)\to (0,-1)$ or $(x',u')\to (0,-1)$. The question is to know if this makes the operator $M(s)$ unbounded,
for a purely imaginary $s=iS$.

To investigate this problem, 
we construct a simplified functional that focuses on the diverging part of the matrix kernel (\ref{eq:noyau_matriciel}), replacing each 
non-zero limit expression by its limit, and replacing the vanishing expressions by their leading order (here quadratic) approximations:
\be
u+\ch x \simeq \frac{1}{2} (x^2 + y^2) \ \mbox{with}\ \ y\equiv\pi-\theta.
\ee
Dropping numerical factors and other bounded pieces (for example the bit raised to the power $s$, of modulus one),  
we obtain the mean-$\Omega$ functional
\be
\langle \Omega\rangle = \frac{\int dx du \int dx' du' \frac{\Phi^*(x,u)\Phi(x',u')}{[(x^2+y^2)(x^{'2}+y^{'2})]^{1/4}(x^2+y^2+x'^2+y'^2)}}
{\int dx du |\Phi(x,u)|^2}
\label{eq:fes}
\ee
where the integrals are taken over some convenient neighborhood of $(x,u)=(0,-1)$.
It is convenient to use the angle $\theta$ rather than $u=\cos\theta$ as integration variable, which pulls out a Jacobian $\sin\theta \simeq y$;
we absorb it in the integral in the denominator of Eq.~(\ref{eq:fes}) with the change of function
\be
\check{\Phi}=(\sin\theta)^{1/2} \Phi(x,u)
\label{eq:chgtdefonc}
\ee
A factor $(\sin\theta\sin\theta')^{1/2}\simeq (yy')^{1/2}$ remains in the integrand in the numerator. We restrict the integration
over $(x,y)$ to the upper half $y>0$ of the disk $x^2+y^2<1$. Then it is natural to move to polar coordinates:
\be
(x,y)=(\rho \cos \phi,\rho \sin \phi)
\ee
so that $x^2+y^2=\rho^2$, $x'^2+y'^2=\rho'^2$ and $(yy')^{1/2}=(\rho \rho')^{1/2} (\sin \phi \sin \phi')^{1/2}$.
The occurrence of the Jacobians $\rho$ and $\rho'$ in the elements $\rho d\rho$ and $\rho' d\rho'$ motivates the change of variable in the radial integration:
\be
X=\rho^2 \ \mbox{and}\ X'=\rho'^2.
\ee
Then, considering $\check\Phi$ as a function of $X$ and $\phi$, we obtain
\be
\langle\Omega\rangle=\frac{\int_0^1 \frac{dX dX'}{X+X'} \int_0^\pi \!d\phi d\phi' 
(\sin\phi\sin\phi')^{1/2} \check{\Phi}^*(X,\phi) \check{\Phi}(X',\phi')}
{2\int_0^1 dX \int_0^\pi d\phi \,|\check{\Phi}(X,\phi)|^2}
\label{eq:depend_de_phi}
\ee
To get rid of the polar angle $\phi$, we introduce
\be
{\Phi}_a(X)\equiv \int_0^\pi \!d\phi\, (\sin \phi)^{1/2} \check{\Phi}(X,\phi)
\ee
so that the integral over $\phi$ and $\phi'$ in the numerator of Eq.~(\ref{eq:depend_de_phi}) reduces to  the product
${\Phi}_a^*(X) {\Phi}_a(X')$.
In that numerator, we use the fact that the modulus of the integral over $X$ and $X'$  is less than the integral of the modulus, and
that $\displaystyle\frac{1}{X+X'}\leq \frac{1}{(X^2+X'^2)^{1/2}}$.
In the denominator of Eq.~(\ref{eq:depend_de_phi}),
at fixed $X$, we apply over the interval $\phi\in [0,\pi]$ 
the Cauchy-Schwarz inequality $|\langle f|g\rangle|^2\leq \langle f|f\rangle  \, \langle g|g\rangle$ 
(in Dirac's notation) with $f(\phi)=(\sin \phi)^{1/2}$ and $g(\phi)=\check{\Phi}(X,\phi)$; after
integration of the resulting inequality over $X$, we get:
\be
\int_0^1 dX \, |{\Phi}_a(X)|^2 \leq  2 \int_0^1 dX \int_0^\pi d\phi \, |\check{\Phi}(X,\phi)|^2
\ee
whose right-hand side is the denominator of Eq.~(\ref{eq:depend_de_phi}). We arrive at
\be
|\langle\Omega\rangle| \leq \frac{\int_0^1 dX \int_0^1 dX' \frac{|{\Phi}_a(X)|\,|{\Phi}_a(X')|}
{(X^2+X'^2)^{1/2}}}{\int_0^1 dX \, |{\Phi}_a(X)|^2}
\label{eq:pas_loin}
\ee
We again move to polar coordinates
\be
(X,X')=(r\cos\psi,r\sin \psi)
\ee
so as to simplify the factor $\displaystyle\frac{1}{(X^2+X'^2)^{1/2}}=\frac{1}{r}$ with the Jacobian and to obtain
\be
|\langle\Omega\rangle| \leq \frac{\int_0^{\pi/2} d\psi \int_0^{R(\psi)} dr |{\Phi}_a(r\cos\psi)| |{\Phi}_a(r\sin\psi)|}
{\int_0^1 dX \, |{\Phi}_a(X)|^2}
\label{eq:majpf}
\ee
Since the domain of integration over $(X,X')$ is the square $[0,1]^2$, $\psi$ runs over $[0,\pi/2]$ and, at fixed $\psi$,
$r$ runs over $[0,R(\psi)]$ with
\be
R(\psi) = \min\left(\frac{1}{\cos\psi},\frac{1}{\sin\psi}\right).
\ee
In the integral over $r$ at fixed $\psi$, we again use the Cauchy-Schwarz inequality over the interval $r\in [0,R(\psi)]$
with $f(r)=|{\Phi}_a(r\cos\psi)|$ and $g(r)=|{\Phi}_a(r\sin\psi)|$:
\begin{multline}
\int_0^{R(\psi)} \!\!\!\!\!\!\!\! dr |{\Phi}_a(r\cos\psi)| |{\Phi}_a(r\sin\psi)| \leq  \\
\left[\int_0^{R(\psi)}\!\!\!\!\!\!\!\! dr |{\Phi}_a(r\cos\psi)|^2\right]^{1/2} 
\left[\int_0^{R(\psi)}\!\!\!\!\!\!\!\! dr |{\Phi}_a(r\sin\psi)|^2\right]^{1/2}
\label{eq:CSexpli}
\end{multline}
In the first factor in the right-hand side of Eq.~(\ref{eq:CSexpli}), we perform the change of variable $X=r\cos\psi$,  so that
\begin{multline}
\int_0^{R(\psi)}\!\!\!\!\!\!\!\! dr |{\Phi}_a(r\cos\psi)|^2 = \frac{1}{\cos\psi} \int_0^{R(\psi)\cos\psi} dX |{\Phi}_a(X)|^2\\
\leq \frac{1}{\cos\psi} \int_0^{1} dX |{\Phi}_a(X)|^2
\label{eq:dernier_pas}
\end{multline}
where we used $R(\psi)\cos\psi\leq 1$ and the non-negativeness of $|{\Phi}_a|^2$. The last integral in Eq.~(\ref{eq:dernier_pas})
is nothing but the denominator in the right-hand side of Eq.~(\ref{eq:majpf})!
We proceed similarly in the second factor in the right-hand side of Eq.~(\ref{eq:CSexpli}), except that $\cos\psi$ is replaced with $\sin \psi$.
Finally the denominator in Eq.~(\ref{eq:majpf}) cancels out, so that
\be
|\langle \Omega\rangle| \leq \int_0^{\pi/2} \frac{d\psi}{(\cos\psi \sin\psi)^{1/2}} < +\infty
\ee
and the spectrum of $M(s)$ is bounded, when $s\in i \mathbb{R}$.

\section{Enunciating and testing a conjecture for $b_4$}
\label{appen:conjecture}

\subsection{The cluster expansion} 

Consider a spatially uniform spin 1/2 Fermi gas at thermal equilibrium in the grand canonical ensemble in the thermodynamic limit, with a temperature $T$,
and a single chemical potential $\mu$ since the gas is unpolarized.  
The well-known cluster expansion is a series expansion of its pressure in powers of the fugacity $z=\exp(\beta \mu)$ in the
non-degenerate limit $\mu\to -\infty$ for a fixed temperature $T$, with $\beta=1/(k_B T)$ \cite{Huang}. 
For our gas, it is generally written as
\be
\frac{P\lambda^3}{k_B T} = 2\sum_{n\geq 1} b_n z^n
\ee
where the overall factor $2$ accounts for the number of spin components and $\lambda$ is the thermal de Broglie wavelength
\be
\lambda = \left(\frac{2\pi \hbar^2}{m k_B T}\right)^{1/2}
\ee
When reexpanded in terms of the small degeneracy parameter $\rho \lambda^3$,
where $\rho$ is the total density, the cluster expansion gives rise to the virial expansion with virial coefficients $a_n$ \cite{Huang}.
In practice, one rather considers the deviation $\Delta b_n$ of $b_n$ from its ideal Fermi gas value, that is (for $n>1$) from the mere
effect of Fermi statistics:
\be
b_n = \frac{(-1)^{n+1}}{n^{5/2}} + \Delta b_n
\ee

While the cluster expansion has been studied for a long time and the second
cluster coefficient $b_2$ was obtained analytically in reference \cite{BethUl}
(note that $b_1=1$ according to the ideal gas law), there is a renewed interest in the cluster coefficients for $n>2$.
First, the new challenge is to
calculate the $b_n$ for resonant $s$-wave interactions (with a scattering length $a$ much larger in absolute value than the interaction range), whereas
previous studies were concentrating on the hard sphere model \cite{durdur}. Second, the $b_n$ have been extracted up to $n=4$ in the unitary limit,
from a measurement of the equation of state of ultracold atomic Fermi gases \cite{Navon,Zwierlein}. The
two independent groups have reported consistent values of the fourth cluster coefficient:
\be
\Delta b_4^{\rm ENS} = 0.096 (15) \ \ \mbox{and}\ \ \Delta b_4^{\rm MIT}=0.096 (10)
\label{eq:valb4exp}
\ee

\subsection{In the unitary limit}

For zero-range interactions with infinite $s$-wave scattering length $a^{-1}=0$, i.e.\ in the unitary limit,
the harmonic regulator method used in \cite{ComtetOuvry}, that introduces an isotropic harmonic trapping potential, 
is quite efficient, due to the $\mathrm{SO(2,1)}$ dynamical symmetry
resulting from the scale invariance \cite{PitaRosch,CastinCRAS} and the subsequent separability of Schr\"odinger's equation in hyperspherical coordinates
\cite{Werner3corps,WernerCastinPRA} in the trap. The value of $b_n$ can be deduced
from the canonical partition functions, that is from the energy spectra, of all the possible $k$-body problems in the trap, with $k\leq n$.
One has the following expansion of the grand potential $\Omega$ of the thermal equilibrium gas in the trap:
\be
\frac{-\Omega}{k_B T Z_1} = \sum_{(n_\uparrow,n_\downarrow)\in \mathbb{N}^{2*}} B_{n_\uparrow,n_\downarrow}(\omega) 
z_\uparrow^{n_\uparrow} z_\downarrow^{n_\downarrow}
\ee
where $Z_1$ is the canonical partition function for one particle in the trap, and
it is convenient at this stage to be general and introduce independent chemical potentials $\mu_\sigma$ for the various spin components
$\sigma$, so that $z_\sigma=\exp(\beta \mu_\sigma)$.
Then, from the asymptotically exact local density approximation \cite{Drummond} (see also \cite{ComtetOuvry}), and introducing also the deviations
$\Delta B_{n_\uparrow,n_\downarrow}(\omega)$ of $B_{n_\uparrow,n_\downarrow}(\omega)$ from the ideal Fermi gas value
\footnote{Note that $\Delta B_{n_\uparrow,n_\downarrow}$
is actually equal to $B_{n_\uparrow,n_\downarrow}$ as soon as the two indices differ from zero, since the ideal gas grand potential is the sum of the
grand potential of each spin component, and no $\uparrow$-$\downarrow$ crossed term can appear in the resulting cluster expansion.}, one has
\be
2 \Delta b_n =  n^{3/2} \sum_{n_\uparrow=1}^{n-1} \Delta B_{n_\uparrow,n_\downarrow=n-n_\uparrow}(0^+) 
\label{eq:DeltabnefdB}
\ee
where $\Delta B(0^+)=\lim_{\omega\to 0^+} \Delta B(\omega)$ and where
we could restrict the sum to $n_\sigma\neq 0$, $\sigma=\uparrow, \downarrow$, since the fully polarized configurations are non-interacting
and have zero deviations from the ideal gas.

For $n=3$, extending to fermions the technique initially developed for bosons \cite{CastinWerner}, the following analytical expression was
obtained \cite{ChaoEndoCastin}
\footnote{It can be deduced from Eq.~(7) of reference \cite{ChaoEndoCastin} by integration by parts.}:
\be
\Delta B_{2,1}(0^+)=\sum_{\ell\in \mathbb{N}} \left(\ell+\frac{1}{2}\right) \int_0^{+\infty} \frac{dS}{\pi} S \frac{d}{dS} [\ln \Lambda_l(iS,\alpha)]
\label{eq:DB21analy}
\ee
where the function $\Lambda_l$ is given by Eq.~(\ref{eq:defLam}),
and the mass ratio between the opposite spin component $\alpha$ is equal to one (so that $\Delta B_{2,1}=\Delta B_{1,2}$). It gives
\be
\Delta b_3 \simeq -0.355103
\ee
in agreement with previous numerical studies \cite{Drummond,Blumeb3} and with the experimental values \cite{Navon}.

For $n=4$, the problem is still open. A numerical attempt \cite{Blumeb4}, 
with brute force calculation of the 4-body unitary spectrum in the trap, has produced the value
\be
\Delta b_4^{\rm Blume}=-0.016(4)
\ee
The disagreement with the experimental results (\ref{eq:valb4exp}) is attributed to uncertainties in extrapolating  to $\omega\to 0$
the numerical values of $\Delta B_{n_\uparrow,n_\downarrow}(\omega)$, in practice obtainable only for $\hbar \omega \gtrsim k_B T$.
An approximate diagrammatic theory \cite{JesperLevinsen} (keeping {\sl even} in the unitary limit only the diagrams that have leading contribution
in the perturbative regime of a large effective range or a small scattering length) gives an estimate closer to the experimental values (\ref{eq:valb4exp}),
\be
\Delta b_4^{\rm Levinsen} \approx 0.06
\ee
Extending the analytical method of reference \cite{CastinWerner} to the fermionic four-body problem is technically challenging and goes beyond
the scope of the present work. On the contrary, it is reasonable here to propose and test a guess by direct transposition
of Eq.~(\ref{eq:DB21analy}): the transcendental function $\Lambda_l(s)$ of the three-body problem is formally replaced by
$\det M^{(\ell)}(s)$ for the four-body problem, where $\det$ is the determinant and the operator $M^{(\ell)}(s)$, acting on the spinor functions 
$\Phi^{(\ell)}_{m_z}(x,u)$ as in the right-hand side of Eq.~(\ref{eq:check}), was introduced and spectrally discussed in section \ref{subsec:le3corpsretrouve}
for the $2+2$ fermionic problem, and has a known equivalent for the $3+1$ fermionic problem, see Eq.~(14) of reference \cite{CMP}.
Indeed, in both cases, the scaling exponents $s$ (purely imaginary in the efimovian channels, real otherwise) allowed by Schr\"odinger's
equation in the unitary Wigner-Bethe-Peierls model are such that
$\Lambda_l(s)=0$ for $n=3$, or such that Eq.~(\ref{eq:check}) has a non-zero solution $\Phi^{(\ell)}_{m_z}(x,u)$, that is $M^{(\ell)}(s)$
admits a zero eigenvalue. Hence our conjecture:
\begin{multline}
\Delta B^{\rm conj}_{n_\uparrow,n_\downarrow}(0^+)=\sum_{\ell\in \mathbb{N}} \left(\ell+\frac{1}{2}\right)  \\
\times \int_0^{+\infty} \frac{dS}{\pi} S \frac{d}{dS} [\ln \det M_{n_\uparrow,n_\downarrow}^{(\ell)}(iS)]
\label{eq:conjecture}
\end{multline}
with $(n_\uparrow,n_\downarrow)\in \{(1,3),(2,2),(3,1)\}$ and $M_{n_\uparrow,n_\downarrow}^{(\ell)}$ is the operator
$M^{(\ell)}$ for the four-body problem with $n_\sigma$ particles in each spin component $\sigma$.

\subsection{Existence of the logarithmic derivative of the determinant}

The conjecture (\ref{eq:conjecture}) is not as innocent as it may look at first sight. The difficulty is that $M^{(\ell)}$ is actually an operator,
and not a finite size matrix: it has a continuous spectrum, constituting an infinite, dense set of ``eigenvalues"; even its discrete spectrum may
present accumulation points, leading to an infinite but countable number of eigenvalues. In other words, the determinant of $M^{(\ell)}(iS)$ is
not finite. Numerically, as we have already done in section \ref{sec:quete}, one of course truncates the unbounded variable
$x$ to the compact interval $[-x_{\rm max},x_{\rm max}]$, which amounts to imposing the boundary conditions to the spinor:
\be
\Phi_{m_z}^{(\ell)}(x=\pm x_{\rm max},u)=0  \ \forall u\in [-1,1], \forall m_z\in \{-\ell, \ldots,\ell\}
\label{eq:emulcoupnum}
\ee
After discretization of the $x$ and $u$ variables, $M^{(\ell)}(iS)$ is then replaced by a matrix,
with a well-defined determinant; still it remains to know if there is convergence of the integrand in Eq.~(\ref{eq:conjecture})
when $x_{\rm max}\to +\infty$. As we now see, the answer is positive.

The key point is that what appears in the integrand of Eq.~(\ref{eq:conjecture}) is not the determinant itself, rather its logarithmic
derivative, which can be written as
\be
\frac{d}{dS} \ln \det M^{(\ell)}(iS)=\Tr\left\{[M^{(\ell)}(iS)]^{-1}\frac{d}{dS} M^{(\ell)}(iS)\right\}
\label{eq:formule_de_trace}
\ee
where $\Tr$ is the trace and $M^{-1}$ the inverse of $M$.

\noindent{\bf Parity $(-1)^{\ell+1}$}: 
In the parity sector $(-1)^{\ell+1}$, the spectrum of $M^{(\ell)}(iS)$ is at non-zero distance from $0$ for a mass ratio $\alpha=1$,
as there is no $4$-body Efimov effect, see Fig.~\ref{fig:spectre}. So the inverse of $M^{(\ell)}(iS)$ is well defined. Also the operator $M^{(\ell)}(iS)$
is local in the $x$ basis, meaning that the off-diagonal matrix elements of the operator
$\mathcal{D}^{-1/2} K^{(\ell)} \mathcal{D}^{-1/2}$ are rapidly decreasing functions of $|x-x'|$, for example there exists a constant $A^{(\ell)}$ such that
\be
\frac{|\langle x,u,\ell,m_z| K^{(\ell)}(iS)|x',u',\ell,m_z'\rangle|}{[d(x,u) d(x',u')]^{1/2}} \leq \frac{A^{(\ell)}}{\ch(x-x')}
\label{eq:local}
\ee
for all $x,x',u,u'$ and all $m_z, m_z'$ of parity opposite to $\ell$, and for all $S\in \mathbb{R}$.
Here, we have used Dirac's notation and singled out as in Eq.~(\ref{eq:check}) a diagonal part and a kernel part,
\be
M^{(\ell)}(iS) = \mathcal{D} + K^{(\ell)}(iS)
\label{eq:decompDK}
\ee
where the operator $\mathcal{D}$ is positive and defined by the diagonal-element function $d(x,u)$,
\begin{multline}
\mathcal{D} |x,u,\ell,m_z\rangle = d(x,u) |x,u,\ell,m_z\rangle  \ \ \mbox{with} \\
d(x,u) = \left[\frac{\alpha}{(1+\alpha)^2}\left(1+\frac{u}{\ch x}\right)+\frac{e^{-x}+\alpha e^x}{2(\alpha+1)\ch x}\right]^{1/2} 
\label{eq:cestquoid}
\end{multline}
This locality is apparent for the first two contributions in the right-hand side of Eq.~(\ref{eq:noyau_matriciel}): each contribution is bounded, and is consistent
with Eq.~(\ref{eq:local}) at the four infinities $(x,x')=(\pm\infty,\pm\infty)$ (see reasoning in section \ref{subsec:le3corpsretrouve}).
We then expect that the inverse of $M^{(\ell)}(iS)$, that can be written as
\be
[M^{(\ell)}(iS)]^{-1} = \mathcal{D}^{-1} + K^{(\ell)}_{\rm inv}(iS)
\label{eq:ecritMDK}
\ee
is also local, from the geometric series expansion:
\begin{multline}
(\mathcal{D} + K)^{-1} = \mathcal{D}^{-1/2} (\openone +\mathcal{D}^{-1/2} K \mathcal{D}^{-1/2})^{-1}\mathcal{D}^{-1/2} \\
= \mathcal{D}^{-1} + \mathcal{D}^{-1/2} \sum_{n\geq 1}(-1)^n (\mathcal{D}^{-1/2} K \mathcal{D}^{-1/2})^n \mathcal{D}^{-1/2}
\label{eq:une_serie_geometrique}
\end{multline}
each term of the series being local (for simplicity, we omit to write the exponent $(\ell)$ and the argument $iS$). 
This holds of course if the operator $\mathcal{D}^{-1/2} K \mathcal{D}^{-1/2}$ 
is small enough. For $\ell=1$ in the $2+2$ fermionic problem, this can be made rigorous: the best constant in Eq.~(\ref{eq:local}) is
\be
\label{eq:vala}
A=\frac{2(2-\sqrt{3})}{3\pi}\simeq 0.05686
\ee
Then \footnote{For $\ell=1$ within the even parity sector, the minimal value of $A$ given by Eq.~(\ref{eq:vala}) corresponds to $x=x'\to +\infty$, $u=u'=0$
in Eq.~(\ref{eq:local}). 
Let us use the notation $O=\langle \ell=1,m_z=0|\mathcal{D}^{-1/2} K^{(\ell=1)} \mathcal{D}^{-1/2}|\ell=1,m_z'=0\rangle$ and introduce the operator $T$ 
in the space of functions of the single variable $x$ such that, in Dirac's notation, $\langle x |T|x'\rangle=A/\ch(x-x')$.
Then Eq.~(\ref{eq:local}) can be rewritten as $|\langle x,u|O|x',u'\rangle|\leq \langle x|T|x'\rangle$.
For any integer $n$ greater than $1$, we inject $n-1$ closure relations and use the triangular inequality to obtain
$|\langle x,u|O^n|x',u'\rangle|\leq \int_\mathbb{R} dx_1\ldots dx_{n-1}\int_{-1}^1 du_1\ldots du_{n-1}
\langle x|T|x_1\rangle \ldots \langle x_{n-1}|T|x'\rangle$ $= 2^{n-1} \langle x|T^n|x'\rangle$. Then
$|\langle x,u|[(\openone+O)^{-1}-\openone]|x',u'\rangle|\leq \langle x|\frac{T}{\openone-2T}|x'\rangle$
$=\int_\mathbb{R} \frac{dk}{2\pi} e^{ik(x-x')} \frac{t_k}{1-2t_k}$, where we have used a series expansion in powers of $O$ 
and where $t_k=\int_\mathbb{R} dy \frac{A}{\ch y} e^{iky}=\pi A/\ch(k\pi/2)$ is the eigenspectrum of
$T$. Calculating the integral over $k$ and combining the result with the identity
$\langle \ell=1,m_z=0|K_{\rm inv}^{(\ell=1)}|\ell=1,m_z'=0\rangle=\mathcal{D}^{-1/2} [(\openone+O)^{-1}-\openone] \mathcal{D}^{-1/2}$ leads to Eq.~(\ref{eq:majexacte}).} 
\begin{multline}
\frac{|\langle x,u,\ell=1,m_z=0|K^{(\ell=1)}_{\rm inv}(iS)|x',u',\ell=1,m_z'=0\rangle|}{[d(x,u) d(x',u')]^{-1/2}} \leq  \\ \frac{2A}{\sqrt{1-(2\pi A)^2}}
\frac{\sh\!\left(\frac{2\delta}{\pi}|x-x'|\right)}{\sh(2|x-x'|)}
\label{eq:majexacte}
\end{multline}
with $\delta=\arccos(-2\pi A) \in ]\pi/2,\pi[$.

This locality {\sl per se} is not enough to ensure the convergence of the trace in Eq.~(\ref{eq:formule_de_trace}). Making the trace explicit
in that equation
and injecting a closure relation leads to the writing
\begin{multline}
\frac{d}{dS} \ln \det M^{(\ell)}(iS)=\int_\mathbb{R} dxdx' \int_{-1}^{1} dudu' {\sum_{m_z,m_z'}}^{\!\!\! (-1)^{\ell+1}} \\
\langle x,u,\ell,m_z| [M^{(\ell)}(iS)]^{-1}|x',u',\ell,m_z'\rangle \\
\times \langle x',u',\ell,m_z' | \frac{d}{dS} M^{(\ell)}(iS) |x,u,\ell,m_z\rangle
\label{eq:traceexpli}
\end{multline}
where the sum is restricted to $m_z$ and $m_z'$ of parity opposite to that of $\ell$ as the exponent $(-1)^{\ell+1}$ indicates.
The locality of $M(iS)^{-1}$, and even of $\frac{d}{dS} M(iS)$, exponentially bounds the excursion of $|x-x'|$ in the integral over $x'$, but still 
there remains the integral over the unbounded variable $x$. One must take advantage of the structure of $K(iS)$ and of its derivative: splitting
\be
K(iS) = K_1(iS) + K_2(iS),
\label{eq:KK1K2}
\ee
where $K_1$ and $K_2$ respectively correspond to the first term and the second term in the right-hand side of Eq.~(\ref{eq:noyau_matriciel}), one finds
\be
\frac{d}{dS} K(iS) = i [D_1,K_1(iS)] + i [D_2, K_2(iS)]
\label{eq:structure_en_commutateur}
\ee
where $[A,B]=AB-BA$ is the commutator of two operators and the diagonal operators $D_j$ are defined by the following diagonal functions
\bea
d_1(x)=\frac{1}{2} \ln \frac{e^{+x}}{2\ch x} \\
d_2(x)=\frac{1}{2} \ln \frac{e^{-x}}{2\ch x} 
\eea
such that 
\be
D_j|x,u,\ell,m_z\rangle = d_j(x) |x,u,\ell,m_z\rangle
\label{eq:defDj}
\ee
Clearly the diagonal term $\mathcal{D}^{-1}$ in $M^{-1}$, see Eq.~(\ref{eq:ecritMDK}), has a zero contribution to the trace, as
$[\mathcal{D},D_j]=0$. Eq.~(\ref{eq:traceexpli}) is correspondingly rewritten as
\begin{multline}
\frac{d}{dS} \ln \det M^{(\ell)}(iS)=\int_\mathbb{R} dxdx' \int_{-1}^{1} dudu' {\sum_{m_z,m_z'}}^{\!\!\! (-1)^{\ell+1}} \\
\langle x,u,\ell,m_z| K^{(\ell)}_{\rm inv}(iS)|x',u',\ell,m_z'\rangle \\
\times \sum_{j=1}^{2} i[d_j(x')-d_j(x)] \langle x',u',\ell, m_z' |  K^{(\ell)}_j(iS) |x,u,\ell, m_z\rangle
\end{multline}
Then $d_1(x)$ tends exponentially rapidly to $0$ when $x\to +\infty$, whereas it diverges linearly with $x$ when $x\to-\infty$;
the contrary holds for $d_2(x)$.
A second property is that there exists a constant $B$ such that
\begin{multline}
\label{eq:majun}
|\langle x,u,\ell,m_z| K_1(iS)|x',u',\ell,m_z'\rangle| \leq \\
\left(\frac{e^{+x+x'}}{4\ch x\ch x'}\right)^{1/4}\!\!\!\! \frac{B}{\ch(x-x')}
\end{multline}
\begin{multline}
\label{eq:majdeux}
|\langle x,u,\ell,m_z| K_2(iS)|x',u',\ell,m_z'\rangle| \leq \\ \left(\frac{e^{-x-x'}}{4\ch x\ch x'}\right)^{1/4}\!\!\!\! \frac{B}{\ch(x-x')}
\end{multline}
This is due to the fact, evident from Eqs.~(\ref{eq:D2423},\ref{eq:D2423bis}), that the denominator in the integral over $\phi$
in Eq.~(\ref{eq:noyau_matriciel}), is always larger than $(\mu_{\uparrow\downarrow}/m_\uparrow) \ch(x-x')$. 
Then, for $|x-x'|=O(1)$, the upper bound for the matrix elements of $K_1$ (respectively $K_2$)
tends exponentially fast to $0$ when $x\to -\infty$ (respectively $x\to +\infty$), due to the first factor in Eqs.~(\ref{eq:majun},\ref{eq:majdeux}),
which suppresses the linear divergence in $d_1(x)$ (respectively in $d_2(x)$).
Then the integral over $x$ and $x'$ in the trace converges exponentially rapidly at infinity, the logarithmic derivative of the determinant
of $M^{(\ell)}(iS)$ in the $(-1)^{\ell+1}$ parity channel is well defined
\footnote{To finish the calculation, one can use the mild hypothesis that the matrix elements of $K_{\rm inv}$
are uniformly bounded, that $|d_1(x)-d_1(x')| \left(\frac{e^{x+x'}}{4\ch x\ch x'}\right)^{1/4}\leq
|d_1(x)| \left(\frac{e^{x}}{2\ch x}\right)^{1/4}+|d_1(x')| \left(\frac{e^{x'}}{2\ch x'}\right)^{1/4}$, due to
$\exp(x')\leq 2\ch x'$ or $\exp(x)\leq 2\ch x$ and to the triangular inequality $|d_1(x)-d_1(x')|\leq |d_1(x)|+|d_1(x')|$. Then one can integrate over $x'$
or over $x$ (depending on the term), using $\int_\mathbb{R} \frac{dx'}{\ch(x-x')} =\pi$, and one finally faces the integral
$\int_\mathbb{R} dx |d_1(x)|\left(\frac{e^{x}}{2\ch x}\right)^{1/4} < +\infty$.
} and its value can be calculated with a rapidly vanishing error in the truncation $x_{\rm max}$ when
$x_{\rm max}\to +\infty$. The numerics agree with this conclusion, and indicate that the surprisingly low value $x_{\rm max}=5$ is sufficient.

\noindent{\bf Parity $(-1)^{\ell}$}: The situation is physically quite different for the $(-1)^{\ell}$ parity sector, at least for even $\ell$:
The third contribution in Eq.~(\ref{eq:noyau_matriciel}) is non zero, and it leads to a continuous part in the spectrum of $M^{(\ell)}(iS)$
that reaches zero for even $\ell$, see Eq.~(\ref{eq:dritte}). Then the spectrum of the inverse $[M^{(\ell)}(iS)]^{-1}$ is no longer bounded, and its matrix elements
are not bounded even if one uses the optimal $(t,\psi)$ representation in which the matrix elements of $M^{(\ell)}(iS)$ are bounded, see Eq.~(\ref{eq:enfin_borne}),
when the lower cut-off $t_{\rm min}$ on the $t$ variable tends to $-\infty$.
Then, as we shall see, there is no exponential locality in the $t$-basis but still the logarithmic derivative of the determinant of
$M^{(\ell)}(iS)$ has a finite limit when $t_{\rm min}\to -\infty$, which is approached with an error vanishing linearly with $1/t_{\rm min}$.

\begin{figure}[tb]
\centerline{\includegraphics[width=8cm,clip=]{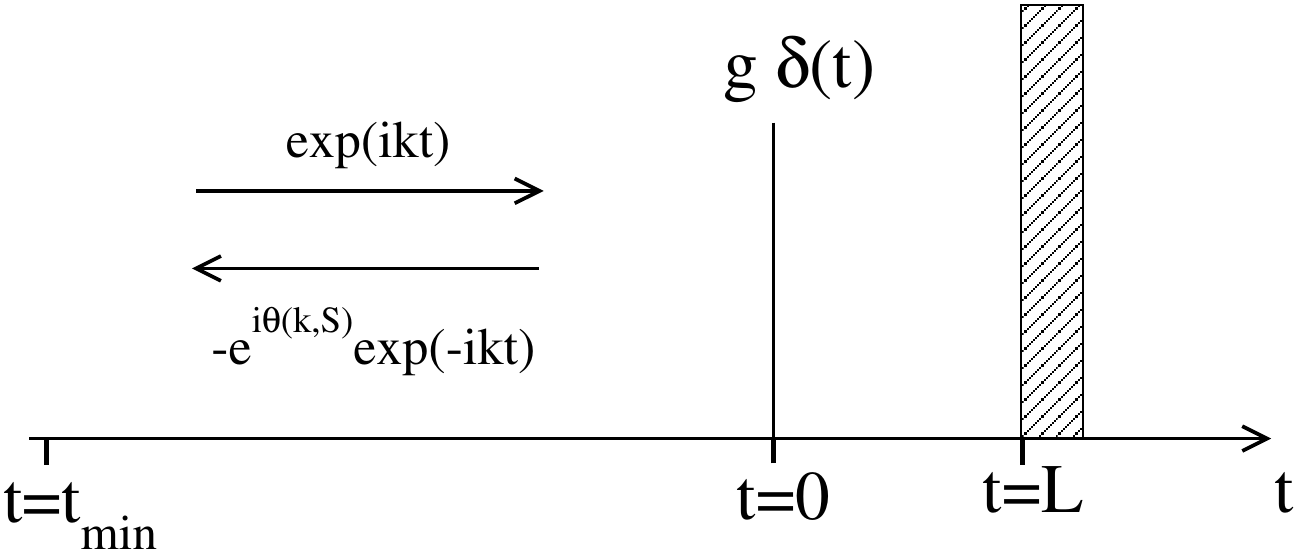}}
\caption{Diagram giving the structure of the eigenstates of the third continuum (\ref{eq:dritte}), in terms of the variable
$t$ of Eq.~(\ref{eq:deft}): $k>0$ is the wave vector of the incoming wave, $-k$ the one of the reflected wave with a phase shift
$\theta(k,S)$. The reflection due to the physics at $t=O(1)$ is, in a toy model, represented by a Dirac scattering potential
at $t=0$ and a hard wall acting as a mirror at $t=L$; the toy model exemplifies the expected low $k$ behavior (\ref{eq:linenk}) of $\theta(k,S)$.}
\label{fig:ondes}
\end{figure}

To derive this property, a spectral or ``Fourier-space" analysis is more appropriate than the ``real"-space analysis of the previous parity case. 
After the gauge transform and the change of functions performed in Eqs.~(\ref{eq:ansatz_pas1},\ref{eq:ansatz_pas2},\ref{eq:deft}), the asymptotic
$t\to -\infty$ part of the eigenstates of $M^{(\ell)}(iS)$ of the continuum (\ref{eq:dritte}) can be written as
\be
\phi_k(t)=e^{ikt}-e^{i\theta(k,S)} e^{-ikt}, \ \ \ k>0
\label{eq:def_theta}
\ee
see Fig.~\ref{fig:ondes}.
The plane waves $e^{ikt}$ and $e^{-ikt}$ are indeed two linearly independent solutions of the eigenvalue problem (\ref{eq:enfin_borne})
with 
\be
\Omega= \Omega_k= \frac{1}{\sqrt{2}}\left[1-\frac{1}{\ch (k\pi/2)}\right]
\ee
[see Eq.~(\ref{eq:dritte}), here $\ell$ is even]. The right solution is some specific superposition of these
two degenerate solutions, with a relative amplitude determined by the physics at $t=O(1)$, that is for $(x,u)$ not extremely close to $(0,-1)$.
Analytically the value of this relative amplitude is an unknown function of $k$ and $S$, but we know that it must be of modulus one,
so that we can express it as in Eq.~(\ref{eq:def_theta}) in terms of a mere phase shift $\theta(k,S)\in \mathbb{R}$: (i) the ``Hamiltonian" $M^{(\ell)}(iS)$
for the spinor is hermitian, so that the corresponding evolution operator is unitary and conserves probability, (ii) the third continuum (\ref{eq:dritte})
is not degenerate with the other continua for the considered mass ratio $\alpha=1$, so the wave $e^{ikt}$ incoming from $t=-\infty$ has
no channel to escape and must fully get out by the incoming channel.

The key property that we shall use is that, as it is common in one-dimensional scattering problems, the phase shift $\theta(k,S)$
vanishes linearly at low $k$:
\be
\theta(k,S) \underset{k\to 0}{=} k b(S) +o(k)
\label{eq:linenk}
\ee
where $b(S)$ is a $S$-dependent effective scattering length.
We present two plausible arguments to establish this. The first argument results from the assumption that the writing (\ref{eq:def_theta}) can be smoothly
extended from $k>0$ to $k<0$: This implies that if one directly replaces $k$ by $-k$ in Eq.~(\ref{eq:def_theta}), the
resulting wave $t\mapsto e^{-ikt}-e^{i\theta(-k,S)} e^{ikt}$ must reproduce the physical solution $e^{ikt} -e^{i\theta(k,S)} e^{-ikt}$
up to a global phase factor, so that $e^{i\theta(k,S)}=e^{-i\theta(-k,S)}$ and there exists an integer $q$ such that
\be
-\theta(-k,S) = \theta(k,S) + 2 q\pi
\label{eq:modulo}
\ee
The fact that, in one dimension, the arbitrarily low energy waves are generically fully retroreflected (no matter how small, but non-zero, the scattering potential is)
leads to $\theta(k,S)\to 0$ for $k\to 0$ and to $q=0$ in Eq.~(\ref{eq:modulo}); then if $\theta(k,S)$ is a smooth function of $k$, Eq.~(\ref{eq:linenk}) holds.
The second argument utilizes some model for the scattering potential in the region $t=O(1)$, introducing on purpose most singular potentials, see
Fig.~\ref{fig:ondes}: a pointlike
fixed scattering center of coupling constant $g=\hbar^2/(2 m_{\rm eff} a)$ placed
at $t=0$ at a distance $L$ from a hard wall (the second element acts as a mirror and ensures that the wave is fully reflected at all energies).
At low $k$, the dispersion relation $\Omega_k$ can be quadratized, $\Omega_k\approx \hbar^2 k^2/(2 m_{\rm eff})$ with $m_{\rm eff}>0$, leading to an effective
Schr\"odinger equation and scattering problem, so that at fixed $S$
\be
e^{i\theta(k)} = \frac{(ka)^{-1}+(\tan kL)^{-1}+i}{(ka)^{-1}+(\tan kL)^{-1}-i}
\ee
The phase shift $\theta(k,S)$ is indeed an odd function of $k$, and at low $k$ one indeed obtains the linear law (\ref{eq:linenk})
with $2/b(S)=1/a(S)+1/L(S)$.

Then the property (\ref{eq:linenk}) leads to the conclusion that the logarithmic derivative of the determinant of $M^{(\ell)}(iS)$
has a finite limit when the lower cut-off value $t_{\rm min}$ tends to $-\infty$, as we now see.  Similarly to Eq.~(\ref{eq:emulcoupnum}), this lower cut-off
corresponds to the boundary condition
\be
\phi(t_{\rm min})=0
\label{eq:condlimtmin}
\ee
which, considering (\ref{eq:def_theta}), leads to the quantization condition for $k$
\footnote{The value $k=0$, that is $n=0$, should not be included. If one directly takes the limit $k\to 0$ in Eq.~(\ref{eq:def_theta}) one gets the absurd
result $\phi_{k=0}(t)=0$. The correct way of taking the limit is to first divide Eq.~(\ref{eq:def_theta}) by $ik$. One then finds
that $\phi_{k=0}(t)$ diverges as $2t-b(S)$  when $t\to -\infty$, so it does not satisfy the boundary condition (\ref{eq:condlimtmin}).}:
\setcounter{notepaszero}{\thefootnote}
\be
2 k |t_{\rm min}| +\theta(k,S) = 2 n \pi, \ \ \forall n\in \mathbb{N}^*
\label{eq:cond_quantif}
\ee
Then the contribution of the corresponding eigenvalues to the logarithmic derivative of the determinant is
\be
\frac{d}{dS} \ln \det M^{(\ell)}(iS)|_{\odot} = \sum_{n>0} \frac{d}{dS} \ln\Omega_k = \sum_{n>0} \Omega_k^{-1} \frac{d\Omega_k}{dk}
\frac{dk}{dS}
\label{eq:dlndetdritte}
\ee
Taking the derivative of Eq.~(\ref{eq:cond_quantif}) with respect to $S$ at fixed $n$ we obtain
\be
\frac{dk}{dS} = - \frac{\partial_S \theta(k,S)}{2|t_{\rm min}|+\partial_k \theta(k,S)}
\ee
In the large $|t_{\rm min}|$ limit, one can neglect $\partial_k \theta(k,S)$ in the denominator and one can replace in Eq.~(\ref{eq:dlndetdritte})
the sum over $n$ by an integral $\int dn$. According to Eq.~(\ref{eq:cond_quantif}),
\be
2 \frac{dk}{dn} |t_{\rm min}| \underset{t_{\rm min}\to -\infty}{\to}  2 \pi
\ee
so that
\be
\frac{d}{dS} \ln \det M^{(\ell)}(iS)|_{\odot} \underset{t_{\rm min}\to -\infty}{\to}\!\!\! -\!\int_0^{+\infty}\!\! \frac{dk}{2\pi}
\frac{1}{\Omega_k} \frac{d\Omega_k}{dk} \partial_S \theta(k,S)
\ee
This is finite even if $\Omega_k$ vanishes quadratically in $k=0$, i.e.\ it is saved from a na\"\i vely expected logarithmic
divergence, because the phase shift $\theta(k,S)$ vanishes linearly with $k$, and so does its derivative with respect to $S$.

This main point being established, there remains a problem of practical interest, the speed of the convergence with $|t_{\rm min}|$. 
The answer is provided by Poisson's summing formula:
\be
\sum_{n\in\mathbb{Z}} f(\lambda n) = \frac{1}{\lambda} \sum_{n\in\mathbb{Z}} \hat{f}(2\pi n/\lambda)
\ee
for any $\lambda>0$ and for an arbitrary function $f(k)$, $\hat{f}(x)=$ $\int_\mathbb{R} dk \exp(-i k x) f(k)$ being its Fourier transform.
For simplicity, we give details in the case where $\theta(k,S)$ is linear in $k$ at all $k$, that is $\theta(k,S)= k b(S)$.
From the quantization condition (\ref{eq:cond_quantif}) one has
\be
k=\lambda n \ \ \mbox{with}\ \ \lambda=\frac{2\pi }{2|t_{\rm min}|+b(S)}
\ee
so that
\be
\frac{dk}{dS}= -\frac{db(S)}{dS} \frac{\lambda }{2\pi} k
\ee
This, together with Eqs.~(\ref{eq:dlndetdritte}), leads to a function $f$ given by
\be
f(k) = \frac{k \frac{db(S)}{dS}}{\Omega_k} \frac{d\Omega_k}{dk}
\ee
such that
\be
\frac{d}{dS} \ln \det M^{(\ell)}(iS)|_{\odot} = -\frac{\lambda }{2\pi} \sum_{n>0} f(\lambda n)
\ee
Then using the fact that the function $f$ is even, one can express the sum over $\mathbb{N}^*$ in terms of the sum over $\mathbb{Z}$,
and then in terms of $f(0)$ and $\hat{f}$:
\begin{multline}
\sum_{n>0} f(\lambda n) = -\frac{1}{2} f(0) + \frac{1}{2} \sum_{n\in\mathbb{Z}} f(\lambda n) \\
= -\frac{1}{2} f(0) + \frac{1}{2\lambda}  \sum_{n\in\mathbb{Z}} \hat{f}(2\pi n/\lambda)
\label{eq:qcbe}
\end{multline}
The function $f$ is a smooth function of $k$, in particular in $k=0$, that rapidly decreases at infinity, so that its Fourier
transform $\hat{f}(q)$ is also rapidly decreasing when $|q|\to +\infty$. In the large $|t_{\rm min}|$ limit, $1/\lambda$ diverges 
linearly in $|t_{\rm min}|$ and one commits an exponentially small error $O[\exp(-C |t_{\rm min}|)]$ ($C$ is some constant) in neglecting
the $n\neq 0$ terms in the last sum over $n$ in Eq.~(\ref{eq:qcbe}).
As a consequence, 
\begin{multline}
\frac{d}{dS} \ln \det M^{(\ell)}(iS)|_{\odot}\underset{t_{\rm min}\to -\infty}{=} 
-\frac{1}{2\pi}  \int_0^{+\infty}\!\!\! dk \frac{k \frac{db(S)}{dS}}{\Omega_k} \frac{d\Omega_k}{dk} \\
+ \frac{\frac{db(S)}{dS}}{2|t_{\rm min}|+b(S)}  + O[\exp(-C |t_{\rm min}|)]
\end{multline}
where we have replaced $\lambda$, $f(0)$ and $\hat{f}(0)$ by their values.  When $\theta(k,S)$ is not a linear function of $k$, we obtain
the general result
\begin{multline}
\frac{d}{dS} \ln \det M^{(\ell)}(iS)|_{\odot}\!\!\underset{t_{\rm min}\to -\infty}{=}\!\!\!  -\frac{1}{2\pi}  
\int_0^{+\infty} \!\!\!\!\! dk \frac{\partial_S \theta(k,S)}{\Omega_k} \frac{d\Omega_k}{dk} \\
+\lim_{k\to 0} \frac{\frac{1}{2}\frac{d\Omega_k}{dk}\partial_S \theta(k,S)}{\Omega_k[2|t_{\rm min}|+\partial_k\theta(k,S)]} + O[\exp(-C |t_{\rm min}|)]
\label{eq:resultat_general}
\end{multline}
In any case, when $t_{\rm min}\to -\infty$, the limiting value
of the logarithmic derivative of the determinant of $M^{(\ell)}(iS)$ is approached with an error that vanishes only polynomially 
with $1/t_{\rm min}$ \footnote{In the numerics, we extrapolate to $1/t_{\rm min}=0$ using a cubic fit in $1/t_{\rm min}$, with data down to
minimal values $1/|t_{\rm min}|=1/{200}$ for $\ell=0$ and $1/|t_{\rm min}|=1/30$ for $\ell>0$. 
For $\ell=0$, as a test of the finite-$t_{\rm min}$ formalism, we have used Eq.~(\ref{eq:resultat_general}) to predict the leading numerical error
on $\Delta B_{2,2}^{\mathrm{conj}(\ell=0)}(0^+)$ due to the $t_{\rm min}$ truncation, that is $(8\pi |t_{\rm min}|)^{-1} \int_\mathbb{R} dS
[b(\infty)-b(S)] = 2.3(1)/(8\pi |t_{\rm min}|)$, which agrees with the direct numerical calculation. To obtain $b(S)$ at any given $S$,  and hence
the integral of $b(\infty)-b(S)$,
we calculated numerically the eigenvectors corresponding to the first few eigenvalues $\Omega_n$ ($n\geq 1$) 
of $M^{(\ell=0)}(s=iS)$, and we fitted the corresponding functions $\phi_n(t)$
[defined as in Eq.~(\ref{eq:deft})] with a three-parameter sine function
$t\mapsto A_n \sin(k_nt-\theta_n/2)$ as suggested by Eq.~(\ref{eq:def_theta}), where $A_n$ is a complex amplitude, $k_n$ an effective wavenumber and $\theta_n$ a phase shift. 
The fits are very good, the obtained values of $k_n$ agree very well
with the dispersion relation (\ref{eq:dritte}); setting $\theta(k_n,S)=\theta_n$, we also find that the quantization condition (\ref{eq:cond_quantif}) is 
well obeyed; finally, extrapolating $\theta_n/k_n$ to $n=0$ linearly in $k_n^2$ gives $b(S)$. 
To be complete, we note that $b(S)$ looks like a negative amplitude Gaussian on a non-zero background 
$b(\infty)$, that is $b(\infty)-b(S) \simeq 1.05 \times \exp(-0.668 S^2)$. 
As the variable $t$ in Eq.~(\ref{eq:deft}) depends on
$\rho_0$, so does $b(\infty)$. In our numerics, $\rho_0=2/5$ and we find $b(\infty)= 3.84(1)$. More analytically,
one expects at large $S$ that the phase shift $\theta(k,S)$ is imposed by the third contribution in Eq.~(\ref{eq:noyau_matriciel}), the first two ones becoming rapidly
oscillating and negligible. Then one can use the analytical results of footnote [\thenotesimon]: 
introducing the phase shift $\theta_\psi(k)$ such
that $\psi(t)=\sin[kt-\theta_\psi(k)/2]+o(1)$ for $t\to-\infty$, 
one expects that $\theta(k,S)\to\theta_\psi(k)-2k\ln(\rho_0/\sqrt{2})$ for $S\to\infty$,
so that $b(\infty)=b_\psi-2 \ln(\rho_0/\sqrt{2})$ with $b_\psi\simeq 1.33$. Our numerics fulfill these expectations, which constitutes a good test.
}

On the contrary, if the dispersion relation $\Omega_k$ nowhere approaches zero, as for an odd $\ell$ in the parity sector $(-1)^{\ell}$, the
$\lim_{k\to 0}$ term in the right-hand side of Eq.~(\ref{eq:resultat_general}) is zero and the 
convergence of the logarithmic derivative of the determinant is exponentially fast with $|t_{\rm min}|$, 
as also observed numerically; this last situation is then similar
to the exponentially fast convergence of $\frac{d}{dS} \ln \det M^{(\ell)}(iS)$ when $x_{\rm max}\to +\infty$, which is always
achieved for the continua (\ref{eq:continuum_droite},\ref{eq:continuum_gauche}), even when the third term
in Eq.~(\ref{eq:noyau_matriciel}) is active.

\begin{figure*}[tb]
\begin{center}
\includegraphics[height=4cm,clip=]{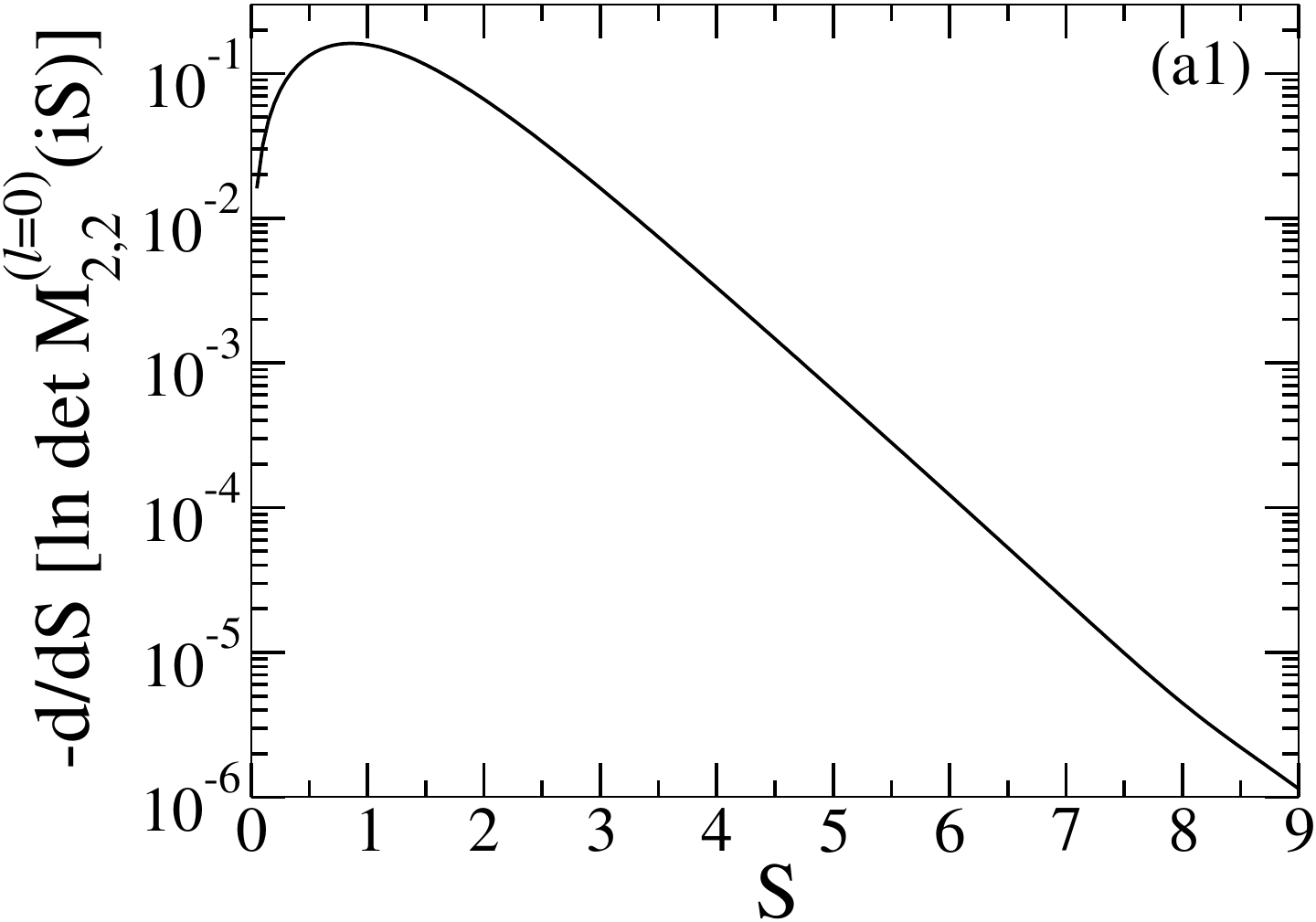} 
\includegraphics[height=4cm,clip=]{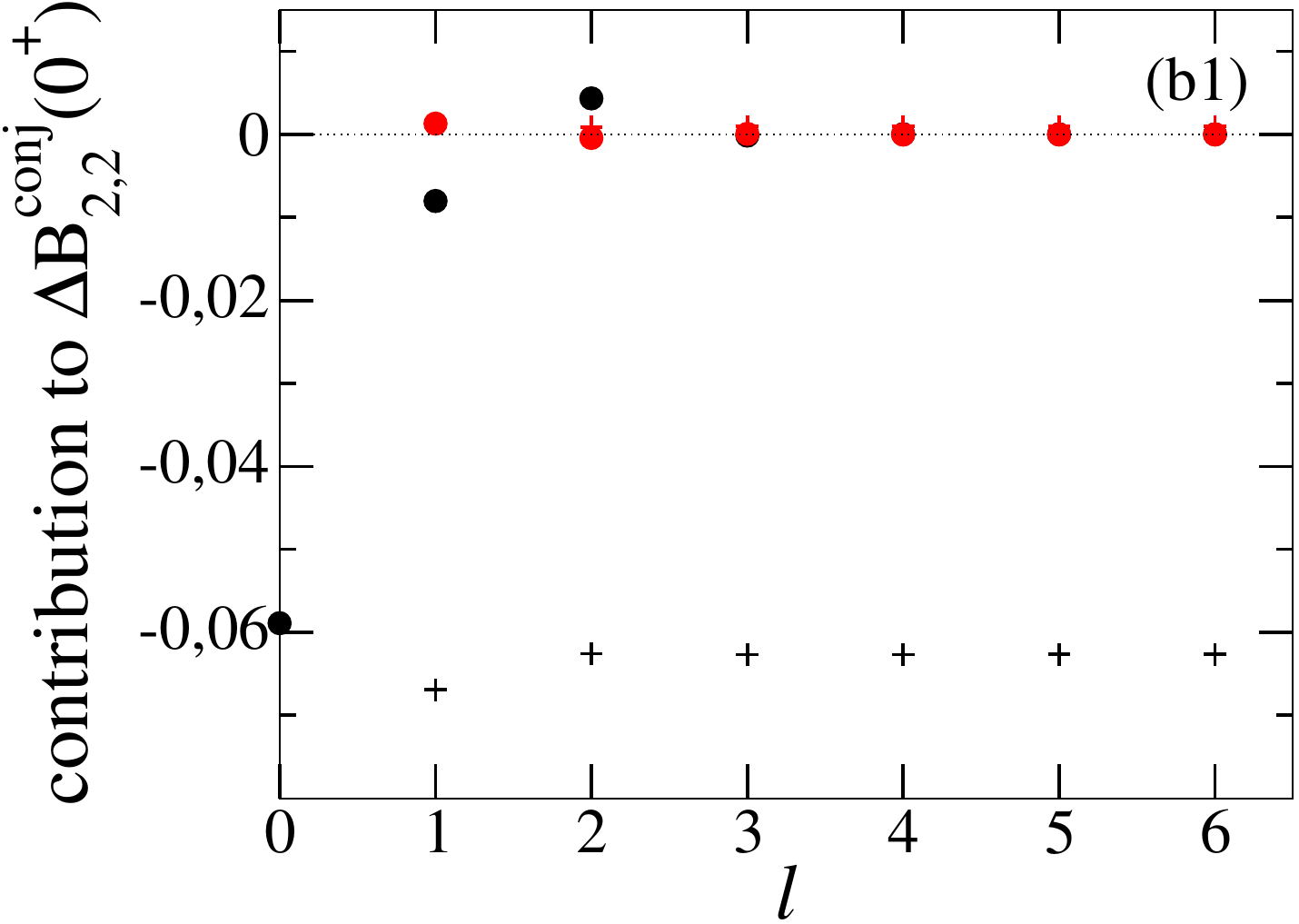} 
\includegraphics[height=4cm,clip=]{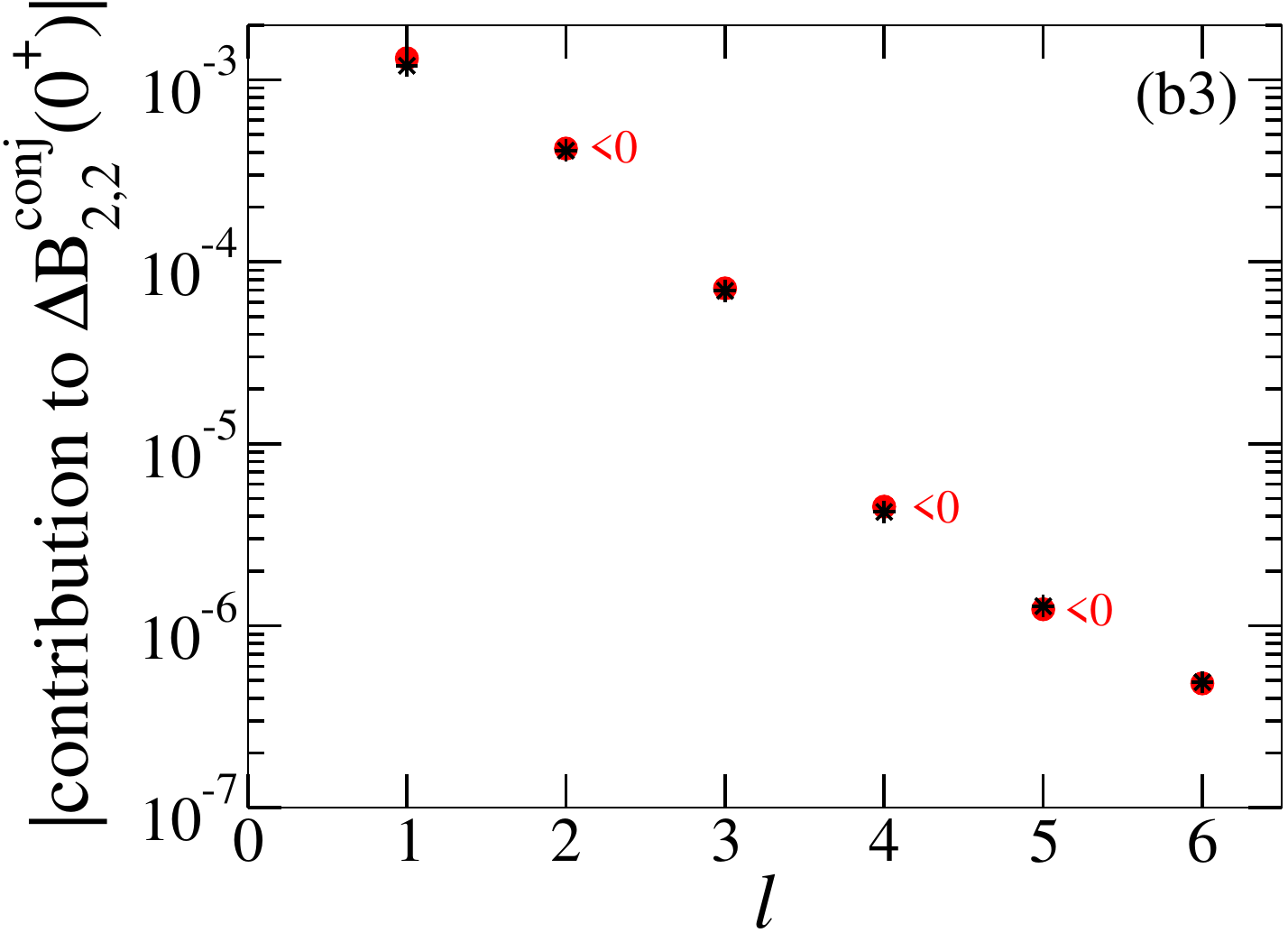}  \\
\includegraphics[height=4cm,clip=]{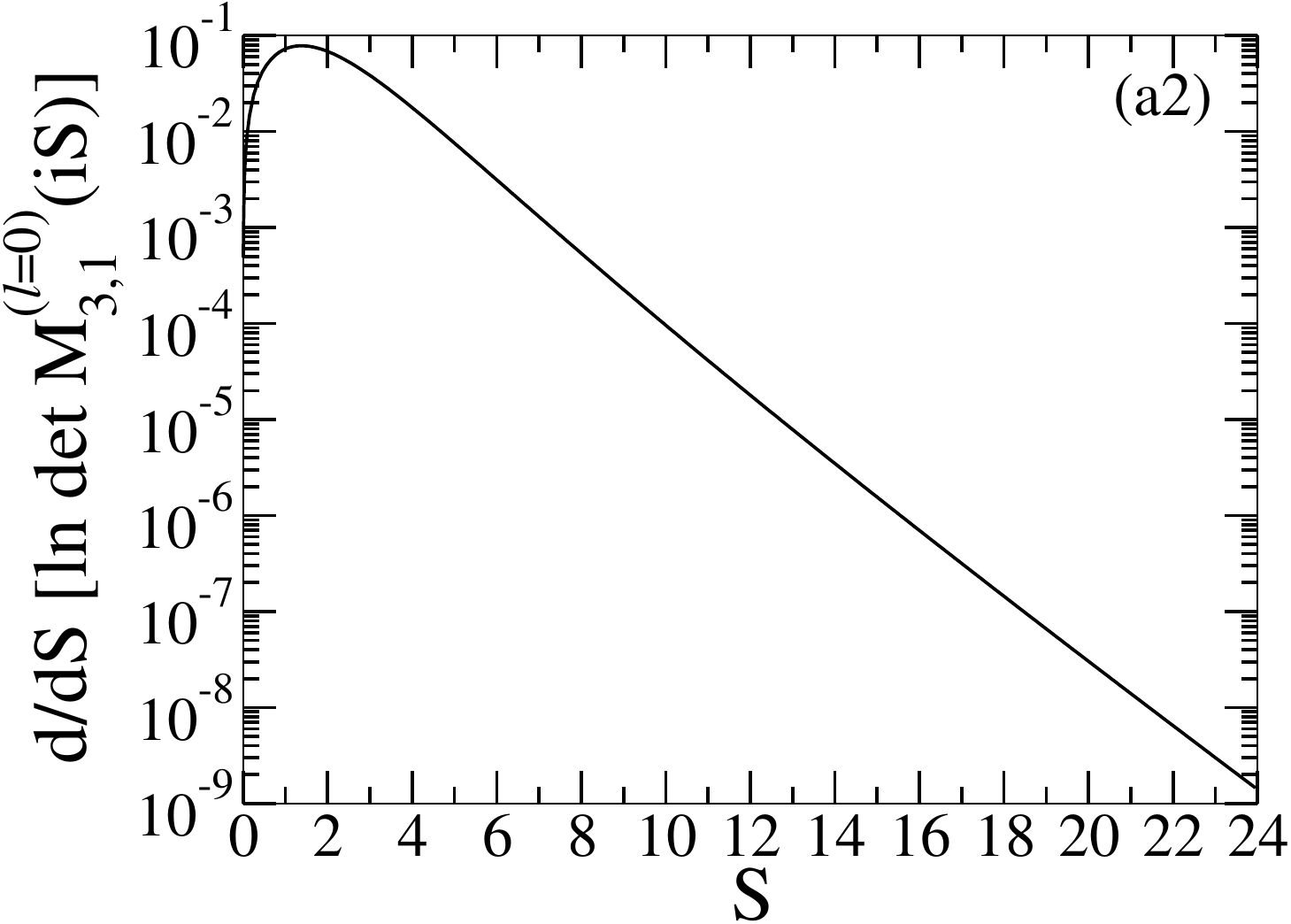}
\includegraphics[height=4cm,clip=]{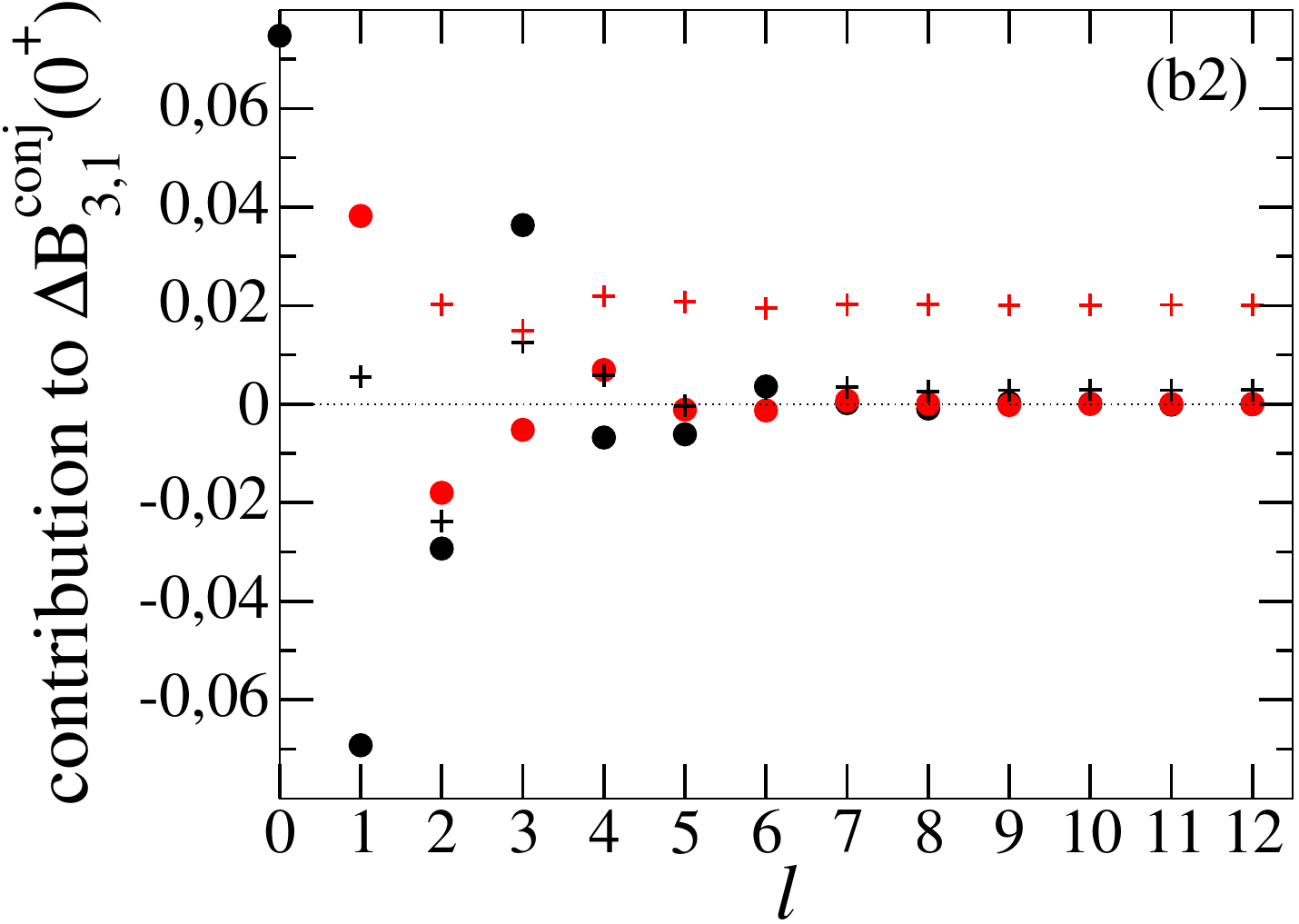}
\includegraphics[height=4cm,clip=]{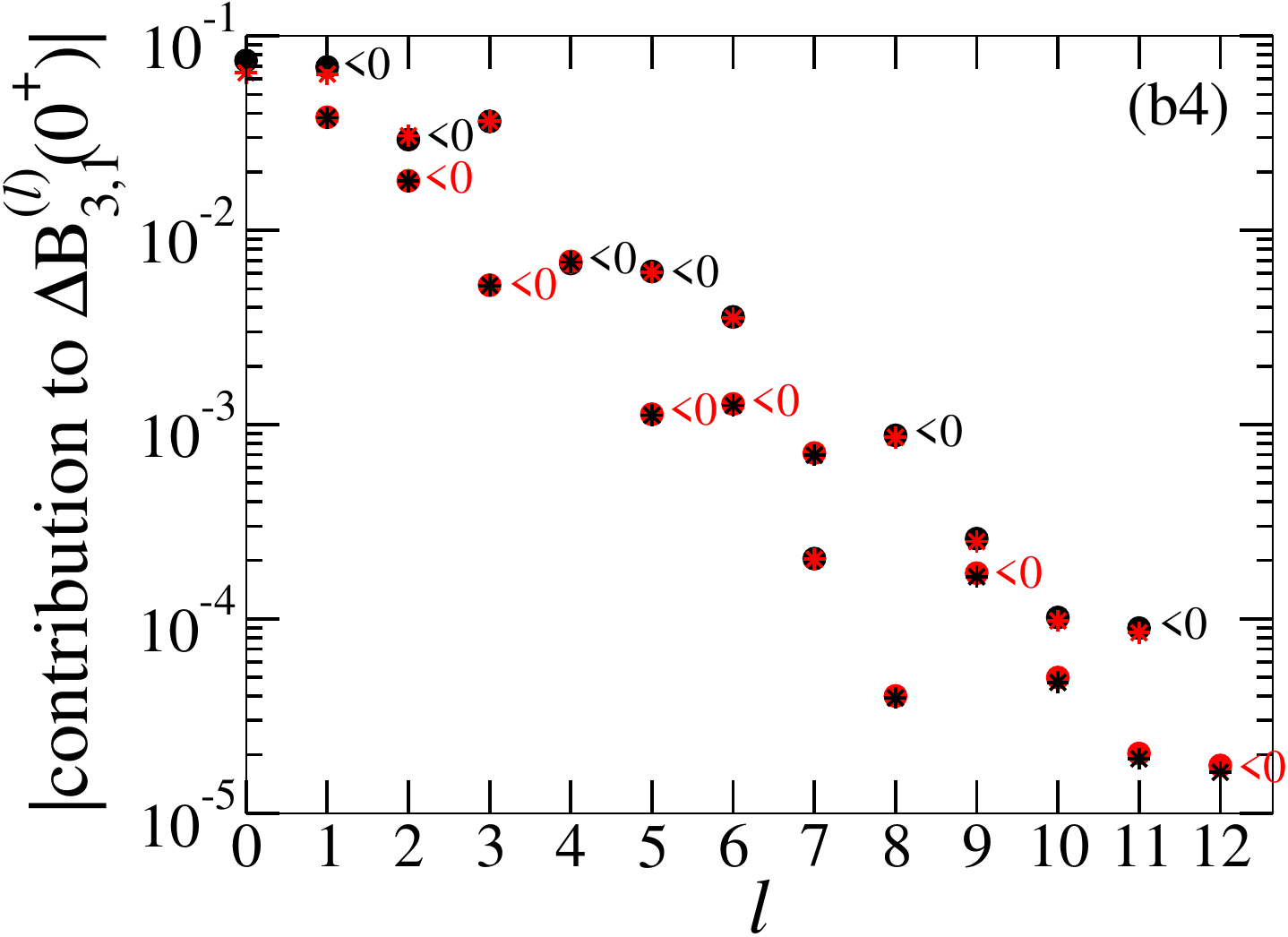}
\end{center}
\caption{{(Color online)} Convergence in the integration over $S$ and summation over $\ell$ in Eq.~(\ref{eq:conjecture}) (we recall that the mass ratio is $\alpha=1$). 
(a) The logarithmic
derivative of the determinant of $M^{(\ell)}(iS)$ is a rapidly decreasing function of $S$; the figure takes as an example
(a1) the $\ell=0$ channel of the $2+2$ fermionic problem (for a numerical cut-off $t_{\rm min}=-9$, thus without extrapolation; 
note the minus sign in the vertical axis), 
(a2) the $\ell=0$ channel of the $3+1$ fermionic problem. 
(b) The sum over $\ell$ also seems to converge well, see the contribution of each
angular momentum channel to the result (\ref{eq:conjecture}) for (b1) the $2+2$ problem and (b2) the $3+1$ problem: black disks for the parity sector $(-1)^\ell$,
red disks for the parity sector $(-1)^{\ell+1}$; the plus signs indicate the corresponding cumulative sums. Convincing evidence
is even shown in (b3) for the $2+2$ problem in the $(-1)^{\ell+1}$ parity sector and in (b4) for the $3+1$ problem in both parity sectors, 
where the numerical results (black disks for parity $(-1)^\ell$, red disks for parity $(-1)^{\ell+1}$) are compared to
the perturbative results (\ref{eq:bel_equivalent}) and (\ref{eq:bel_equivalent3p1})  [red asterisks for parity $(-1)^\ell$, black asterisks for parity $(-1)^{\ell+1}$] 
that extend to four bodies a technique developed for three bodies in reference \cite{CastinWerner} and are expected
to be exact asymptotic equivalents for $\ell\to +\infty$ (what is actually plotted is the absolute value of the results, 
so as to allow for a log scale,
but their sign is indicated with the label ``$<0$" of the same color as the corresponding disks when they are negative: 
{the negative black (red) disks are indicated with a black (red) ``$<0$" label;
note that the black (red) asterisks have always the same sign as the corresponding red (black) disks}).
\label{fig:conv}}
\end{figure*}

\subsection{Other convergence issues}

To show that the conjectured values (\ref{eq:conjecture}) are finite, one must also check that the integral over $S$ is convergent
at infinity, and that the sum over the angular momenta $\ell$ is convergent. This we have initially explored numerically. First, at a given $\ell$, we found that 
the logarithmic derivative of the determinant of $M^{(\ell)}(iS)$ rapidly decreases when $S\to +\infty$, presumably exponentially fast, see Fig.~\ref{fig:conv}(a). 
Second, after the integration over $S$ is taken, one observes also a rapid convergence of the series
over $\ell$, see Fig.~\ref{fig:conv}(b), if one takes the precaution
to be accurate enough in the discretization of the integral over $u$  
\footnote{In practice, we used a Gauss-Legendre scheme with up to $59$ points,
using $\theta$ rather than $u=\cos\theta$ as the integration variable, with the change of function (\ref{eq:chgtdefonc}) and the inclusion
of the extra Jacobian $(\sin\theta \sin \theta')^{1/2}$ in the matrix kernel.}.

These numerical results suggest that the contribution of angular momentum $\ell$ to $\Delta B_{2,2}^{\rm conj}(0^+)$
and to $\Delta B_{3,1}^{\rm conj}(0^+)$ can be obtained, when $\ell$ is large enough, 
from a perturbative calculation in Eq.~(\ref{eq:formule_de_trace}), limited to leading order
in the operators $K_j^{(\ell)}$ defined by Eqs.~(\ref{eq:decompDK},\ref{eq:KK1K2}), at least for the $2+2$ problem in the parity channel $(-1)^{\ell+1}$ where 
$K_{\rm inv}$ in Eq.~(\ref{eq:ecritMDK}) has a chance of being bounded. This idea was implemented with success at
the three-body level in reference \cite{CastinWerner}, treating the integral term in Eq.~(\ref{eq:defLam}) as a perturbation of the constant term
\footnote{The integration over $S\in \mathbb{R}$ of $\Lambda_{\ell}(iS)/\cos\nu -1$ using the first line of Eq.~(\ref{eq:defLam})
leads to $\int_{-1}^{1} du P_\ell(u)/(1+u\sin \nu)$.
The large $\ell$ limit of that integral reproduces exactly Eq.~(42) of reference \cite{CastinWerner},
as we have checked using $P_\ell(u)=(2^\ell \ell !)^{-1} \frac{d^\ell}{du^\ell}[(u^2-1)^\ell]$, then integrating $\ell$ times by parts 
then using Laplace's method.}. 

Let us implement the idea for the $2+2$ problem, in the parity sector $(-1)^{\ell+1}$ of the subspace of angular momentum $\ell$. 
We truncate Eq.~(\ref{eq:une_serie_geometrique}) to order one included in the operator $K$, to obtain
\be
\frac{d}{dS} \ln \det M \simeq \Tr\left[\left(\mathcal{D}^{-1}-\mathcal{D}^{-1} K \mathcal{D}^{-1}\right)\frac{d}{dS} K\right]
\ee
Then we split $K$ as in Eq.~(\ref{eq:KK1K2}) and we use the commutator structure (\ref{eq:structure_en_commutateur}). Using the invariance of the trace
in a cyclic permutation and the fact that the diagonal operators $\mathcal{D}$ of Eq.~(\ref{eq:cestquoid}) and $D_j$ of Eq.~(\ref{eq:defDj}) commute, we find that only the crossed quadratic contributions
in $K_1$ and $K_2$ survive, so that
\begin{multline}
\frac{d}{dS} \ln \det M \simeq \Tr\left[-\mathcal{D}^{-1} K_1 \mathcal{D}^{-1} \frac{d}{dS} K_2- (1\leftrightarrow 2) \right]  \\
= \frac{d}{dS}\Tr \left(-\mathcal{D}^{-1} K_1 \mathcal{D}^{-1} K_2\right)
\end{multline}
Integrating by parts in Eq.~(\ref{eq:conjecture}) and using the fact that the integrand is an even function of $S$ we obtain the approximation
\be
\int_0^{+\infty} \frac{dS}{\pi} S \frac{d}{dS} \ln \det M \simeq \int_\mathbb{R} \frac{dS}{2\pi} \Tr \left(\mathcal{D}^{-1} K_1 \mathcal{D}^{-1} K_2\right)
\ee
Calculating the trace in the $|x,u,\ell,m_z\rangle$ basis (with $\ell+m_z$ odd) and injecting a closure relation in that basis as e.g.\, in Eq.~(\ref{eq:traceexpli}),
we realize that the integrand has a very simple dependence with $S$, due to simplifications as follows:
\be
\left(\frac{e^x\ch x'}{e^{x'}\ch x}\right)^{s/2} \left(\frac{e^{-x'}\ch x}{e^{-x}\ch x'}\right)^{s/2}= e^{i(x-x')S}
\ee
where we wrote the phase factor of the first term as it is in Eq.~(\ref{eq:noyau_matriciel}) and the phase factor of the second term of Eq.~(\ref{eq:noyau_matriciel})
with $x \leftrightarrow x'$, and used $s=iS$ with $S$ real.
So the integral over $x$ or $x'$ takes the form of a Fourier transform with respect to $x$ or $x'$, with $S$ as the conjugate variable; this is the Fourier
transform of a smooth rapidly decreasing function of $x$ or $x'$, so that, as a function of $S$, it is a rapidly decreasing function.
This gives a reason for the numerically observed fast decay of $\frac{d}{dS} \ln \det M$ at large $S$. Also, integration over $S$ is
straightforward due to
\be
\int_\mathbb{R} \frac{dS}{2\pi} e^{i(x-x')S} = \delta(x-x')
\ee
We finally obtain the leading order approximation
\begin{widetext}
\begin{multline}
\label{eq:bel_equivalent}
\Delta B_{2,2}^{\rm conj}(0^+)|^{(\ell)}_{\mathrm{parity} (-1)^{\ell+1}} \simeq
\frac{2\ell+1}{(4\pi)^2} \int_\mathbb{R} dx \int_0^\pi d\theta \int_0^\pi d\theta'
\int_0^{2\pi} \frac{d\phi}{2\pi} \int_0^{2\pi} \frac{d\phi'}{2\pi} \frac{vv'}{d(x,u) d(x,u') \ch x}  
\\
\times \frac{\displaystyle\frac{1}{4}\sum_{n=0}^{1} \sum_{n'=0}^{1} (-1)^{(\ell+1)(n+n')} \mathcal{T}_\ell(\theta+n\pi,\theta'+n'\pi,\phi,\phi')}
{\displaystyle\left\{1+\frac{1}{1+\alpha}\left[(u+e^{-x})(u'+e^{-x})+v v' \cos\phi\right]\right\} 
\left\{1+\frac{\alpha}{1+\alpha}\left[(u+e^{x})(u'+e^{x})+v v' \cos\phi'\right]\right\}}
\end{multline}
In the integrand of Eq.~(\ref{eq:bel_equivalent}), $d(x,u)$ is given by Eq.~(\ref{eq:cestquoid}), we again use the notations $u=\cos\theta$
and $v=\sin\theta$ and the same for $\theta'$, and we introduced the function
\bea
\mathcal{T}_\ell(\theta,\theta',\phi,\phi') &\equiv &
\sum_{m_z, m_z'=-\ell}^{\ell} e^{-i m_z\theta} \langle \ell,m_z|e^{i\phi L_x/\hbar}|\ell,m_z'\rangle  
e^{i m_z' \theta'} \langle \ell,m_z'|e^{i\phi' L_x/\hbar}|\ell,m_z\rangle\\
&=& \Tr_{\ell}\left[e^{-i\theta L_z/\hbar} e^{i\phi L_x/\hbar} e^{i\theta' L_z/\hbar} e^{i\phi' L_x/\hbar}\right] =\frac{\sin[(2\ell+1)\delta/2]}{\sin(\delta/2)}
\eea
where the trace is taken over the whole subspace $\{|\ell,m_z\rangle, -\ell\leq m_z\leq \ell\}$ of angular momentum $\ell$ without any parity restriction
and the angle $\delta \in [0,\pi]$ is such that
\footnote{The product of the four unitary operators under the trace represents in the single-particle Hilbert space
a rotation of angle $\delta$ around some axis. It is easy to explicitly evaluate this trace as a function of the angles
$\theta,\theta',\phi,\phi'$ in the case $\ell=1$, where each operator can be replaced by a well-known $3\times 3$ rotation
matrix in the usual, three-dimensional space. This leads to the expression (\ref{eq:quevautdelta}).}
\be
1+2\cos\delta = uu'(1+\cos\phi \cos\phi')-(u+u')\sin\phi \sin \phi'+vv'(\cos\phi+\cos\phi')+\cos\phi \cos\phi'
\label{eq:quevautdelta}
\ee
\end{widetext}
The sum over $n$ and $n'$ in the numerator of the integrand of Eq.~(\ref{eq:bel_equivalent}) suppresses the contribution to $\mathcal{T}_\ell$
of the states $|\ell,m_z\rangle$ and $|\ell,m_z'\rangle$ of the wrong parity, $(-1)^{m_z}=(-1)^{m_z'}=(-1)^\ell$.
We expect the approximation (\ref{eq:bel_equivalent}) to be an exact asymptotic equivalent for $\ell\to +\infty$, and this is also
what the comparison to the numerical results in Fig.~\ref{fig:conv}(b3) indicates. 
Amazingly it is already good for $\ell=1$, as it deviates from the numerical value by about $9\%$ only.

This perturbative treatment can also be applied to the $3+1$ problem, using the integral equations of reference \cite{CMP}.
The main difference is that the spinor $\Phi^{(\ell)}_{m_z}(x,u)$ is now subjected to a condition reflecting the fermionic exchange
symmetry of the two $\downarrow$ particles that are spectators of the interacting $\uparrow\downarrow$ pair \cite{CRAS4corps},
\be
\Phi^{(\ell)}_{-m_z}(-x,u) = (-1)^{\ell+1} \Phi^{(\ell)}_{m_z}(x,u)
\label{eq:cond_symetrie}
\ee
This means that the kernel $K^{(\ell)}$ must be restricted to the corresponding subspace, hence the occurrence of a projector $P=(1+U)/2$
on that subspace, where the unitary operator $U$ such that in Dirac's notation
\begin{multline}
U |x,u,\ell,m_z\rangle = (-1)^{\ell+1} |-x,u,\ell,-m_z\rangle \\
= - e^{i\pi L_x/\hbar} |-x,u,\ell,m_z\rangle
\end{multline}
is an involution ($U^2=\openone$) \footnote{Furthermore the Hilbert space was limited in reference \cite{CMP} to the kets $|x,u,\ell,m_z\rangle$ with $x>0$, as the
symmetry condition (\ref{eq:cond_symetrie}) allows, which amounts to adding an extra projector $\mathcal{P}_{x>0}$. This complicates
things because $\mathcal{P}_{x>0}$ and $U$ do not commute. Fortunately, in calculating operator traces, one can use the properties $UP=PU=P$, $\mathcal{P}_{x<0}=U \mathcal{P}_{x>0} U$ and $\mathcal{P}_{x<0}
+\mathcal{P}_{x>0}=\openone$, as well as the invariance of the trace under a cyclic permutation of the operators,  so that
$\Tr(\mathcal{P}_{x<0} P A P)= \Tr(U \mathcal{P}_{x>0} U P A P)= \Tr(\mathcal{P}_{x>0} P A P) = \frac{1}{2} \Tr(PAP)= \frac{1}{2} \Tr(AP)$
and $\Tr(A P \mathcal{P}_{x<0} P B)=\Tr(A P \mathcal{P}_{x>0} P B)=\frac{1}{2} \Tr(A P B)$, where $A$ and $B$ are arbitrary operators.}.
The interesting point is now that, even if $\frac{d}{dS} K$ is a sum of commutators as in Eq.~(\ref{eq:structure_en_commutateur}), 
the corresponding $D_j$ do not commute with the projector $P$. As a consequence, when one expands $M^{-1}$ up to first order in $K$,
$\frac{d}{dS} \ln \det M$ contains both a contribution of order one in $K$ and two contributions of order two in $K$.
Here is the result in the subspace of angular momentum $\ell$ and parity $\varepsilon$ \footnote{If one writes the operator $M$ of reference
\cite{CMP} before its restriction to the subspace of symmetry (\ref{eq:cond_symetrie}) as $\mathcal{D}+ K_0 + U K_0 U$ then after implementation
of the symmetry and restriction to the Hilbert space of kets $|x,u,\ell,m_z\rangle$ with $x>0$, it becomes
$\mathcal{P}_{x>0} [\mathcal{D} + (1+U) K_0 (1+U)] \mathcal{P}_{x>0}$. The first, second and third contributions in the right-hand side
of Eq.~(\ref{eq:bel_equivalent3p1}) are respectively given by $-(\ell+1/2)/(2\pi)$ times the integral over $S\in \mathbb{R}$
of $\Tr_{\ell,\varepsilon}(\tilde{K}_0 U)$, of $-\frac{1}{2}\Tr_{\ell,\varepsilon}[\tilde{K}_0 (U \tilde{K}_0 U)]$ and of $-\Tr_{\ell,\varepsilon}(\tilde{K}_0^2 U)$,
where we have set $\tilde{K}_0\equiv \mathcal{D}^{-1} K_0$ and the index $\ell,\varepsilon$ means that the trace is restricted
to the states $|\ell,m\rangle$ with $(-1)^m=\varepsilon$. Note that $U\mathcal{D}U=\mathcal{D}$ and $[\mathcal{P}_{x>0},\mathcal{D}]=0$.}:
\begin{widetext}
\begin{multline}
\label{eq:bel_equivalent3p1}
\Delta B_{3,1}^{\rm conj}(0^+)|^{(\ell)}_{\mathrm{parity}\, \varepsilon} \simeq
\frac{2\ell+1}{2\pi\sqrt{2}} \int_0^{\pi} d\theta \int_0^{2\pi} \frac{d\phi}{2\pi}\, \frac{v}{d_{31}(0,u)}\,
\frac{\displaystyle\frac{1}{2} \sum_{n=0}^1 \varepsilon^n \mathcal{T}_\ell(\theta+n\pi,0,\phi+\pi,0)}
{\displaystyle 3+\frac{2\alpha}{1+\alpha} (2u+u^2+v^2\cos\phi)} \\
+\frac{2\ell+1}{8 \pi^2} \int_\mathbb{R} dx \int_0^\pi d\theta \int_0^\pi d\theta'
\int_0^{2\pi} \frac{d\phi}{2\pi} \int_0^{2\pi} \frac{d\phi'}{2\pi}\, \frac{vv'}{d_{3,1}(x,u) d_{3,1}(x,u') \ch x}  
\\
\times \frac{\displaystyle\frac{1}{4}\sum_{n=0}^{1} \sum_{n'=0}^{1} \varepsilon^{n+n'} \mathcal{T}_\ell(\theta+n\pi,\theta'+n'\pi,\phi,\phi')}
{\displaystyle\left\{2+e^{-2x}+\frac{2\alpha}{1+\alpha}\left[e^{-x}(u+u')+u u'+v v' \cos\phi\right]\right\} 
\left\{2+e^{2x}+\frac{2\alpha}{1+\alpha}\left[e^x (u+u')+u u'+v v' \cos\phi'\right]\right\}} \\
-\frac{2\ell+1}{4\pi^2} \int_\mathbb{R} dx' \int_0^\pi d\theta \int_0^\pi d\theta' \int_0^{2\pi} \frac{d\phi}{2\pi} \int_0^{2\pi} \frac{d\phi'}{2\pi}
\, \left(\frac{e^{x'}}{\ch x'}\right)^{1/2} \frac{v v'}{d_{31}(0,u)d_{31}(x',u')}  \\
\times\frac{\displaystyle\frac{1}{4}\sum_{n=0}^{1} \sum_{n'=0}^{1} \varepsilon^{n+n'} \mathcal{T}_\ell(\theta+n\pi,n'\pi,\phi,\phi'+\pi)}
{\displaystyle\left[2e^{-x'}+e^{x'}+\frac{2\alpha}{1+\alpha} (u e^{-x'}+u'+u u'+v v'\cos\phi)\right]
\left[2e^{-x'}+e^{x'}+\frac{2\alpha}{1+\alpha}(u e^{-x'}+u'+u u'+v v'\cos\phi')\right]}
\end{multline}
\end{widetext}
where $d_{31}(x,u)$ defines the diagonal part $\mathcal{D}$ of the operator $M$ for the $3+1$ problem (as $d(x,u)$ did for the $2+2$ problem),
see reference \cite{CMP}:
\be
d_{31}(x,u)=\left[\frac{1+2\alpha}{(1+\alpha)^2}+\frac{\alpha u}{(1+\alpha)^2\ch x}\right]^{1/2}
\ee
As can be checked in Fig.~\ref{fig:conv}(b4), this approximation is in good agreement with the numerical results even for $\ell=0$, where
it deviates from the exact result only by $\simeq 13\%$. In the large $\ell$ limit the first contribution in the right-hand side
of Eq.~(\ref{eq:bel_equivalent3p1}) rapidly dominates over the other two;  summing over the two parity sectors $\varepsilon=\pm 1$
and restricting for simplicity to a mass ratio $\alpha=1$, one can integrate it over $\theta$ and $\phi$ at
fixed $\delta\in [0,\pi]$, where $1+2\cos\delta=u+\cos\phi + u\cos\phi$ as shown by Eq.~(\ref{eq:quevautdelta}) taken with $\theta'=\phi'=0$, 
to obtain the rapidly decreasing large-$\ell$ equivalent 
\footnote{The $\ell$ independent function in factor of the sine function in the integrand of Eq.~(\ref{eq:equiv_plus_expli}) is a smooth function of $\delta$
over $[0,\pi]$, with all its even order derivatives (including the zeroth order) vanishing at $\delta=0$ and all its odd order derivatives
vanishing at $\delta=\pi$. Under repeated integration by parts (always integrating the sine function), the fully-integrated term vanishes
at the boundaries and one pulls out at each step a factor $(\ell+1/2)^{-1}$. So Eq.~(\ref{eq:equiv_plus_expli}) is $O[(\ell+1/2)^{-n}]$
when $\ell\to +\infty$, for all integers $n$.}
\begin{multline}
\Delta B_{3,1}^{\rm conj}(0^+)|^{(\ell)} \underset{\ell\to +\infty}{\sim} 
\frac{2\ell+1}{2\pi^2} \int_0^{\pi} d\delta  \sin[(\ell+1/2)\delta] \\
\times \frac{\displaystyle\arccos \frac{8\cos^2\delta +5\cos\delta-1}{3(3+\cos\delta)}}
{[(5+4\cos\delta)(1+\cos\delta+\cos^2\delta)]^{1/2}} 
\label{eq:equiv_plus_expli}
\end{multline}

\begin{figure}[tb]
\includegraphics[width=8cm,clip=]{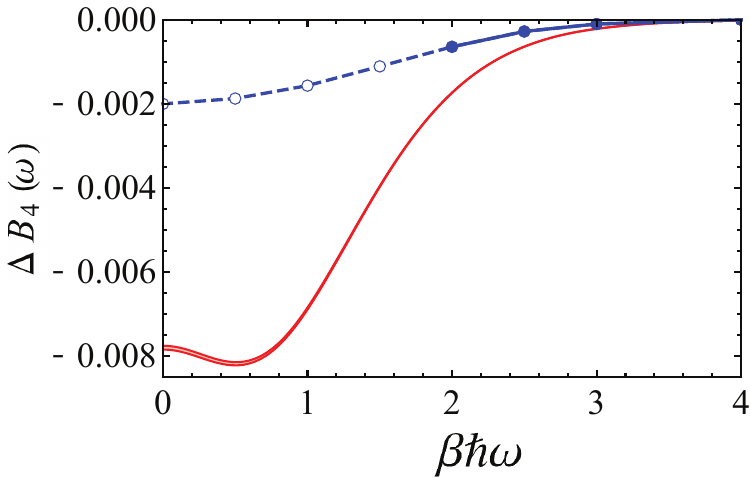}
\caption{(Color online) Fourth order cluster coefficient (\ref{eq:defDB4}) for a harmonically trapped unpolarized spin $1/2$ unitary Fermi gas at temperature $T$, 
as a function of $\beta \hbar \omega$, with $\beta=1/(k_B T)$, $\omega$ the angular oscillation frequency of a fermion in the trap and 
$\alpha=m_{\uparrow}/m_{\downarrow}=1$.
Blue line with symbols: results of reference \cite{Blumeb4} obtained by brute force numerical calculation of the up-to-four-body spectra
in the trap (disks: actually calculated values; circles: values resulting from an extrapolation). Red lines: our conjecture
(\ref{eq:conjecture_etendue}) (the values slightly differ depending on the linear or cubic extrapolation to the numerical cut-off limit $1/t_{\rm min}\to 0^{-}$).
\label{fig:vsBlume}
}

\end{figure}

\subsection{The verdict}

The numerical results for our conjecture (\ref{eq:conjecture}) are
\bea
\Delta B^{\rm conj}_{2,2}(0^+) &=& -0.0617(2) \\
\Delta B^{\rm conj}_{3,1}(0^+) &=& +0.02297 (4) 
\eea
leading, after use of Eq.~(\ref{eq:DeltabnefdB}), to
\be
\Delta b_4^{\rm conj}= -0.063 (1)
\ee
This clearly disagrees with the experimental values (\ref{eq:valb4exp}). {Remarkably, for $\Delta B_{3,1}(0^+)$ our conjectured value
is very close to the approximate diagrammatic result $0.025$ of \cite{JesperLevinsen}, whereas for $\Delta B_{2,2}(0^+)$ it widely differs from 
the (still approximate) result $-0.036$ of \cite{JesperLevinsen} {(these values were communicated to us privately by Jesper Levinsen)}.}

A useful complementary test is to compare to the theoretical results of reference \cite{Blumeb4}.
As mentioned above and in that reference, these results, obtained with the harmonic regulator technique, are trustable
at {\sl non-zero} values of $\beta \hbar \omega$ without extrapolation to $\beta \hbar \omega=0$ ($\omega$ is the angular oscillation frequency
in the trap and $\beta=1/(k_B T)$). It is actually
straightforward to extend with the same notations the conjecture (\ref{eq:conjecture}) to a non-zero value of $\omega$, see Eq.~(38) of reference \cite{CastinWerner}:
\begin{multline}
\Delta B^{\rm conj}_{n_\uparrow,n_\downarrow}(\omega)=\sum_{\ell\in \mathbb{N}} \left(\ell+\frac{1}{2}\right)  \\
\times \int_0^{+\infty} \frac{dS}{\pi} \frac{\sin (S\beta\hbar\omega)}{\sh(\beta\hbar\omega)} \frac{d}{dS} [\ln \det M_{n_\uparrow,n_\downarrow}^{(\ell)}(iS)]
\label{eq:conjecture_etendue}
\end{multline}
Since $|\sin(S\beta\hbar\omega)/\sh(\beta\hbar\omega)|\leq S$, this does not raise new convergence issues and the numerical evaluation of
$\Delta B^{\rm conj}_{2,2}(\omega)$ and $\Delta B^{\rm conj}_{3,1}(\omega)$ is straightforward once the logarithmic derivatives of the determinant
of $M$ are known. The resulting value of the fourth in-trap cluster coefficient
\be
\Delta B_4(\omega) \equiv \frac{1}{2} [\Delta B_{3,1}(\omega)+\Delta B_{2,2}(\omega) + \Delta B_{1,3}(\omega)]
\label{eq:defDB4}
\ee
(with $\Delta B_{3,1}=\Delta B_{1,3}$ for the mass ratio $\alpha=1$) is plotted as a function of $\beta \hbar \omega$ in Fig.~\ref{fig:vsBlume}.
It clearly disagrees with the results of reference \cite{Blumeb4}, not only with the ones resulting from the extrapolation
to $\beta\hbar\omega=0$ but also with the actually calculated ones.

The conjecture is thus invalidated, and more theoretical work is needed to derive
the correct analytical expression for $\Delta b_4$ of the unitary Fermi gas.

\end{document}